%% file: main.tex
\RequirePackage{fix-cm}
\documentclass[aps,pra,reprint,superscriptaddress,longbibliography,floatfix]{revtex4-2}
\usepackage[T1]{fontenc}
\usepackage{amsmath,amssymb,amsthm,bm,graphicx,booktabs,mathptmx,xcolor}
\usepackage[colorlinks=true,linkcolor=blue!45!black,citecolor=blue!45!black,urlcolor=blue!45!black]{hyperref}
\newcommand{\STAB}{\operatorname{STAB}}
\newcommand{\id}{\operatorname{id}}
\newcommand{\tr}{\operatorname{Tr}}
\newcommand{\norm}[1]{\left\|#1\right\|}
\newcommand{\ket}[1]{|#1\rangle}
\newcommand{\bra}[1]{\langle#1|}
\newtheorem{theorem}{Theorem}
\hypersetup{pdftitle={Autonomous steady-state quantum magic under local noise},pdfauthor={Mingdi Xu and Lei Pan}}

\begin{document}
\title{Autonomous steady-state quantum magic under local noise}
\author{Mingdi Xu}
\affiliation{School of Physics, Nankai University, Tianjin 300071, China}
\author{Lei Pan}
\email{panlei@nankai.edu.cn}
\affiliation{School of Physics, Nankai University, Tianjin 300071, China}
\date{November 15, 2026}
\begin{abstract}
We determine the local reset cost of sustaining quantum magic under continuous depolarization. Static spin interactions convert reset-induced polarization into nonstabilizer correlations. For interactions involving at most $k$ spins and local coherent strength $J$, the optimal strong-noise threshold obeys $K_{\rm on}^{(k)}/\gamma=\Theta_k[(\gamma/J)^{1/(k-1)}]$, where $\gamma$ is the noise rate and $K$ the reset cap. The exponent is uniform in device size, and the two-body leading coefficient is $1/(2\sqrt2)$. The hierarchy follows from a coherent transition competing with simultaneous local defects. In weak noise, population amplification across growing devices lets the normalized onset cost vanish with static two-body interactions. Entropy balance separately constrains the size and total reset rate required for a fixed global resource. Explicit two-, three-, and four-body dynamics generate magic from free inputs and converge to a unique mixed steady state. A physical two-body construction connects this stationary resource to its coherent budget, control precision, and preparation time.
\end{abstract}
\maketitle

\section{Introduction}
Quantum magic supplies the nonstabilizer states needed to extend Clifford gates, Pauli measurements, and stabilizer ancillas to universal quantum computation~\cite{Gottesman1997,AaronsonGottesman2004,BravyiKitaev2005}. Magic-state distillation identifies usable states and noise thresholds~\cite{Reichardt2005,CampbellBrowne2009,CampbellBrowne2010,BravyiHaah2012,VanDamHoward2009}. Resource theories quantify states and operations~\cite{Veitch2014,ChitambarGour2019,HowardCampbell2017,SeddonCampbell2019,WangWildeSu2019}, while quasiprobability and stabilizer decompositions relate magic to classical simulation costs~\cite{Pashayan2015,BravyiGosset2016,Bravyi2019,Seddon2021}. For mixed states, the free set is the convex hull of all pure stabilizer states, including entangled ones~\cite{HeinrichGross2019,Warmuz2025}. Many-body studies examine its extensivity and the behavior of mixed-state monotones~\cite{LiuWinter2022,Leone2022,Rubboli2024mixedstate}. A continuously operating device must generate this resource while noise removes it.

The response of magic to noise depends on the state and channel. Hypergraph-state families have size-dependent noise thresholds~\cite{WeiLiu2024}, magic-breaking channels send every input into the stabilizer hull~\cite{Patra2024}, and dynamical studies find loss and revival of magic~\cite{Cao2026,Lyu2026}. Symmetry constrains the nonstabilizerness of X-states~\cite{DeOliveira2026Symmetry}. Measurements and injected resources add further dynamical settings: monitored circuits exhibit magic transitions~\cite{Niroula2024}, stabilizer entanglement changes injection performance~\cite{Hou2026}, and quantum trajectories admit perturbative stationary-response descriptions~\cite{Zhen2026WeakInjection}. For autonomous maintenance, a static Hamiltonian generates correlations while an engineered environment replenishes the polarization lost to noise.

Engineered dissipation prepares attractive entangled states~\cite{Kraus2008,Diehl2008,Verstraete2009,TicozziViola2012} and supports autonomous memories~\cite{Pastawski2011,Kapit2016}. Local reset models sustain noisy correlations~\cite{Hartmann2006,Brask2015,HaackJoye2021}. Steady magic has been obtained with thermal and spin baths~\cite{Mukhopadhyay2018}, non-Hermitian and Lindblad dynamics~\cite{MartinezAzcona2026}, and conversion of athermality~\cite{DeOliveira2026Thermal}. A local reset budget asks how much replenishment bounded interactions require. Increasing the size adds configurations over which population bias can accumulate, together with additional noise. Increasing the reset rate removes defects and damps resource-carrying coherence.

We study static $k$-body Hamiltonians with local strength budget $J$, depolarization rate $\gamma$ on every physical and auxiliary qubit, and stabilizer resets capped at $K$. The earlier onset analysis established the interaction-order law and sharp two-body coefficients in fixed-size or pure-target, common-rate growing classes~\cite{Xu2026Onset}. We place the complete size-uniform two-body bound alongside weak-noise size advantages, finite-rate constructions, and resource-cost bounds. A direct $k$-spin transition competes with $k$ simultaneous defects in strong noise. In weak noise, every fixed size retains a positive normalized threshold, while growing finite devices can make it vanish. Entropy supply and physical implementation then determine the amount maintained. Three dynamical examples establish autonomous generation from free inputs.

\section{Magic and dissipative spin dynamics}
\subsection{The resource in a mixed state}
A pure stabilizer state is a common eigenstate of a complete commuting set of Pauli operators~\cite{Gottesman1997}. Their convex hull is
\begin{equation}
 \STAB_N=\operatorname{conv}\{\ket{\phi_s}\bra{\phi_s}\}.
 \label{eq:free}
\end{equation}
A density operator is magic outside this set. For one qubit the Bloch condition is $|r_x|+|r_y|+|r_z|>1$: $\ket+=(\ket0+\ket1)/\sqrt2$ is free, while $(\ket0+e^{i\pi/4}\ket1)/\sqrt2$ is magic. Bell states are entangled and free, so coherence or entanglement alone does not diagnose magic.

For the dynamical examples we use the robustness of magic~\cite{HowardCampbell2017},
\begin{equation}
 \mathcal R(\rho)=\min_{\rho=\sum_s a_s\ket{\phi_s}\bra{\phi_s}}
                    \sum_s|a_s|,\qquad \mathcal M(\rho)=\mathcal R(\rho)-1.
 \label{eq:robustness}
\end{equation}
The real coefficients sum to one; hence $\mathcal M=0$ exactly on $\STAB_N$, including for mixed inputs. Mixed-state faithfulness is needed under noise~\cite{Leone2022,Warmuz2025}.

For resource bounds we also use the trace distance
$d_N(\rho)=\frac12\inf_{\sigma\in\STAB_N}\norm{\rho-\sigma}_1$.
The trace distance has an operational distinguishability meaning~\cite{FuchsVandeGraaf1999}. We plot $\mathcal M$ in the dynamics and use a fixed target $d_N\ge d_0$ for resource costs; the two measures have different numerical and size dependence.

\subsection{Hamiltonian, noise, and resets}
Consider $N$ spin-$1/2$ degrees of freedom with a static Hamiltonian
\begin{equation}
 H=\sum_{1\le|X|\le k}h_X,\qquad h_X=h_X^\dagger,
 \qquad \max_i\sum_{X\ni i}\norm{h_X}_\infty\le J.
 \label{eq:H}
\end{equation}
The support $X$ contains at most $k\ge2$ sites. Terms with the same support are grouped before charging their norm. The per-site budget $J$ allows arbitrary graphs, including nearest-neighbor chains. Every auxiliary spin counts toward $N$ and experiences the same noise.

Every spin undergoes natural depolarization at rate $\gamma>0$. An engineered reset at site $i$ replaces it by a fixed $\tau_i\in\STAB_1$ at rate $\kappa_i\ge0$. The density operator obeys the Lindblad equation~\cite{GKLS1976,Lindblad1976}
\begin{equation}
 \dot\rho=\mathcal L\rho=-i[H,\rho]
 +\sum_i\bigl[\gamma(\mathcal T_i-\id)+\kappa_i(\mathcal R_i-\id)\bigr]\rho,
 \label{eq:model}
\end{equation}
where $\mathcal T_i\rho=(I_i/2)\otimes\tr_i\rho$ and $\mathcal R_i\rho=\tau_i\otimes\tr_i\rho$, with the tensor factors returned to their original positions. Equivalently, the depolarizing jumps are $\sqrt{\gamma/4}\,X_i$, $\sqrt{\gamma/4}\,Y_i$, and $\sqrt{\gamma/4}\,Z_i$. For $\tau_i=\ket0\bra0$, reset jumps are $\sqrt{\kappa_i}\ket0\bra0_i$ and $\sqrt{\kappa_i}\ket0\bra1_i$. These jump operators fix the rate convention used below.

The local cap is $K=\max_i\kappa_i$, the total reset rate is $R=\sum_i\kappa_i$, and $y=\gamma/J$, $u=K/\gamma$. The cap bounds each site's reset bandwidth; the total rate also counts the number of resetting spins.

The stationary state satisfies $\mathcal L\rho_*=0$. Natural depolarization on every site gives a unique, full-rank $\rho_*$ at every finite size, and
\begin{equation}
 \lim_{t\to\infty}e^{t\mathcal L}\rho_0=\rho_*
 \quad\text{for every initial density operator }\rho_0.
 \label{eq:attraction}
\end{equation}
Irreducibility and relaxation criteria give Eq.~\eqref{eq:attraction}~\cite{Evans1977,Spohn1977,Frigerio1978}; the Supplemental Material gives a direct proof~\cite{SupplementalMaterial}. It holds for every density operator, with the convergence time determined separately for each finite device.

With no resets, the unique stationary state is $I/2^N$, independent of $H$. With $H=0$, it is the free product
\begin{equation}
 \rho_*^{(0)}=\bigotimes_i\theta_i,
 \qquad \theta_i=\frac{\gamma I/2+\kappa_i\tau_i}{\gamma+\kappa_i}.
 \label{eq:reference}
\end{equation}
Noise and resets preserve the stabilizer set. Their competition with coherent dynamics determines whether the stationary state leaves that set.

\section{Conditions for a magic steady state}
A Hamiltonian commuting with Eq.~\eqref{eq:reference} retains a free steady state. We therefore seek exclusion conditions covering every admissible controller and sufficient conditions attained by explicit Hamiltonians.

\begin{theorem}[Finite-rate exclusion and construction]
\label{thm:finite}
For the model in Eqs.~\eqref{eq:H} and \eqref{eq:model}, put
$c_k=[64k(21e)^k]^{-1}$. Every finite device has a free stationary state if
\begin{equation}
 \frac J\gamma\left(1+\frac K\gamma\right)^{k-1}\le c_k.
 \label{eq:finite-free}
\end{equation}
Conversely, a $k$-spin module with common resets to $\ket0$, interaction strength $0<h\le J$, and
\begin{equation}
 H_k=\frac h{\sqrt2}(X_1+Y_1)X_2\cdots X_k
 \label{eq:module}
\end{equation}
has a magic stationary state whenever
\begin{equation}
 G_k=\frac{\sqrt2h}{k(\gamma+\kappa)}(p_0-p_k)-p_k>0.
 \label{eq:module-condition}
\end{equation}
Here $p_0$ and $p_k$ are its exact stationary populations of $\ket{0^k}$ and $\ket{1^k}$, given below.
\end{theorem}

The exclusion covers mixed targets, unequal or zero reset rates, and arbitrary finite graphs. Its conservative constant gives a guaranteed free regime; the module gives a sufficient magic condition. The proof combines local perturbative expansions~\cite{KirkwoodThomas1983,Bravyi2008,Yarotsky2006,DeRoeckSchutz2016,Crawford2015} with connected-cluster estimates~\cite{KoteckyPreiss1986,BovierZahradnik2000,FernandezProcacci2007,ScottSokal2005,MannMinko2024}. Related work controls locality and stability of open-system steady states~\cite{BarthelKliesch2012,KastoryanoEisert2013,Cubitt2015,Bertini2025,Temme2010}, quasi-local Gibbs descriptions of weakly interacting channels~\cite{Arad2025}, and high-temperature Gibbs-state and high-noise dynamical thresholds~\cite{Kliesch2014,Bakshi2024,Putterman2026,Kiani2026,TrivediCirac2022,GonzalezGarcia2025}.

The module is directly computable at finite rates. With $a=\gamma+\kappa$, its stationary $Z$-moments obey
\begin{equation}
 m_0=1,\qquad
 m_\ell=\frac{\ell\kappa\,m_{\ell-1}}
 {\ell a+[4h^2/(ka)]\mathbf1_{\ell\ {\rm odd}}},
 \label{eq:moments}
\end{equation}
and $p_0=2^{-k}\sum_{\ell=0}^k\binom{k}{\ell}m_\ell$,
$p_k=2^{-k}\sum_{\ell=0}^k(-1)^\ell\binom{k}{\ell}m_\ell$.
The endpoint coherence is proportional to $(p_0-p_k)/[ka]$, with equal real and imaginary quadratures. Clifford syndrome decoding converts $G_k>0$ into a stabilizer-facet violation while all intermediate configurations remain populated. Decoding diagnoses the state. The stationary elimination giving Eq.~\eqref{eq:moments} applies to the steady state.

For two spins this condition reduces to
\begin{equation}
 \kappa>\frac{\gamma^2+2h^2}{2\sqrt2h}.
 \label{eq:pair-threshold}
\end{equation}
Optimizing $h\le J$ gives the attainable upper envelope $\gamma$ for $J\ge\gamma/\sqrt2$ and $(\gamma^2+2J^2)/(2\sqrt2J)$ otherwise.

To compare controllers, let $K_{\rm on}^{(k,N)}(J,\gamma)$ be the infimum of $K$ for which an allowed $N$-spin steady state has $d_N>0$, and let $K_{\rm on}^{(k)}=\inf_N K_{\rm on}^{(k,N)}$. The infimum allows the interaction graph, reset targets, and rate allocation to vary. Equation~\eqref{eq:finite-free} gives
\begin{equation}
 K_{\rm on}^{(k)}\ge\gamma\bigl[(c_k\gamma/J)^{1/(k-1)}-1\bigr]_+.
 \label{eq:onset-lower}
\end{equation}
For each fixed finite $N$, purity balance~\cite{Sauer2013} gives a Hamiltonian-independent lower bound $K_{\rm on}^{(k,N)}\ge\gamma a_N$ with $a_N>0$. Complete two-qubit stabilizer geometry~\cite{Reichardt2009} strengthens this to $K/\gamma>13/20$, regardless of coherent strength. The Supplemental Material specifies $a_N$ and the exact geometric certificate.

\section{Noise limits and their physical origin}
The limiting parameter is the noise-to-coherent-budget ratio $\gamma/J$.

\begin{theorem}[Strong- and weak-noise onset]
\label{thm:strong}\label{thm:weak}
For each fixed interaction order $k\ge2$, the optimal onset over all finite sizes satisfies
\begin{equation}
 K_{\rm on}^{(k)}=
 \Theta_k\!\left(\gamma^{k/(k-1)}J^{-1/(k-1)}\right),
 \qquad \gamma/J\to\infty.
 \label{eq:strong}
\end{equation}
For two-body interactions the coefficient is sharp and uniform in size:
$K_{\rm on}^{(2)}=[1/(2\sqrt2)+o(1)]\gamma^2/J$.
For fixed $N$ and $k\ge2$, weak noise instead gives
\begin{equation}
 K_{\rm on}^{(k,N)}=\gamma[c_{k,N}+o(1)],
 \qquad c_{k,N}>0,
 \label{eq:weak}
\end{equation}
whereas optimizing over finite sizes gives
$K_{\rm on}^{(k)}/\gamma\to0$ as $\gamma/J\to0$.
\end{theorem}

A $k$-spin module attains the size-uniform exclusion exponent, including when the optimization permits arbitrarily many noisy auxiliaries. The full-class coefficient is sharp for $k=2$; the $k=3,4$ module coefficients are constructive upper bounds. Two-body weak-noise constructions are admissible for every $k\ge2$, establishing the vanishing size-optimized threshold throughout these classes.

\begin{figure}[tbp]
\centering\includegraphics[width=\columnwidth]{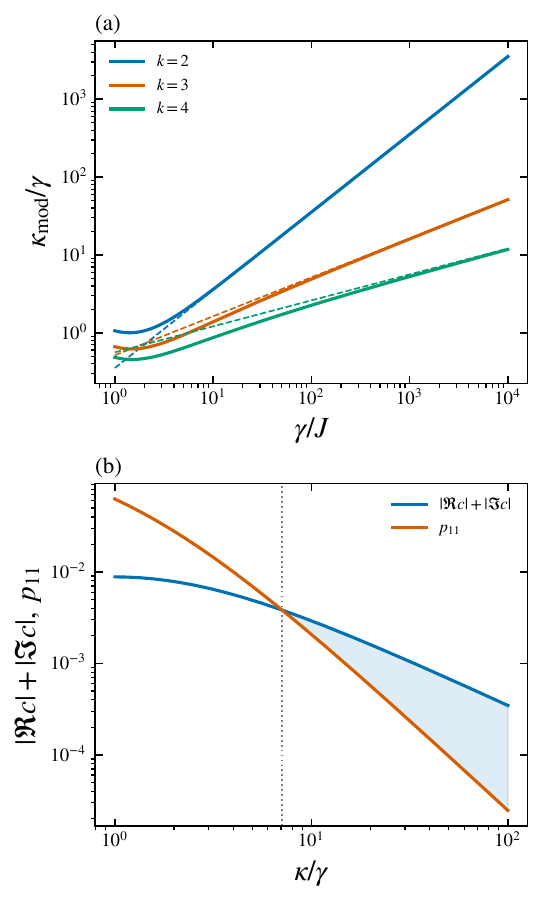}
\caption{Strong-noise mechanism. (a) Exact endpoint-witness onset of the modules in Eq.~\eqref{eq:module}, with $h=J$, for $k=2,3,4$. Dashed curves are their large-noise power laws, with exponents $1$, $1/2$, and $1/3$. They are attainable family thresholds; only the two-body leading coefficient is a full-class optimum. (b) Exact two-body stationary coherence quadratures and minority population at $J/\gamma=0.05$. Both shrink as the reset rate grows, but their crossing produces a positive stabilizer-facet violation. Shading marks the interval with $|\Re c|+|\Im c|>p_{11}$. AI-assisted plotting code renders the analytical results.}
\label{fig:strong}
\end{figure}

\input{physical_sections.tex}

Within the two-body class, $c_{2,2}\ge13/20$ and an endpoint construction gives $c_{2,3}<0.589$. The lower bound covers every two-spin Hamiltonian in its budget. A finite three-spin controller sustains magic at $K/\gamma=0.64$, $J/\gamma=25$; its Hamiltonian and exact certificate are supplied. These are attained values, with the complete three-spin optimum remaining open.

\begin{figure}[tbp]
\centering\includegraphics[width=\columnwidth]{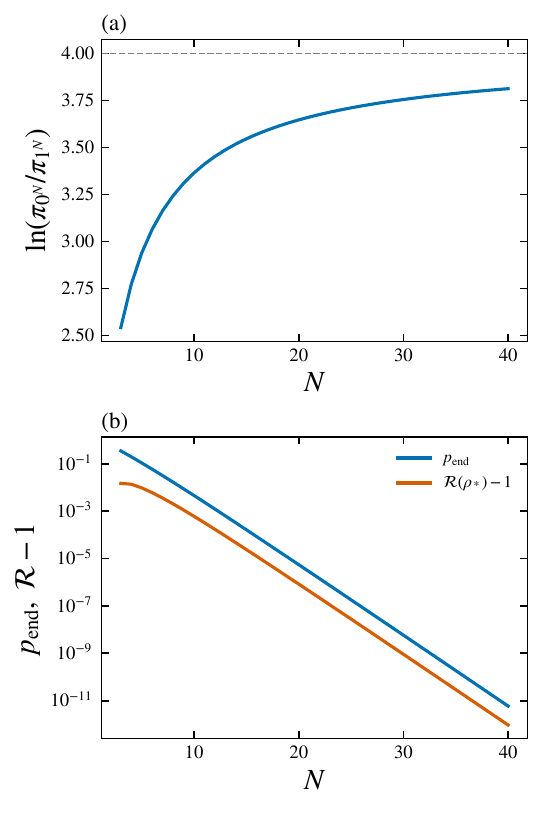}
\caption{Population contrast and resource weight are different costs. The endpoint construction uses $\kappa=2\gamma/N$. (a) The uncoupled endpoint population ratio approaches the finite value $e^4$. (b) The probability of the resonant endpoint sector and the exact robustness excess of the driven stationary state both decrease exponentially with size. A larger logical population contrast therefore need not imply a larger unconditional resource output. The curves use the exact endpoint solution with its phase and witness-optimized coupling; all intermediate populations are retained. AI-assisted plotting code renders the saved data.}
\label{fig:weak}
\end{figure}

\section{Autonomous generation from initially free states}
We evolve three devices at $\gamma=J$, with common resets $\kappa=2\gamma$ to $\ket0$, to resolve how a stationary resource develops from free inputs.

The two-body example is a connected, open three-spin chain,
\begin{equation}
 H_2=\frac{0.9J}{\sqrt2}(X_1X_2+Y_1X_2)+0.1JZ_2Z_3.
 \label{eq:chain-example}
\end{equation}
Its site budgets are $(0.9J,J,0.1J)$: the first bond generates pair coherence and the second connects it to a third noisy spin. The other devices are Eq.~\eqref{eq:module} at $N=k=3,4$, $h=J$, with contiguous support and norm $J$. Their different sizes and Hamiltonians illustrate generation in each interaction class.

\begin{figure}[tbp]
\centering\includegraphics[width=\columnwidth,height=0.55\textheight,keepaspectratio]{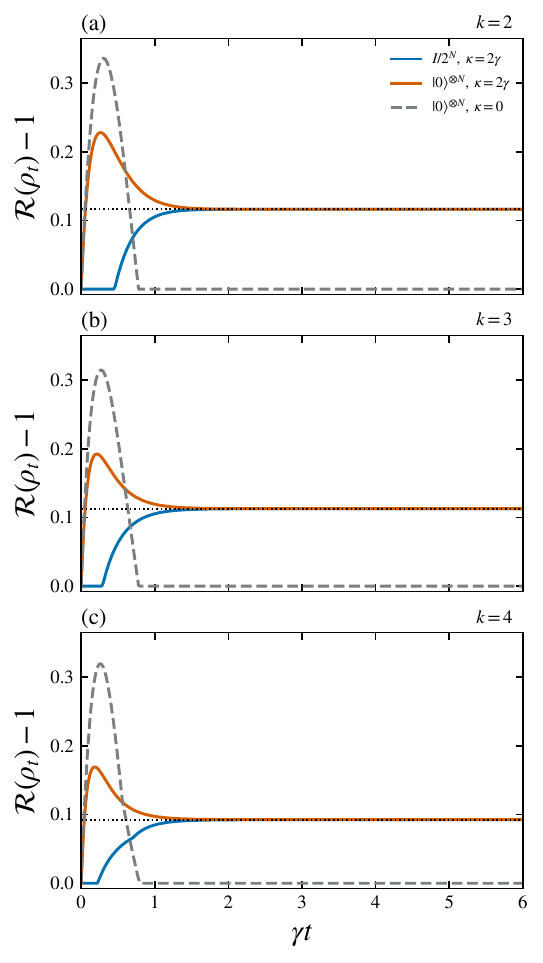}
\caption{Magic generated from free initial states by the complete Lindblad dynamics. (a) Nearest-neighbor two-body chain, Eq.~\eqref{eq:chain-example}, with $N=3$. (b,c) Three- and four-body modules, Eq.~\eqref{eq:module}, with $N=k$. All panels have $\gamma=J$. At $\kappa=2\gamma$, inputs $I/2^N$ (blue) and $\ket0^{\otimes N}$ (orange) approach the same mixed magic steady state (black dotted). With no resets (gray dashed), magic is transient. The ordinate is the full mixed-state robustness excess, Eq.~\eqref{eq:robustness}. AI-assisted plotting code renders the calculated trajectories.}
\label{fig:dynamics}
\end{figure}

Both $I/2^N$ and $\ket{0}\bra{0}^{\otimes N}$ start with zero magic. The mixed input first polarizes; the pure stabilizer input can produce an early overshoot. They approach the same mixed state, with $\mathcal M(\rho_*)\simeq0.116$, $0.113$, and $0.093$ in the two-, three-, and four-body examples.

With $\kappa=0$, coherent motion generates transient magic from the pure input, followed by decay to $I/2^N$. Resets sustain the polarization supply and a positive stationary resource. Removing the interactions leaves local stabilizer-preserving dynamics. The nonmonotone traces reflect coherent generation together with noise and reset-induced coherence loss.

The three-spin chain uses all $1080$ stabilizer vertices. For the modules, diagonal populations and complement coherences give an exact robustness formula by Clifford decoding into syndrome-labelled qubits, cross-checked against the complete $1080$- and $36720$-vertex optimizations. The Supplemental Material contains the formulas, convergence distances, and longer traces. Equation~\eqref{eq:attraction} establishes initial-state independence. OpenAI Codex assisted with analytical development, numerical implementation, and manuscript preparation under author direction; numerical validation procedures and exact certificates are specified with the accompanying calculations.

\section{Resource amount, reset expenditure, and control costs}
\subsection{Entropy supply limits a stationary resource}
Onset requires any separation from the stabilizer set. A fixed amount also requires entropy removal: the coherent evolution conserves von Neumann entropy. Heat-bath algorithmic cooling~\cite{Boykin2002,Schulman2005} and finite-size Landauer bounds~\cite{ReebWolf2014} provide related entropy-accounting settings.

Combining depolarization and reset into the replacement with target $\theta_i$ in Eq.~\eqref{eq:reference} gives an exact stationary entropy balance. Together with strong subadditivity it bounds the relative-entropy magic $M_{\rm rel}(\rho)=\inf_{\sigma\in\STAB_N}D(\rho\Vert\sigma)$, using natural logarithms. At small local reset cap $u=K/\gamma$,
\begin{equation}
 \frac{M_{\rm rel}(\rho_*)}{N}\le\frac{u^2}{2}+O(u^3).
 \label{eq:density}
\end{equation}
The bound is independent of coherent strength and geometry. It follows from reset entropy-production identities and product-reference tensorization~\cite{HaackJoye2024,Capel2018}. Coherence measures~\cite{Baumgratz2014,Scalici2024Coherence} and skew information~\cite{WignerYanase1963,Lieb1973,Yu2017,Takagi2019} describe complementary coherent constraints; Eq.~\eqref{eq:density} isolates dissipative supply.

For a fixed target distance $d_N\ge d_0$ with $0<d_0<1$, the same argument requires
\begin{equation}
 N=\Omega(u^{-2}),\qquad R/\gamma=\Omega(u^{-1})
 \quad (u\to0).
 \label{eq:necessary-costs}
\end{equation}
At common rates $u=C/N$, $M_{\rm rel}\le C^2/(2N)+O(N^{-2})$ and $d_N\le C/[2\sqrt N]+O(N^{-3/2})$, ruling out fixed global distance along this single-device path. The endpoint source in Fig.~\ref{fig:weak} has the stronger, construction-specific exponential loss.

\subsection{Typical configurations and a fixed global target}
For the complementary-configuration source of Sec.~\ref{sec:physics-weak}, take $v=u/(1+u)$, $N=O(u^{-2}\ln(1/u))$, and magnetizations $M\ge Nv-2\sqrt N$. These sectors retain typical probability while reducing contamination from reversed populations, giving witness size $\Omega(v/\sqrt N)$.

An inhomogeneous nearest-neighbor Ising chain realizes the source with the first transverse-field phase fixed at $\pi/4$. Exact Gaussian dressing controls the complement coupling and the change of every original local dissipator. Staggered bonds separate directly coupled sectors, and the first dressing transformation is local. Effective-Hamiltonian constructions~\cite{BravyiSW2011,JordanFarhi2008,OliveiraTerhal2008} and open-system effective-operator methods~\cite{ReiterSorensen2012} provide related implementation tools; here the physical support norms and original noise are charged explicitly.

Independent copies then attain fixed global distance. A sequential test includes stabilizer states entangled across copies, since conditioning on stabilizer measurements leaves unused registers free. Restricted measurements and generalized hypothesis testing~\cite{Piani2009,Lami2025} and concentration bounds~\cite{Hoeffding1963,Azuma1967} provide the testing framework. Table~\ref{tab:costs} lists simultaneous sufficient costs; $N$ counts one source and $N_{\rm tot}$ counts all copies.

\begin{table}[t]
\caption{One explicit two-body construction for a fixed global target $d_{N_{\rm tot}}\ge d_0$, $0<d_0<1$, as $u=K/\gamma\to0$. Write $\ell=\ln(1/u)$. All entries are sufficient costs and tolerances for this construction. The preparation time starts from the free state $I/2^N$.}
\label{tab:costs}
\begin{ruledtabular}\begin{tabular}{ll}
Quantity & Sufficient scaling\\
Source size $N$ & $O(u^{-2}\ell)$\\
Independent copies & $O(u^{-4}\ell)$\\
Total source qubits $N_{\rm tot}$ & $O(u^{-6}\ell^2)$\\
Total reset rate $R_{\rm tot}/\gamma$ & $O(u^{-5}\ell^2)$\\
Preparation time $\gamma T$ & $O(\ell^2)$\\
Coherent budget $\ln(J/\gamma)$ & $2N\ln N+O(N\ln\ln N)$\\
Sitewise longitudinal tolerance & $\Theta(\gamma\ell^{-3/4})$
\end{tabular}\end{ruledtabular}
\end{table}

Complement coherence has linewidth $N(\gamma+\kappa)$, and the phase-aligned witness responds quadratically to longitudinal detuning. The resulting absolute tolerance is listed in Table~\ref{tab:costs}; resolving it relative to the large Ising scale requires $O(N\ln N)$ binary digits. Physical noise, precision, and finite preparation time obey the stated costs simultaneously.

For the ideal complementary-configuration source, a diagonal free state has fidelity at least $1-O(s^2)$, where $s=\lambda/[N(\gamma+\kappa)]$. Hence $d_{mN}(\rho_*^{\otimes m})\le s\sqrt m$ requires $m=\Omega(s^{-2})$, matching the sufficient copy order for this source. This lower bound applies to the specified ideal source; the witness-preserving physical approximation requires its own copying analysis.

The necessary bounds and Table~\ref{tab:costs} leave a gap in optimal size and total reset expenditure. A more efficient source could reduce it. High-fidelity single-qubit outputs and extraction flux are further operational tasks beyond fixed global distance.

\section{Conclusion and outlook}
Static interactions and local stabilizer resets generate and maintain a mixed magic steady state under depolarization on every qubit. Every initial state reaches this unique finite-device attractor. General finite-rate exclusion and explicit constructions delimit its onset.

Interaction order fixes the strong-noise exponent through the competition between coherent transitions and local defects. In weak noise, size amplifies population bias and lowers the local onset cost. Sector weight controls the amount produced, with entropy supply setting separate size and total-rate requirements.

Improving the physical source, reducing its coherent budget, and quantifying extraction into useful magic-state outputs would bring autonomous maintenance closer to continuous non-Clifford processing.

\section*{Data availability}
The dynamical data, exact certificates, and computational code supporting this article accompany the Supplemental Material.
\bibliography{references}
\end{document}

\title{\texorpdfstring{Supplemental Material\\[3pt]\large Autonomous steady-state quantum magic under local noise}{Supplemental Material: Autonomous steady-state quantum magic under local noise}}
\author{Mingdi Xu}\author{Lei Pan}\affiliation{School of Physics, Nankai University, Tianjin 300071, China}
\date{November 15, 2026}
\maketitle

We use the full stabilizer hull $\STAB_N$, the signed robustness $\mathcal R$, and $d_N(\rho)=\frac12\inf_{\sigma\in\STAB_N}\|\rho-\sigma\|_1$. Each finite device includes all physical and auxiliary spins in the noise and local coherent budgets. The interaction order $k$ and size $N$ are independent. In the two-body sections, $F_N(y)=K_{\rm on}^{(N)}/\gamma$ and $F_\infty(y)=\inf_{N\ge2}F_N(y)$, with $y=\gamma/J$. General-order optimization is denoted by $K_{\rm on}^{[\le k]}$. References use the numbering of the accompanying article.

\input{proofs/general_k.tex}
\input{proofs/finite_k_modules.tex}
\input{proofs/optimization.tex}
\input{proofs/strong_two_body.tex}
\input{proofs/weak_fixed_size.tex}
\input{proofs/two_qubit_certificate.tex}
\input{proofs/endpoint_threshold.tex}
\input{proofs/endpoint_implementation.tex}
\input{proofs/weak_size_advantage.tex}
\input{proofs/finite_triangle_certificate.tex}
\input{proofs/finite_ratio.tex}
\input{proofs/entropy_budget.tex}
\input{proofs/endpoint_resources.tex}

\section{Continuous-time dynamical examples and their resource measure}
\label{sec:sm-dynamics}
\input{proofs/dynamics_k2.tex}
\input{proofs/dynamics_k34.tex}

\input{proofs/allsector_source.tex}
\input{proofs/boundary_source.tex}
\input{proofs/copy_lower_bound.tex}
\input{proofs/witness_poisson.tex}

\section{Exact Gaussian normal forms and the physical two-body controller}
\label{sec:sm-gaussian-device}
\input{proofs/gaussian_foundation.tex}
\input{proofs/staggered_bridge.tex}

The static controller for either Gaussian normal form is
\begin{align}
 H_{\rm phys}&=\Lambda V_ZH_\eta V_Z^\dagger\\
 &=\Lambda\left[-\sum_iJ_iZ_iZ_{i+1}
 +\eta\left(\frac{X_1+Y_1}{\sqrt2}+\sum_{i=2}^NX_i\right)\right],\\
 V_Z&=e^{-i\pi Z_1/8},\qquad \Lambda=\lambda/s_0,
 \qquad U_Z=V_ZUV_Z^\dagger.
 \label{eq:sm-actual-controller}
\end{align}
Then $U_Z^\dagger H_{\rm phys}U_Z=F+H_s$, with
$F=-\Lambda\sum_i\widetilde J_iZ_iZ_{i+1}$ and
$H_s=\lambda V_ZG_NV_Z^\dagger$. Its complement matrix elements have
modulus $\lambda$ and odd-$\pi/4$ phases, and $F_x=F_{\bar x}$.
The first transverse-field direction supplies the phase as part of the
static Hamiltonian. Natural depolarization and the physical $Z$ resets
are retained; they are covariant under $V_Z$.

\input{proofs/local_noise_bridge.tex}
\input{proofs/static_robustness.tex}
\input{proofs/preparation_time.tex}
\input{proofs/dressing_barrier.tex}

%% file: physical_sections.tex
\subsection{Strong noise: coherent transitions and local defects}
\label{sec:physics-strong}

At $\gamma\gg J$, the attaining modules approach a product of pure stabilizer states through resets faster than the natural noise. Resets remove minority populations and erase Hamiltonian-generated coherence. The two quantities approach the stabilizer boundary with different powers of the reset rate.

With common resets to $|0\rangle$, the uncoupled minority occupation is $\epsilon=\gamma/[2(\gamma+\kappa)]$. A configuration with $k$ minority spins has probability $\epsilon^k$. A $k$-body term connects it to the majority configuration in one transition, generating coherence of order $J/[k(\gamma+\kappa)]$. Thus
\begin{equation}
 \frac{J}{\kappa}\sim\left(\frac{\gamma}{\kappa}\right)^k,
 \qquad
 \frac{K_{\rm on}}{\gamma}
 =\Theta_k\!\left[\left(\frac{\gamma}{J}\right)^{1/(k-1)}\right].
 \label{eq:physics-support-balance}
\end{equation}
For fixed $k=2,3,4$, the powers are $1$, $1/2$, and $1/3$. Repeated lower-body transitions incur extra coherent factors under the same local budget. At onset, $K\gg\gamma\gg J$; both coherence and minority occupation vanish, with their ratio crossing a stabilizer facet. This is an onset mechanism at vanishing absolute resource.

For $H=J(X_1X_2+Y_1X_2)/\sqrt2$, equal transverse coherence quadratures optimize the octahedral condition after even-parity projection and Clifford decoding:
\begin{equation}
 K_{\rm on}=\left[\frac{1}{2\sqrt2}+o(1)\right]
 \frac{\gamma^2}{J}.
 \label{eq:physics-sharp-pair}
\end{equation}
The projection is a diagnostic. All incident pair interactions share the budget $J$. Below this coefficient, their leading correlations admit a positive mixture of entangled stabilizer states, while connected corrections carry extra defect factors. This yields a size-uniform two-body converse for arbitrary mixed targets and heterogeneous rates. For every fixed $k$, exclusion and construction match the exponent; the sharp coefficient is established for $k=2$.

\subsection{Weak noise: population amplification and resource weight}
\label{sec:physics-weak}

At $\gamma/J\ll1$, resonances can involve many configurations. A small reset bias then accumulates across the device. The coherent coupling converts population asymmetry into logical coherence and also mixes the populations supplying it, giving a finite useful coupling strength.

For a drive between $|0^N\rangle$ and $|1^N\rangle$, the uncoupled product distribution has
\begin{equation}
 \frac{\pi_{0^N}}{\pi_{1^N}}=(1+2u)^N
 \longrightarrow e^{2C},\qquad u=C/N.
 \label{eq:physics-endpoint-ratio}
\end{equation}
A suitable resonant phase turns this finite imbalance into a stabilizer-facet violation, with all intermediate populations retained. At fixed $C$ above onset, the sector weight is $2^{-N}[2\cosh C+o(1)]$. Its conditioned logical magic remains finite, while unconditional trace-distance magic is $\Theta(2^{-N})$. A syndrome measurement extracts that logical state with exponentially small probability, and fixed global resource requires $\Theta(2^N)$ independent modules in this family.

Driving many complement pairs $\{|x\rangle,|\bar x\rangle\}$ uses typical weight. The reference magnetization has mean $Nv$, width $O(\sqrt N)$, and $v=u/(1+u)$. Its typical complement ratio is $\exp[O(Nu^2)]$, since magnetization is of order $Nu$. A source of size $N=O(u^{-2}\ln(1/u))$ therefore retains substantial aggregate probability in useful sectors.

In the soluble source driving every complement pair with modulus $\lambda$, coherence decays at $Na$, $a=\gamma+\kappa$. Stationary elimination gives a global-complement jump with rate
\begin{equation}
 w=\frac{2\lambda^2}{Na}.
 \label{eq:physics-global-flip}
\end{equation}
Choosing $\lambda=O(av\sqrt N)$ keeps this magnetization-reversing rate of order $av^2$. Selecting sectors within $O(\sqrt N)$ of the positive mean suppresses recently reversed populations and gives an aggregate witness of order $v/\sqrt N$. A static two-body Ising chain realizes the analytical source by spectrum pairing and Gaussian dressing, retaining every physical noise channel. The construction has polynomial size and reset expenditure for fixed global magic, with a coherent budget exponential in $N\ln N$.

%% file: proofs/general_k.tex
% Condensed from the supplied 2026-11-08 proof. Fixed-point and positivity
% estimates are retained; related intermediate identities share equation groups.
\section{Global stabilizer decomposition}
\label{s:global:section}
\subsection{Model and statement}
\label{s:global:setup}
For finite \(N\), write
\(\mathcal R^{(i)}_\omega(\eta)=\omega_i\otimes\operatorname{Tr}_i\eta\)
with factors restored to their original order, and
\(\mathcal T_i=\mathcal R^{(i)}_{I/2}\). The model and budgets are
\begin{equation}
\begin{aligned}
\mathcal L&=-i[H,\cdot]+\gamma\sum_i(\mathcal T_i-\mathrm{id})\\
&\quad+\sum_i\kappa_i(\mathcal R^{(i)}_{\tau_i}-\mathrm{id}),\\
H&=\sum_{1\le |X|\le k}h_X,\qquad
\max_i\sum_{X\ni i}\|h_X\|_\infty\le J,\\
\gamma&>0,\quad 0\le\kappa_i\le K<\infty,\quad
\tau_i\in\STAB_1,\quad k\ge2.
\end{aligned}
\label{s:global:replacement}
\end{equation}
% alias: s:global:model -> s:global:replacement
% alias: s:global:reset-assumptions -> s:global:replacement
% alias: s:global:hamiltonian-budget -> s:global:replacement

The terms \(h_X\) are Hermitian, with arbitrary support geometry and coordination.
All auxiliary spins are included in \(N\) and charged identically.
Define
\begin{equation}
\begin{gathered}
x=J/\gamma,\quad A=1+K/\gamma,\quad z=xA^{k-1},\\
c_k=[64k(21e)^k]^{-1},\qquad Q_{a,s}=(I+sP_a)/2,
\end{gathered}
\label{s:global:parameters}
\end{equation}
% alias: s:global:pauli-projectors -> s:global:parameters

where \(a=x,y,z\) and \(s=\pm1\).

\begin{theorem}[Size-uniform free stationary region]
\label{s:global:free}
Every finite instance has a unique full-rank attracting stationary
state. If \(z\le c_k\), it has the fully separable stabilizer decomposition
\begin{equation}
\begin{gathered}
\rho_*=\sum_{\boldsymbol\sigma}p(\boldsymbol\sigma)\bigotimes_iQ_{\sigma_i},\\
p(\boldsymbol\sigma)>0,\qquad\sum_{\boldsymbol\sigma}p=1.
\end{gathered}
\label{s:global:positive-decomposition}
\end{equation}
Here \(\sigma_i\) runs over the six Pauli states; the constant is
uniform in size, graph, mixed targets and heterogeneous rates.
\end{theorem}

\subsection{Primitivity at every finite size}
\label{s:global:relaxation}
Use \(\mathcal D[L](\eta)=L\eta L^\dagger-
\{L^\dagger L,\eta\}/2\). A jump representation is
\begin{equation}
\begin{aligned}
\gamma(\mathcal T_i-\mathrm{id})
 &=\frac\gamma4\sum_{a=x,y,z}\mathcal D[P_a^{(i)}],\\
L^{\rm res}_{iuv}&=\sqrt{\kappa_i\lambda_{iu}}
 |u_i\rangle\langle v_i|\otimes I_{\bar i},
\end{aligned}
\label{s:global:dissipator}
\end{equation}
% alias: s:global:pauli-jumps -> s:global:dissipator
% alias: s:global:reset-jumps -> s:global:dissipator

where \(\tau_i=\sum_u\lambda_{iu}|u_i\rangle\langle u_i|\).

\begin{lemma}[Positivity improvement and relaxation]
\label{s:global:primitive}
For every \(t>0\), \(e^{t\mathcal L}\) maps each nonzero positive
operator to a strictly positive one. The stationary state is unique
and full rank, and every density operator converges to it.
\end{lemma}
\begin{proof}
Let \(G=-iH-\frac12\sum_\ell L_\ell^\dagger L_\ell\). Its positive,
norm-convergent jump expansion is
\begin{equation}
\begin{aligned}
e^{t\mathcal L}(\eta)
 &=\sum_{n,\boldsymbol\ell}\int_{0<t_1<\cdots<t_n<t}
 K_{\boldsymbol\ell}\eta K_{\boldsymbol\ell}^\dagger\,d\boldsymbol t,\\
K_{\boldsymbol\ell}
 &=e^{(t-t_n)G}L_{\ell_n}e^{(t_n-t_{n-1})G}\\
 &\quad\cdots L_{\ell_1}e^{t_1G},
\end{aligned}
\label{s:global:no-jump-generator}
\end{equation}
% alias: s:global:dyson -> s:global:no-jump-generator
% alias: s:global:trajectory -> s:global:no-jump-generator

For nonzero \(v,w\), the Pauli orbit of \(v\) spans the Hilbert
space. Invertibility of \(e^{tG}\) gives a Pauli string with
\begin{equation}
\langle w|e^{tG}P|v\rangle\ne0.
\label{s:global:pauli-orbit}
\end{equation}
Its at most \(N\) local factors are available natural jumps. Taking
their ordered times near zero gives a nonzero matrix element on a
positive-measure set; the corresponding positive integral has strictly
positive \(w\) expectation. Use the no-jump term if \(P=I\).
A general nonzero positive input dominates a rank-one input, proving
positivity improvement.

Time averages \(T^{-1}\int_0^T e^{t\mathcal L}(\rho)dt\) have stationary
limit points by compactness. Fix \(\mathcal E=e^{t_0\mathcal L}\),
\(D=2^N\). Positivity improvement and compactness give
\begin{equation}
\begin{gathered}
\varepsilon=\min_{\rho\ {\rm density}}
 \lambda_{\min}(\mathcal E\rho)>0,\qquad
\alpha=D\varepsilon/2\in(0,1/2],\\
\mathcal U(\eta)=\operatorname{Tr}(\eta)I/D,\qquad
\mathcal R=(\mathcal E-\alpha\mathcal U)/(1-\alpha).
\end{gathered}
\label{s:global:finite-minimum}
\end{equation}
% alias: s:global:positive-contraction-map -> s:global:finite-minimum

The map \(\mathcal R\) is positive and trace preserving, hence contracts
the trace norm of Hermitian inputs by their Jordan decomposition.
Consequently, for \(t=nt_0+s\), \(0\le s<t_0\),
\begin{equation}
\begin{aligned}
\|\mathcal E\rho-\mathcal E\sigma\|_1
 &\le(1-\alpha)\|\rho-\sigma\|_1,\\
\|e^{t\mathcal L}\rho-\rho_*\|_1
 &\le(1-\alpha)^n\|\rho-\rho_*\|_1\longrightarrow0.
\end{aligned}
\label{s:global:finite-contraction}
\end{equation}
% alias: s:global:finite-relaxation -> s:global:finite-contraction

This proves uniqueness and attraction; \(\rho_*=\mathcal E\rho_*>0\).
The constant \(\alpha\) depends on the finite instance.
\end{proof}
For trace-one \(\omega\), the traceless part of the stationary state is
\(-\mathcal L_{\rm tr=0}^{-1}\mathcal L\omega\). Relaxation makes this
inverse well defined, giving continuity in finite generator parameters
and analyticity along analytic families.

\subsection{Weighted creation coordinates and the exact equation}
\label{s:global:coordinates}\label{s:global:bch-section}
Combine replacements using
\begin{equation}
\begin{gathered}
a_i=\gamma+\kappa_i,\quad r_i=\gamma/a_i\in[A^{-1},1],\\
\theta_i=r_iI/2+(1-r_i)\tau_i,\quad\theta=\bigotimes_i\theta_i,\\
\Pi_i(\eta)=\theta_i\otimes\operatorname{Tr}_i\eta,\quad
\mathcal L_0=\sum_i a_i(\Pi_i-\mathrm{id}),\quad
\mathcal V=-i[H,\cdot].
\end{gathered}
\label{s:global:reference-state}
\end{equation}
% alias: s:global:unperturbed -> s:global:reference-state

Use \(B_{i0}=\theta_i\), \(B_{i\beta}=r_iP_\beta/2\).
For a colored support \((M,\boldsymbol\beta)\), let
\(B_{M\boldsymbol\beta}=\bigotimes_{i\in M}B_{i\beta_i}
\otimes\bigotimes_{i\notin M}\theta_i\). Its coefficient is denoted
\([\eta]_{M\boldsymbol\beta}\), with
\([\eta]_\varnothing=\operatorname{Tr}\eta\). Then
\begin{equation}
\begin{gathered}
r_M=\prod_{i\in M}r_i,\quad a_M=\sum_{i\in M}a_i,\\
\mathcal L_0B_{M\boldsymbol\beta}=-a_MB_{M\boldsymbol\beta},
\qquad a_M\ge\gamma|M|.
\end{gathered}
\label{s:global:weighted-basis}
\end{equation}
% alias: s:global:global-basis -> s:global:weighted-basis
% alias: s:global:decay-denominator -> s:global:weighted-basis

The local creator \(b_{i\beta}(\eta)=B_{i\beta}\operatorname{Tr}\eta\),
extended by the identity off \(i\), creates from the vacuum and kills
every excitation. Thus
\begin{equation}
\begin{gathered}
b_{i\beta}b_{i\delta}=0,\qquad[b_{i\beta},b_{j\delta}]=0,\\
b_{M\boldsymbol\beta}=\prod_{i\in M}b_{i\beta_i},\\
S=\sum_{M\ne\varnothing,\boldsymbol\beta}
s_{M\boldsymbol\beta}b_{M\boldsymbol\beta},\qquad S^{N+1}=0.
\end{gathered}
\label{s:global:creator-local}
\end{equation}
% alias: s:global:creator-algebra -> s:global:creator-local
% alias: s:global:creation-polynomial -> s:global:creator-local

\begin{lemma}[Creation parametrization and locality]
\label{s:global:parametrization}\label{s:global:bch}
Every trace-one operator has a unique representation \(e^S\theta\),
and \(\operatorname{Tr}(e^{\pm S}\eta)=\operatorname{Tr}\eta\).
In a nonzero nested commutator of \(\mathcal V_X\) with creators on
\(T_1,\ldots,T_n\), every \(T_j\) touches \(X\), creator supports
on each side are disjoint, and all supports are disjoint outside \(X\).
Its output obeys
\begin{equation}
M\setminus X=\bigcup_j(T_j\setminus X),\qquad
M\subseteq X\cup\bigcup_jT_j.
\label{s:global:output-support}
\end{equation}
The BCH series terminates at degree \(2|X|\).
\end{lemma}
\begin{proof}
For \(\eta=(\mathrm{id}+C_\eta)\theta\), the commutative nilpotent
creation algebra has the inverse logarithm
\begin{equation}
S=\log(\mathrm{id}+C_\eta)
=\sum_{j=1}^N(-1)^{j+1}C_\eta^j/j.
\label{s:global:nilpotent-log}
\end{equation}
Injectivity of \(C\mapsto C\theta\) proves uniqueness. Every nonempty
creator is traceless, giving the trace identity. A creator disjoint
from \(X\) commutes with the interaction and can be moved innermost
to annihilate the nested commutator. Expand the rest into left--right
products. Nilpotence permits at most \(|X|\) disjoint creators on
either side. Outside \(X\), any overlap vanishes and every surviving
excitation is created once, proving the support identities.
\end{proof}
Use \(\operatorname{ad}_C(D)=[C,D]\). Since
\([\mathcal L_0,b_M]=-a_Mb_M\), the exact conjugations are
\begin{equation}
\begin{aligned}
e^{-S}\mathcal L_0e^S&=\mathcal L_0-\sum_{M,\boldsymbol\beta}
a_Ms_{M\boldsymbol\beta}b_{M\boldsymbol\beta},\\
e^{-S}\mathcal Ve^S&=\sum_{n=0}^{2k}\operatorname{ad}_{-S}^n(\mathcal V)/n!.
\end{aligned}
\label{s:global:unperturbed-commutator}
\end{equation}
% alias: s:global:unperturbed-bch -> s:global:unperturbed-commutator
% alias: s:global:finite-bch -> s:global:unperturbed-commutator

Thus the nonempty stationary rows are
\begin{equation}
s_{M\boldsymbol\beta}=a_M^{-1}
[e^{-S}\mathcal Ve^S\theta]_{M\boldsymbol\beta}.
\label{s:global:kt-equation}
\end{equation}
The scalar row vanishes by trace preservation.

\subsection{Heterogeneous-rate and rooted estimates}
\label{s:global:rate-estimates}\label{s:global:rooted-section}
The dual functionals are \(d_{i0}=\operatorname{Tr}\) and
\begin{equation}
d_{i\beta}(\eta)=r_i^{-1}\operatorname{Tr}[(P_\beta-u_{i\beta}I)\eta],
\quad u_{i\beta}=\operatorname{Tr}(P_\beta\theta_i).
\label{s:global:dual-basis}
\end{equation}
Their norms are at most \(1,2/r_i\). A local input on \(T\subseteq X\)
has trace norm \(r_T\), so
\begin{equation}
|(\mathcal V_X)_{U\boldsymbol\beta,T\boldsymbol\delta}|
\le2^{|U|+1}\|h_X\|_\infty r_T/r_U.
\label{s:global:local-matrix-element}
\end{equation}

\begin{lemma}[Rate saving and rooted bound]
\label{s:global:purity-saving}\label{s:global:rooted}
For a nonempty source output \(U\subseteq X\),
\begin{equation}
r_U^{-1}/a_U\le A^{|U|-1}/\gamma\le A^{k-1}/\gamma.
\label{s:global:source-saving}
\end{equation}
At positive degree the local absolute column sum, before the \(2^n\)
left--right count, is at most
\begin{equation}
v'_X=2\,7^kA^{k-1}\|h_X\|_\infty.
\label{s:global:nonlinear-column}
\end{equation}
Define
\begin{equation}
\|S\|_\mu=\max_i\sum_{M\ni i,\boldsymbol\beta}
 |s_{M\boldsymbol\beta}|\mu^{|M|},\qquad\mu\ge1,
\label{s:global:creator-norm}
\end{equation}
and the multilinear maps
\begin{equation}
\begin{aligned}
\mathfrak B_n(S_1,\ldots,S_n)
&=\sum_{M\ne\varnothing,\boldsymbol\beta}
\frac{[\operatorname{ad}_{-S_1}\cdots\operatorname{ad}_{-S_n}
(\mathcal V)\theta]_{M\boldsymbol\beta}}{a_M}b_{M\boldsymbol\beta},\\
\mathfrak F(S)&=\sum_{n=0}^{2k}\mathfrak B_n(S,\ldots,S)/n!,
\end{aligned}
\label{s:global:multilinear-map}
\end{equation}
% alias: s:global:fixed-point-map -> s:global:multilinear-map

using \(\mathcal V\theta\) at degree zero. For
\(\epsilon_\mu=2(7\mu)^kz\),
\begin{equation}
\begin{aligned}
\|\mathfrak B_0\|_\mu&\le\epsilon_\mu,\\
\|\mathfrak B_n(S_1,\ldots,S_n)\|_\mu
&\le\epsilon_\mu2^n[k^n+n(k+1)k^{n-1}]
       \prod_j\|S_j\|_\mu.
\end{aligned}
\label{s:global:source-norm}
\end{equation}
% alias: s:global:rooted-bound -> s:global:source-norm

All constants are independent of size.
\end{lemma}
\begin{proof}
For the source, put \(b_i=a_i/\gamma\in[1,A]\).
Then \(r_U^{-1}/a_U=\prod_Ub_i/(\gamma\sum_Ub_i)\le
A^{|U|-1}/\gamma\); the empty output vanishes by the commutator trace.
For a surviving nonlinear term let \(R,L\subseteq X\) be right and
left occupied sets and \(U\) the output before left creation. Then
\begin{equation}
U\cap L=\varnothing,\qquad
\frac{r_Rr_L}{r_{U\cup L}}=\frac{r_R}{r_U}\le A^{|X|-1}.
\label{s:global:left-vacancy}
\end{equation}
% alias: s:global:exact-scale-ratio -> s:global:left-vacancy

If \(R\ne\varnothing\), the exponent is at most \(|U\setminus R|\);
otherwise \(L\ne\varnothing\) forces \(|U|\le|X|-1\).
Outside factors cancel. Summing the local bound over output colors
and supports costs \(2\sum_{U\subseteq X}6^{|U|}=2\,7^{|X|}\),
proving the rate saving.

For the source anchored at \(i\), \(X\ni i\), hence
\begin{equation}
\|\mathfrak B_0\|_\mu
\le2(1+6\mu)^kA^{k-1}J/\gamma\le\epsilon_\mu.
\label{s:global:source-norm-proof}
\end{equation}
For positive degree, support locality gives
\begin{equation}
\mu^{|M|}\le\mu^k\prod_j\mu^{|T_j|},\qquad
\sum_{T\cap X\ne\varnothing,\boldsymbol\beta}
 |s_{T\boldsymbol\beta}^{(j)}|\mu^{|T|}\le k\|S_j\|_\mu.
\label{s:global:weight-submultiplicativity}
\end{equation}
% alias: s:global:touching-sum -> s:global:weight-submultiplicativity

Let \(v'_* =\max_i\sum_{X\ni i}v'_X\). If the anchor is in \(X\),
the interaction sum costs \(v'_*\), each input costs \(k\|S_j\|_\mu\),
and \(a_M^{-1}\le\gamma^{-1}\). Otherwise the anchor belongs to some
\(T_\ell\setminus X\). Summing over these choices only overcounts;
the support identity gives
\begin{equation}
\begin{gathered}
v'_*\le2\,7^kA^{k-1}J,\qquad
|T_\ell|\le|M|+k\le(k+1)|M|,\\
a_M^{-1}\le\frac{k+1}{\gamma|T_\ell|},\qquad
\sum_{X\cap T_\ell\ne\varnothing}v'_X\le|T_\ell|v'_*.
\end{gathered}
\label{s:global:local-interaction-sum}
\end{equation}
% alias: s:global:root-denominator -> s:global:local-interaction-sum
% alias: s:global:root-interaction-sum -> s:global:local-interaction-sum

The touching-support factor cancels the denominator. The anchored
input costs \(\|S_\ell\|_\mu\), the others \(k\|S_j\|_\mu\).
The two anchor cases give \(k^n+n(k+1)k^{n-1}\); multiply by the
\(2^n\) left--right count and use \(\mu^kv'_*/\gamma\le\epsilon_\mu\).
\end{proof}

\subsection{The stationary fixed point}
\label{s:global:fixed-point-section}
\begin{lemma}[Uniform local fixed point]
\label{s:global:fixed-point}
If \(z\le c_k\), the exact stationary map has a unique real fixed
point in \(\|S\|_{3e}\le r_0=1/(8k)\).
\end{lemma}
\begin{proof}
For \(\|S\|_\mu,\|T\|_\mu\le r\), the rooted estimate and its
multilinear telescoping difference give
\begin{equation}
\begin{aligned}
\|\mathfrak F(S)\|_\mu
 &\le\epsilon_\mu e^{2kr}[1+2(k+1)r],\\
\|\mathfrak F(S)-\mathfrak F(T)\|_\mu
 &\le\epsilon_\mu e^{2kr}
 [4k+2+4k(k+1)r]\|S-T\|_\mu.
\end{aligned}
\label{s:global:ball-bound}
\end{equation}
% alias: s:global:lipschitz-bound -> s:global:ball-bound

The exponential bounds extend the finite nonnegative degree sums.
Choose \(\mu=3e\), \(r=r_0=1/(8k)\), and
\(\epsilon=2(21e)^kz\le r_0/4\). Since \(e^{1/4}<4/3\),
\begin{equation}
\|\mathfrak F(S)\|_{3e}\le\frac{11}{6}\epsilon
\le\frac{11}{24}r_0<r_0,\qquad
\operatorname{Lip}(\mathfrak F)\le\frac{23}{96}<\frac14.
\label{s:global:fixed-point-parameters}
\end{equation}
% alias: s:global:contraction-constants -> s:global:fixed-point-parameters

The contraction theorem applies in the finite-dimensional ball.
The Hermitian basis makes the coefficient map real; iteration from
zero gives real \(S\). The exact BCH equation and scalar trace identity
make \(e^S\theta\) Hermitian, trace one and stationary.
\end{proof}

\subsection{Positive reconstruction by a finite recursion}
\label{s:global:section-map}\label{s:global:polymer-section}
Choose \(\tau_i=\sum_{a,s}w_i(a,s)Q_{a,s}\), with \(w_i\ge0\) and
\(\sum w_i=1\). Put
\begin{equation}
p_i^0(a,s)=r_i/6+(1-r_i)w_i(a,s)\ge r_i/6>0.
\label{s:global:reset-decomposition}
\end{equation}
% alias: s:global:reference-probabilities -> s:global:reset-decomposition

The six-state linear section is
\begin{equation}
\begin{aligned}
\mathcal F_i(\eta)(a,s)
&=(\operatorname{Tr}\eta)p_i^0(a,s)\\
&\quad+\frac s2\operatorname{Tr}[P_a(\eta-\theta_i\operatorname{Tr}\eta)].
\end{aligned}
\label{s:global:linear-section}
\end{equation}
Summing over \(s\) gives the trace identity; using
\(\sum_s sQ_{a,s}=P_a\) gives reconstruction. Substitution gives
\begin{equation}
\begin{gathered}
\sum_{a,s}\mathcal F_i(\eta)=\operatorname{Tr}\eta,\\
\sum_{a,s}\mathcal F_i(\eta)Q_{a,s}=\eta,\\
\mathcal F_i(\theta_i)=p_i^0,\qquad
f_{i\beta}(a,s)=\frac{sr_i\delta_{a\beta}}{2p_i^0(a,s)},\qquad
|f_{i\beta}|\le3.
\end{gathered}
\label{s:global:section-lemma}
\end{equation}
% alias: s:global:section-identities -> s:global:section-lemma
% alias: s:global:section-excitation -> s:global:section-lemma

For a fixed configuration \(\boldsymbol\sigma\), define activities on
arbitrary nonempty supports and their hard-core polynomial by
\begin{equation}
\begin{aligned}
w_M&=\sum_{\boldsymbol\beta}s_{M\boldsymbol\beta}
             \prod_{i\in M}f_{i\beta_i}(\sigma_i),\\
Z_U(w)&=\sum_{\mathcal C\ {\rm disjoint\ in}\ U}
             \prod_{M\in\mathcal C}w_M,\qquad Z_\varnothing=1.
\end{aligned}
\label{s:global:polymer-activities}
\end{equation}
% alias: s:global:polymer-polynomial -> s:global:polymer-activities

Disjoint creator orderings cancel the exponential factorial, so the
tensor section \(\mathcal F=\bigotimes_i\mathcal F_i\) satisfies
\begin{equation}
\begin{gathered}
\mathcal F(e^S\theta)(\boldsymbol\sigma)
=\left(\prod_i p_i^0(\sigma_i)\right)Z_{[N]}(w),\\
\max_i\sum_{M\ni i}|w_M|e^{|M|}\le\|S\|_{3e}\le r_0.
\end{gathered}
\label{s:global:polymer-identity}
\end{equation}
% alias: s:global:exact-polymer -> s:global:polymer-identity
% alias: s:global:activity-bound -> s:global:polymer-identity

\begin{lemma}[Positive hard-core weights]
\label{s:global:polymer-positive}
If \(\max_i\sum_{M\ni i}|w_M|e^{|M|}\le r<1/2\), then all
\(Z_U\ne0\) and \(|Z_U/Z_{U\setminus i}-1|\le r\).
Real activities give \(Z_U>0\).
\end{lemma}
\begin{proof}
Splitting by the polymer covering \(i\) gives
\begin{equation}
Z_U=Z_{U\setminus i}+\sum_{M\subseteq U:i\in M}w_MZ_{U\setminus M}.
\label{s:global:polymer-smallness}
\end{equation}
% alias: s:global:polymer-recursion -> s:global:polymer-smallness

Induct over all subsets and distinguished sites, starting from
\(Z_\varnothing=1\). Successively deleting \(|M|-1\) sites from
\(U\setminus i\) and using the inductive ratio bound gives
\begin{equation}
\left|\frac{Z_{U\setminus M}}{Z_{U\setminus i}}\right|
\le(1-r)^{-(|M|-1)}\le2^{|M|-1}\le e^{|M|}.
\label{s:global:polymer-inverse-ratio}
\end{equation}
Divide the recurrence by \(Z_{U\setminus i}\) and apply the activity
bound to obtain
\begin{equation}
\left|Z_U/Z_{U\setminus i}-1\right|\le r<1.
\label{s:global:polymer-ratio}
\end{equation}
This closes nonvanishing. For real activities each ratio is at least
\(1-r>0\), proving positivity for the original activity family
restricted to every subset.
\end{proof}

\subsection{Completion}
\label{s:global:completion}
For the real fixed point, \(r_0=1/(8k)<1/2\), so every reconstructed
weight is positive. The section identities give
\begin{equation}
\begin{gathered}
p(\boldsymbol\sigma)=\left(\prod_i p_i^0(\sigma_i)\right)
 Z_{[N]}(w)>0,\qquad\sum_{\boldsymbol\sigma}p=1,\\
e^S\theta=\sum_{\boldsymbol\sigma}p(\boldsymbol\sigma)
 \bigotimes_iQ_{\sigma_i}.
\end{gathered}
\label{s:global:final-weights}
\end{equation}
% alias: s:global:final-reconstruction -> s:global:final-weights

This positive trace-one stationary operator equals \(\rho_*\) by
uniqueness. Every constant used only \(k\), the rate interval
\(r_i\in[A^{-1},1]\), and the per-site interaction budget. The weights
can correlate all sites; their existence and the finite-instance
relaxation bound hold throughout the stated control class.

%% file: proofs/finite_k_modules.tex
\section{Exact finite-rate modules for general interaction order}
Take $N=k$, common $|0\rangle$ resets at $\kappa$, and one support of norm $h\le J$:
\begin{equation}
 H_k=\frac h{\sqrt2}(X_1+Y_1)X_2\cdots X_k.
 \label{kb:module}
\end{equation}
Put $u=\kappa/\gamma$, $\zeta=h/\gamma$, $r=u/(1+u)$, $a=\gamma+\kappa$, and $\chi=4\zeta^2/[k(1+u)^2]$. Stationary coherence elimination gives upward/downward rates $\gamma/2,\gamma/2+\kappa$ and complement-flip rate $\nu=2h^2/(ka)$. The stationary moment equation $0=-\ell am_\ell+\ell\kappa m_{\ell-1}-2\nu\mathbf1_{\ell\ {\rm odd}}m_\ell$ solves to
\begin{align}
 m_0&=1,\qquad
 m_\ell=r^\ell\prod_{\substack{1\le j\le\ell\\j\ {\rm odd}}}\frac j{j+\chi},
 \label{kb:moments}\\
 P_0&=2^{-k}\sum_{\ell=0}^k\binom k\ell m_\ell,\nonumber\\
 P_k&=2^{-k}\sum_{\ell=0}^k(-1)^\ell\binom k\ell m_\ell,\nonumber\\
 b_k&=\frac{\zeta(P_0-P_k)}{k(1+u)}.
 \label{kb:endpoints}
\end{align}
The populations follow by expanding $\prod_j(I\pm Z_j)/2$; complement coherence decays at $ka$. Its equal quadratures give the exact finite-rate sufficient test
\begin{equation}
 G_k(\zeta,u)=\sqrt2b_k-P_k>0.
 \label{kb:finite-sufficient}
\end{equation}
For a conservative closed test, couple independent populations through the first reset of every spin. Integrating $1-(1-e^{-as})^k$ bounds the population inverse by $\mathsf H_k/a$, $\mathsf H_k=\sum_{j=1}^k1/j$. With $q=\gamma/(2a)$ and $\delta=4\mathsf H_kh^2/(ka^2)$, perturbation by the complement flip gives total population $\ell_1$ error at most $\delta$. Thus
\begin{equation}
 \frac{\sqrt2h}{ka}[(1-q)^k-q^k-2\delta]-q^k-\delta>0
 \label{kb:explicit-sufficient}
\end{equation}
is sufficient and holds at sufficiently large finite $\kappa$ for fixed $h,\gamma>0$.
For strong noise, set $t=x^{1/(k-1)}$, $h=J$, $\kappa=c\gamma/t$. The exact solution gives $P_k=t^k/(2^kc^k)+O_{k,c}(t^{k+1})$ and $b_k=t^k/(kc)+O_{k,c}(t^{k+1})$. Every $c^{k-1}>k/(2^k\sqrt2)$ therefore gives a positive witness for sufficiently small $x>0$, proving
\begin{equation}
 \limsup_{x\downarrow0}x^{1/(k-1)}\frac{K_{\rm on}^{[\le k]}}\gamma
 \le\left(\frac{k}{2^k\sqrt2}\right)^{1/(k-1)}.
 \label{kb:constructive-coefficient}
\end{equation}
The total-rate coefficient is $k$ times this bound.

%% file: proofs/optimization.tex
\section{Optimization functions}
Rescaling frequencies gives $K_{\rm on}^{(N)}(sJ,s\gamma)=sK_{\rm on}^{(N)}(J,\gamma)$. Budget inclusion makes $F_N(y)$ nondecreasing in $y$. Appending an independent noisy idle spin preserves a strict violation by partial trace, so $F_{N+1}\le F_N$.
For right continuity at finite $y_0>0$, choose a finite controller within $\varepsilon$ of the fixed-size or all-size infimum, with strict magic. Replacing $H$ by $(y_0/y)H$ ensures feasibility for nearby $y>y_0$; stationary-state continuity preserves the violation. Monotonicity and $\varepsilon\downarrow0$ give right continuity of both $F_N$ and $F_\infty$.

%% file: proofs/strong_two_body.tex
% Source: review/package1/supplement.tex
% Selected section; old numerical-control and old entropy/skew sections omitted. A nonessential creation-map background sentence is omitted; proof retained.
\section{Uniform strong-noise onset for the complete two-body class}
\label{strong:section}

Define
\begin{equation}
 \begin{gathered}
 a_i=\gamma+\kappa_i,\quad b_i=a_i/\gamma,\quad r_i=b_i^{-1},\\
 x=J/\gamma,\qquad q=JK/\gamma^2.
 \end{gathered}
 \label{strong:parameters}
\end{equation}
All targets belong to $\operatorname{STAB}_1$, all rates are nonnegative,
and every physical qubit has natural depolarization rate $\gamma>0$.
The Hamiltonian is static and obeys
$H=\sum_{1\le |X|\le2}h_X$ and
$\max_i\sum_{X\ni i}\|h_X\|_\infty\le J$.

\begin{theorem}[Uniform two-body converse]
\label{strong:theorem}
For every $q_0<q_*=1/(2\sqrt2)$ there is $x_0(q_0)>0$, independent
of size, graph, target mixtures and rate ratios, such that
\begin{equation}
 0<x<x_0(q_0),\qquad q\le q_0
 \quad\Longrightarrow\quad\rho_*\in\operatorname{STAB}_N.
 \label{strong:exclusion}
\end{equation}
Consequently
\begin{equation}
 \lim_{x\downarrow0}\sup_{N\ge2}
 \left|\frac{JK_{\rm on}^{(N)}}{\gamma^2}-q_*\right|=0.
 \label{strong:uniform-limit}
\end{equation}
The total-rate cost has the uniform leading coefficient
$R_{\rm on}^{(N)}=[1/\sqrt2+o(1)]\gamma^2/J$.
\end{theorem}

\subsection{Exact coordinates for mixed targets}
Write $\tau_i=\sum_v p_{iv}\tau_{iv}$ using the six pure Pauli states.
Zero-probability labels are omitted. Introduce a classical label only
for the proof, and jointly replace its label and qubit at rate $a_i$ by
\begin{equation}
 \Xi_i=\sum_v p_{iv}|v\rangle\langle v|\otimes\theta_{iv},
 \qquad \theta_{iv}=r_iI/2+(1-r_i)\tau_{iv}.
 \label{strong:cq-reference}
\end{equation}
The quantum marginal is the physical generator. With
$\Xi=\bigotimes_i\Xi_i$ and whole-site replacement $\Pi_i$, the CQ
reference dissipator is $\widehat{\mathcal L}_0=\sum_i a_i(\Pi_i-\operatorname{id})$.

Use local Clifford axes in each label block. The local coefficient
spaces are the vacuum $p_{iv}\theta_{iv}$, classical excitations
$p_{iv}f(v)\theta_{iv}$ with $\sum_vp_{iv}f(v)=0$, and quantum
excitations $p_{iv}r_if(v)P_{\alpha,v}/2$.
Equivalently, for conditional blocks $O_v$, the three projections are
\begin{align}
 (P^{\rm v}O)_v&=\theta_v\bar t,\qquad
 (P^{\rm c}O)_v=\theta_v(\operatorname{tr}O_v-\bar t),\nonumber\\
 (P^{\rm q}O)_v&=O_v-\theta_v\operatorname{tr}O_v,
 \qquad\bar t=\sum_vp_v\operatorname{tr}O_v.
 \label{strong:vcq}
\end{align}
The conditional trace-norm bounds are $1,2,2$. Taking the label
supremum before summing Pauli colors keeps estimates uniform at
vanishing label probabilities.

Creators trace their input and insert nonvacuum excitations on $T$.
Their label functions are separately centered in classical variables.
Overlapping products vanish, and disjoint creators commute, giving the
unique trace-one expansion $e^S\Xi$, $S=\log(1+C)$, with zero scalar
part. A nonempty support decays at $a_T=\sum_{i\in T}a_i$, so
\begin{equation}
 S_T=\frac1{a_T}\left[
 \sum_{n=0}^{4}\frac1{n!}
 \operatorname{ad}_{-S}^{,n}(-i[H,\cdot])\Xi\right]_T
 =:\mathfrak F(S)_T.
 \label{strong:stationary-map}
\end{equation}
The scalar row vanishes. A creator disjoint from an interaction
commutes through it. On a two-site interaction, at most two right
creators and one left creator survive, so the displayed finite BCH
sum is exact. 

\subsection{A uniform dynamical estimate}
Let $A=\max_i b_i$, $z=xA=x+q$, and choose a cutoff $B\ge1$.
Call a site strong if $b_i\ge B$ and weak otherwise. For a nonempty
support $T$ set $n_T=|T|$, $p_T=|T\cap\mathcal S|$ and
\begin{equation}
 w_\mu(T)=(2\mu)^{n_T}\frac{n_T}{\binom{n_T}{p_T}},
 \qquad \mu\ge1.
 \label{strong:weight}
\end{equation}
The anchored norm $\|S\|_{w,\lambda}$ sums coefficients with weight
$w_\mu(T)\lambda^{|C|}$ over every C/Q pattern and Pauli color on
supports containing an anchor, taking label suprema first. Its maximum
over anchors controls $\|S\|_\mu$, since $w_\mu(T)\ge\mu^{|T|}$.

For strong $i,j$ retain only the even double-transverse source.
In their conditional pure-target axes its off-diagonal element is
\begin{equation}
 (C_{ij})_{00,11}
 =\frac{i(2-r_i-r_j)\langle00|h_{ij}|11\rangle}{2(a_i+a_j)}.
 \label{strong:pair-source}
\end{equation}
Denote the corresponding all-quantum creator polynomial by $S_p$.

\begin{lemma}[Uniform CQ remainder]
\label{strong:remainder}
Fix $Q<\infty$ and $\mu$. For $z\le Q$, choose
$B=x^{-1/2}$. There are a finite $\lambda$ and $x_0>0$, independent
of all physical rates, labels and size, such that
\begin{equation}
 S=S_p+E,\qquad S=\mathfrak F(S),\qquad
 \|E\|_{w,\lambda}\le C_{\mu,Q,\lambda}\sqrt x.
 \label{strong:remainder-result}
\end{equation}
\end{lemma}
\begin{proof}
For disjoint nonempty \(D,F\), a deleted \(i\in D\), and
\(M=(D\cup F)\setminus\{i\}\), Vandermonde's identity gives
\begin{align}
 \frac{w_\mu(D\cup F)}{w_\mu(D)w_\mu(F)}&\le2,\nonumber\\
 \frac{w_\mu(M)}{w_\mu(D)w_\mu(F)}
 &\le\frac{|M|}{2\mu d_D|F|},\nonumber\\
 \frac{w_\mu(D\cup F)}{w_\mu(D)w_\mu(F)}
 &\le\frac{p_{D\cup F}}{|D|p_F}\quad(p_F\ge1).
 \label{strong:combinatorics}
\end{align}
Here \(d_D\) counts the sites of \(D\) with \(i\)'s strong/weak type,
including \(D=\{i\}\). For weak deletion the second bound follows from
\(\binom{|M|}{p_D+p_F}\ge\binom{|D|-1}{p_D}\binom{|F|}{p_F}\);
the third follows from
\(\binom{|D|+|F|-1}{p_D+p_F-1}\ge
\binom{|D|}{p_D}\binom{|F|-1}{p_F-1}\).
A single overlap costs at most \((2\mu)^{-1}\). Strong and weak births
have ratios \(2\mu(p+1)/n\) and \(2\mu(n-p+1)/n\); deletion of a
type represented \(d\) times has ratio \((n-1)/(2\mu d)\).

Let \(R_q,R_c\) be right quantum/classical occupations on \(X\), and
\(U_q,U_c\) the intermediate output. Left factors cancel against their
output factors, leaving exactly
\begin{equation}
 \frac{b_{U_q}}{b_{R_q}},\qquad
 \lambda^{|U_c|-|R_c|},\qquad
 \frac1{\gamma\sum_{j\in M}b_j},
 \label{strong:local-ratio}
\end{equation}
Here \(b_D=\prod_{i\in D}b_i\). Conditional trace preservation gives
\(U_q\ne\varnothing\); after extracting inverse \(r_i\), every Pauli
dual has norm at most two. C/Q centering and two-site color sums cost
a fixed constant. Nonempty output and nilpotence give the exhaustive
support list
The occupied sets \((L,R)\) are
\((\varnothing,i)\), \((\varnothing,ij)\), \((i,\varnothing)\),
\((i,j)\), \((i,i)\), and \((i,ij)\).
They give, respectively, identity/transfer/birth; union/deletion;
birth; disjoint union; overlap/birth; and overlap with optional union.
For terms without classical loss, the support/rate sums are
\(O_\mu(\epsilon_B)\), where
\begin{equation}
 \epsilon_B=xB+z/B+2x.
 \label{strong:epsilon}
\end{equation}
Indeed, a strong birth obeys
\((p+1)b_j/(\sum_{i\in T}b_i+b_j)\le1+A/B\), and a weak birth
costs at most \(B\). Their \(1/|T|\) cancels the touching budget
\(J|T|\). Weak-to-strong transfer costs at most \(A/(Bq_T)\), charged
only to its \(q_T\) weak endpoints; reverse transfer costs at most
\(1/p_T\). The deletion bound cancels one decay denominator, followed
by budget \(Jd_D\) or \(J|F|\) for an anchor in \(D\) or \(F\).
Identity, union and overlap use \(\sum_Mb_i\ge|M|\); anchors in
\(X\) use the direct per-site budget.

Classical loss supplies \(1/\lambda\). At most one unmatched quantum
output then costs \(O_\mu(z/\lambda)\). The exceptional case
\(U_q=X,R_q=\varnothing\) retains all inputs, consumes a classical
input and has factor \(b_ib_j\). For an outside anchor,
\begin{align}
 &\sum_{X:X\cap T_c\ne\varnothing}\|h_X\|_\infty
 \frac{b_ib_j}{\gamma\sum_M b}\nonumber\\
 &\qquad\le\frac{A\sum_{u\in T_c}b_u
                 \sum_{X\ni u}\|h_X\|_\infty}
                {\gamma\sum_{u\in T}b_u}
 \le z.
 \label{strong:energy-touch}
\end{align}
Its classical loss again supplies \(1/\lambda\).
The remaining retained-classical conversions are
\(C_j\to C_iQ_j\), costing \(x\) from its retained endpoint, and
\(Q_iC_j\to C_iQ_j\). For the latter, an anchor in its classical input
also costs \(x\). At an anchor in a disjoint quantum input \(D\), weak
\(j\) costs \(xB\); for strong \(j\in F\),
\begin{equation}
 \frac{w(M)}{w(D)w(F)}
 \frac{b_j/b_i}{\sum_M b}
 \le\frac{p_M}{|D|p_F}\frac{A}{Bp_M}
 \le\frac{A}{B|D|}.
 \label{strong:retained-transfer}
\end{equation}
The touching budget \(J|D|\) gives \(z/B\), completing the C/Q split.
Absolute multilinearity, rooted convolution and polarization give,
with \(\mathcal P(v)=v+v^2+v^3+v^4\),
\begin{align}
 \operatorname{Lip}_{\|S\|\le L}\mathfrak F
 &\le C_\mu(\lambda^2\epsilon_B+z/\lambda)\mathcal P'(L),\nonumber\\
 \|\mathfrak F(S_p)-S_p\|_{w,\lambda}
 &\le D_{\mu,Q}\lambda^2\epsilon_B.
 \label{strong:kernel-bounds}
\end{align}
The source obeys \(\|S_p\|_{w,\lambda}\le P_Q=64\mu^2Q\),
independently of \(\lambda\). Weak-site source terms cost
\(O_\mu(xB)\), and \(S_p\) has no classical-loss residual.
Choose \(\lambda\) with
\(C_\mu Q\mathcal P'(P_Q+1)/\lambda\le1/4\), then choose \(x\)
so the remaining Lipschitz term is at most \(1/4\) and the center
residual at most \(1/2\). The radius-one ball around \(S_p\) is
invariant and contractive. Since
\(\epsilon_B=(1+z)\sqrt x+2x\), its real fixed point has the stated
remainder bound.
\end{proof}

\subsection{A positive stabilizer reference for the pair source}
For pure conditional targets put $t_i=r_i/2$ and
$\theta=\bigotimes_i\operatorname{diag}(1-t_i,t_i)$ in their local
Clifford axes. Define the complex symmetric edge matrix and its
positive envelope by
\begin{equation}
 \mathsf R_{ij}=(C_{ij})_{11,00}/(t_it_j),\qquad
 w_{ij}\ge|\operatorname{Re}\mathsf R_{ij}|
             +|\operatorname{Im}\mathsf R_{ij}|.
 \label{strong:edge}
\end{equation}
One envelope works for every label. The physical source gives
\begin{align}
 \max_i\sum_jw_{ij}&\le2\sqrt2q,\nonumber\\
 \max_i\sum_jw_{ij}\max(t_i,t_j)&\le2\sqrt2x.
 \label{strong:row-bound}
\end{align}
The second bound retains the local-population factor.

\begin{lemma}[Forest reference]
\label{strong:forest}
Fix a row bound $\nu<1$ and an exponential assigned-support weight
$e^{b|T|}$. For sufficiently small $t=\max_i t_i$ there is a
positive stabilizer mixture $\Sigma$, possibly unnormalized, with
\begin{equation}
 \Sigma/\operatorname{tr}\Sigma=e^{S_p+\Delta}\theta,
 \qquad\|\Delta\|_\mu\le C_{\mu,b,\nu}t.
 \label{strong:forest-result}
\end{equation}
The underlying defect activities obey the same $O(t)$ bound with
the stated assigned-support weight, including activities whose
actual creator support is smaller than their assigned support.
\end{lemma}
\begin{proof}
Start from an even stabilizer \(\psi\) with nonzero vacuum amplitude
and coordinates \(f_U=\psi_U/\psi_\varnothing\). Add a site \(i\)
with older parent \(j\) using
\begin{equation}
 |A\rangle=(|0\rangle_i\psi+\omega|1\rangle_iX_j\psi)/\sqrt2,
 \qquad |B\rangle=Z_i\operatorname{CZ}_{ij}|A\rangle,
 \label{strong:extension}
\end{equation}
where \(\omega\in\{\pm1,\pm i\}\). Averaging the coordinates keeps
the old ones, gives \(\omega f_{U\setminus\{i,j\}}\) when \(i,j\in U\),
and gives zero when \(i\in U,j\notin U\).
Decompose \(\overline{\mathsf R}_{ij}\) into these phases with
nonnegative total weight \(w_{ij}\). Choose each older parent with
those probabilities and pin \(i\) to \(|0\rangle\) otherwise. The
averaged coordinates are perfect-matching sums, or hafnians, on even
principal sets. Conjugating \(\mathsf R\) gives the density-matrix
phase of Eq.~\eqref{strong:edge}.

A uniform vertex order generates a parent forest \(F\). Restriction
to a center set \(D\) has the same recursion, with outside parents
becoming pinning choices. Shift its center to string \(D\), obtaining
a stabilizer \(\phi_{D,F[D]}\) of center probability
\(2^{-|E(F[D])|}\). The positive mixture is
\begin{equation}
 \begin{aligned}
 \Sigma&=\sum_Dp_D\mathbb E\!\left[
 2^{|E(F[D])|}|\phi_{D,F[D]}\rangle\langle\phi_{D,F[D]}|\right],\\
 p_D&=\prod_{i\in D}t_i\prod_{i\notin D}(1-t_i).
 \end{aligned}
 \label{strong:forest-mixture}
\end{equation}
Write \(\Sigma=F_\Sigma\theta\) in creator coordinates and use the
submultiplicative polynomial norm
\begin{equation}
 |P|_\mu=\sum_{M,\alpha}|P_{M\alpha}|\mu^{|M|},
 \label{strong:polynomial-norm}
\end{equation}
Expand parent choices by finite differences. A center omitting a
vertex of an assigned connected set \(T\) cancels its incident-edge
difference. Thus a connected birth tree \(G\) carries
\((\prod_{i\in T}t_i)K_G\), with newest-leaf recursion
\begin{align}
 K_G&=\mathcal L_{ij,\omega}(K_{G\setminus i}),\nonumber\\
 \mathcal L_{ij,\omega}(O)
 &=P_{0,i}X_j\mathcal D_{Z_j}(O)X_j\nonumber\\
 &\quad+\omega E_{01,i}E_{01,j}O
   +\bar\omega E_{10,i}OE_{10,j},
 \label{strong:forest-map}
\end{align}
Here \(K_{\{j\}}=P_{1,j}\), \(\mathcal D_Z(O)=(O+ZOZ)/2\),
\(P_s=|s\rangle\langle s|\), and \(E_{ab}=|a\rangle\langle b|\).
Subtract the pinned branch of Eq.~\eqref{strong:extension} and use
\((I+Z)X/2=E_{01}\), \(X(I+Z)/2=E_{10}\) to obtain this recursion.
Finite differences and induced relative orders factor over disjoint
components, making \(F_\Sigma\) a hard-core gas of tree activities
\(U_T\). Split an edge as \(U_e=L_e+V_e\), with pair creator \(L_e\) and
\(V_e=w_et_it_jP_{0,i}P_{0,j}\); put \(V_T=U_T\) for \(|T|>2\).

The weighted matrix-unit bounds are
\(t_i|P_0|_\mu\le t(1+\mu)\), \(t_i|P_1|_\mu\le\mu\), and
\(t_i|E_{01}|_\mu=t_i|E_{10}|_\mu=\mu\).
Assign \(a\ge1+\mu\) per vertex, absorbing \(e^{b|T|}\).
Uniform birth times give
\(\int\prod_{p\to c}\mathbf1_{s_p<s_c}\,ds_T\).
Fixing the oldest root at zero enlarges this integral. Condition on
the first child, retain all parent/child interleavings, and drop
distinct-label restrictions to obtain
\begin{equation}
 \begin{gathered}
 F_0'=a\nu F_1F_0,\qquad F_1'=a\nu(F_0^2+2F_E^2),\\
 F_E'=a\nu F_0F_E,\qquad(F_0,F_1,F_E)(0)=(t,1,1).
 \end{gathered}
 \label{strong:tree-ode}
\end{equation}
At \(t=0\), \(F_0=0,F_E=1,F_1=1+2a\nu s\) for \(0\le s\le1\).
For small \(t\), bootstrap \(F_1\le2+2a\nu\),
\(F_0\le h=t e^{a\nu(2+2a\nu)}\), and
\(F_E\le e^{a\nu h}\); the bounds close when
\(h\le[16(1+a\nu)^2]^{-1}\). Smooth dependence of this polynomial
ODE gives an \(O(t)\) defect beyond isolated vertices and single pairs.
Marking a vertex anchors the label sum; symmetry of \(w\) charges
each stripped edge by \(\nu\) without changing birth order. Absorb
the marking factor \(|T|\) into \(a\), giving
\begin{equation}
 \sup_i\sum_{T\ni i}e^{b|T|}|V_T|_\mu=O_{\mu,b,\nu}(t).
 \label{strong:defect-bound}
\end{equation}
Overlapping pair creators multiply to zero, so exact source extraction
gives
\begin{equation}
 e^{-S_p}F_\Sigma
 =\sum_{\mathcal D\ {\rm disjoint}}\left(\prod_{T\in\mathcal D}V_T\right)
 \exp\!\left[-\sum_{e:e\cap\cup\mathcal D\ne\varnothing}L_e\right].
 \label{strong:source-extraction}
\end{equation}
Group connected decorated families into activities \(A_M\). Every
family contains a defect, and external endpoints are leaves, hence
\(|M|\le2\sum_{T\in\mathcal D}|T|\). Put
\(W=\sup_i\sum_j|L_{ij}|_\mu\le\mu^2\nu\) and
\(v_T=|V_T|_\mu e^{(8+W)|T|}\). Then
\(\delta=\sup_i\sum_{T\ni i}v_Te^{|T|}=O(t)\).
The defect-joining kernel is
\(K(T,U)=\sum_{i\in T,j\in U}|L_{ij}|_\mu\).
The rooted majorant
\(H_T=v_T\exp(\sum_UK(T,U)H_U)\) has supersolution
\(H_T\le v_Te^{|T|}\) when \(W\delta\le1\). An internal anchor costs
\(\delta\), an external anchor \(W\delta\), yielding
\begin{equation}
 \sup_i\sum_{M\ni i}|A_M|_\mu e^{4|M|}
 \le(1+W)\delta.
 \label{strong:decorated-bound}
\end{equation}
Scalar defects retain their assigned supports in this count.
The hard-core Mayer logarithm is bounded by overlap spanning
trees~\cite{KoteckyPreiss1986,FernandezProcacci2007}: partition
connected graphs by breadth-first trees and sum optional edges with
their alternating signs. Its rooted recursion
\(Q_M=|A_M|_\mu\exp(\sum_{U\cap M\ne\varnothing}Q_U)\) has
supersolution \(Q_M\le|A_M|_\mu e^{|M|}\) when
\((1+W)\delta\le1/8\). Marking an anchored member overcounts the
series, bounding the nonempty logarithm by \(O(t)\) uniformly in size.
Absolute convergence identifies it with the finite scalar-plus-nilpotent
logarithm; normalization removes its scalar part.
\end{proof}

\subsection{CQ lifting and a positive replacement buffer}
Apply the forest construction conditionally, average the product label
weights, and normalize by the full CQ trace. An assigned activity on
$T$ depends only on its labels. Splitting each vacuum into mean and
centered parts costs at most two, so CQ lifting costs
$(1+2\mu\lambda)^{|T|}$. Absorb this into the assigned-support
exponential of Eq.~\eqref{strong:defect-bound}. The normalized CQ
forest has logarithm $S_p+\Delta$ with rooted error $O(t)$; its
intermediate label marginal may change.

\begin{lemma}[CQ replacement buffer]
\label{strong:buffer}
Let $\sigma_0$ be a CQ state whose quantum conditional states are
stabilizer mixtures. For $0<d<1$ put
$D_i=(1-d)\operatorname{id}+d\Pi_i$.
If $E$ is real, has zero scalar part and $\|E\|_{6/d}\le1$, then
$e^E(\prod_iD_i)\sigma_0$ is a CQ stabilizer mixture.
\end{lemma}
\begin{proof}
Expand over erased sets \(D\) of probability
\(p(D)=d^{|D|}(1-d)^{N-|D|}\) and put
\(\sigma_D=\Xi_D\otimes\operatorname{tr}_D\sigma_0\).
Since \(b_i\Pi_i=b_i\),
\begin{equation}
 b_T(\prod_iD_i)\sigma_0
 =d^{-|T|}\sum_{D\supseteq T}p(D)b_T\sigma_D.
 \label{strong:buffer-identity}
\end{equation}
Disjoint products in \(e^E\) split the inverse-\(d\) factors.
On erased sites the exact conditional Pauli section has reference
weights \(p_i^0(a,s|v)=r_i/6+(1-r_i)\delta_{a,z}\delta_{s,+}\).
Its quantum relative factor has magnitude at most three; classical
factors have magnitude at most one after the label supremum.
Thus each fixed reconstruction has hard-core activities obeying
\begin{equation}
 \sup_i\sum_{T\ni i}|w_T|2^{|T|}\le\|E\|_{6/d}\le1.
 \label{strong:positive-activity}
\end{equation}
The exact deletion recurrence
\(Z_U=Z_{U\setminus i}+\sum_{T\subseteq U:i\in T}w_TZ_{U\setminus T}\)
gives, by induction, ratios in \([1/2,3/2]\): the deviation is bounded
by \(\sum_{T\ni i}|w_T|2^{|T|-1}\le1/2\).
All reconstructed weights are positive, with trace one. Combine them
with the conditional free decompositions on unerased sites.
\end{proof}

\subsection{Completion and matching upper sequence}
Fix $q_0<q_*$ and choose $d>0$ such that
$2\sqrt2q_0/(1-d)^2<1$, and set $\mu=6/d$.
Fix $Q>q_0$ and choose $\lambda$ as in Lemma~\ref{strong:remainder}.
Choose the assigned-support exponential to absorb the CQ lift.
On strong sites $r_i\le\sqrt x$. Apply the forest lemma to
$S_p/(1-d)^2$ on those sites, tensor it with the weak-site
reference, and perform the CQ lift just described.
Applying $\prod_iD_i$ multiplies a nonempty creator on $T$ by
$(1-d)^{|T|}$. This is an algebra homomorphism because surviving
products have disjoint supports. The resulting free reference is
\begin{equation}
 \sigma_{\rm ref}=e^{S_p+\Delta_d}\Xi,
 \qquad\|\Delta_d\|_{6/d}=O(\sqrt x).
 \label{strong:buffered-reference}
\end{equation}
Lemma~\ref{strong:remainder} gives the exact stationary operator
$e^{S_p+E}\Xi$, with $\|E\|_{6/d}=O(\sqrt x)$.
For sufficiently small $x$,
$\|E-\Delta_d\|_{6/d}\le1$.
Lemma~\ref{strong:buffer} makes
$e^{E-\Delta_d}\sigma_{\rm ref}$ a positive CQ stabilizer mixture.
Its physical marginal is stationary for the original generator.
Finite-system uniqueness identifies it with $\rho_*$, proving
Eq.~\eqref{strong:exclusion}.
All choices depend on $q_0$, uniformly in size, target weights and rates.

For the total rate let $s_i=\kappa_i/\gamma$. Retaining the natural-noise
subtraction in the edge estimate gives
\begin{align}
 w_{ij}&\le\frac{2\sqrt2\|h_{ij}\|_\infty}{\gamma}
 \left(\frac{2b_ib_j}{b_i+b_j}-1\right)\nonumber\\
 &\le\frac{\sqrt2\|h_{ij}\|_\infty}{\gamma}(s_i+s_j),\qquad
 \max_i\sum_jw_{ij}\le\frac{\sqrt2JR}{\gamma^2}.
 \label{strong:total-rate}
\end{align}
The same proof applies for $JR/\gamma^2<1/\sqrt2$; $K\le R$ keeps
$z$ bounded.

Finally reset two sites to $|0\rangle$ at the common rate
\begin{equation}
 H=\frac J{\sqrt2}(XX+YX),\qquad
 \kappa=\frac{\gamma^2+3J^2}{2\sqrt2J}.
 \label{strong:attaining-pair}
\end{equation}
The full steady state has endpoint witness
\begin{equation}
 |\operatorname{Re}\rho_{00,11}|+|\operatorname{Im}\rho_{00,11}|
 -\rho_{11,11}
 =\frac{J^2}{4[(\gamma+\kappa)^2+2J^2]}>0.
 \label{strong:strict-witness}
\end{equation}
Its scaled costs are $(1+3x^2)/(2\sqrt2)$ for $K$ and
$(1+3x^2)/\sqrt2$ for $R$. Adding $N-2$ unreset idle qubits,
each subject to natural noise, preserves the witness after partial
trace and adds no engineering cost. This matches both uniform
converses and completes Theorem~\ref{strong:theorem}.

%% file: proofs/weak_fixed_size.tex
\section{Fixed-size weak-noise thresholds}
\label{w:fixed-size}
For targets in $\STAB_1$, rates $0\le\kappa_i\le K$, and $\gamma>0$, expand $\rho=2^{-N}\sum_Pa_PP$, $a_I=1$, and set
\begin{equation}
 q=\sum_{P\ne I}a_P^2,\quad A=\sum_{P\ne I}|P|a_P^2,\quad
 s_i=\sum_{P:P_i\ne I}a_P^2.
 \label{w:purity-definitions}
\end{equation}
Here $q\le A=\sum_i s_i\le Nq$. Hamiltonian motion conserves purity; replacement supplies $v_{i,P_i}a_{P\setminus i}$. Cauchy--Schwarz and $\|\boldsymbol v_i\|_2\le1$ give
\begin{align}
 \tfrac12\dot q
 &=-\gamma A+\sum_i\kappa_i
 \left(\sum_{P:P_i\ne I}v_{i,P_i}a_Pa_{P\setminus i}-s_i\right)
 \nonumber\\
 &\le-\gamma A+\sum_i\kappa_i
 [\sqrt{s_i(1+q-s_i)}-s_i].\label{w:purity-flow}
\end{align}
For $q\le1$ the brackets are nonnegative. Replacing $\kappa_i$ by $K$ and applying Cauchy--Schwarz again yields
\begin{align}
 \tfrac12\dot q&\le-(\gamma+K)A
       +K\sqrt{A[N(1+q)-A]},\label{w:purity-cap}\\
 t&=\frac{NK^2}{(\gamma+K)^2+K^2}.\label{w:purity-t}
\end{align}
For $t<1/2$, equivalently $\gamma+K>K\sqrt{2N-1}$, flow points inward at $q=1$. The ball reached from $I/2^N$ contains the unique attractor. Squaring the stationary bound and applying Cauchy--Schwarz give
\begin{align}
 \sum_{P\ne I}(|P|-t)a_P^2&\le t,\label{w:weighted-ellipsoid}\\
 \left(\sum_{P\ne I}|a_P|\right)^2
 &\le t\sum_{w=1}^N\frac{\binom Nw3^w}{w-t}.\label{w:pauli-ball}
\end{align}
The increasing right side equals one at the unique root
\begin{equation}
 t_N\sum_{w=1}^N\frac{\binom Nw3^w}{w-t_N}=1.
 \label{w:tN-root}
\end{equation}
Its weight-one term gives $t_N\le(3N+1)^{-1}<1/2$. For $t\le t_N$, $L=\sum_{P\ne I}|a_P|\le1$ and, with $d=2^N$,
\begin{equation}
 \rho_*=(1-L)\frac I d+
 \sum_{P\ne I}|a_P|\frac{I+\operatorname{sgn}(a_P)P}{d}.
 \label{w:explicit-free-decomposition}
\end{equation}
Each summand is a mixture of product Pauli eigenstates. Solving $t\le t_N$ proves
\begin{equation}
 K_{\rm on}^{(N)}\ge\gamma\ell_N,\quad
 \ell_N=\frac1{\sqrt{N/t_N-1}-1}>0.
 \label{w:fixed-N-lower}
\end{equation}
Equality is free, for arbitrary Hamiltonians and mixed, heterogeneous, or zero-rate resets.

For the upper bound, $H_h=h(XX+YX)/\sqrt2$, $h\le J$, with common pure $Z$ resets has witness
\begin{equation}
 |\operatorname{Re}c|+|\operatorname{Im}c|-p_{11}
 =\frac{2\sqrt2h\kappa-\gamma^2-2h^2}
 {4[(\gamma+\kappa)^2+2h^2]}.
 \label{w:pair-upper}
\end{equation}
Taking $h=\gamma/\sqrt2$ gives onset at most $\gamma$ for $J\ge\gamma/\sqrt2$; noisy idle sites preserve magic by partial trace. Budget monotonicity implies
\begin{equation}
 K_{\rm on}^{(N)}=\gamma[c_N+o(1)],\qquad \ell_N\le c_N\le1.
 \label{w:fixed-N-existence}
\end{equation}
The purity bound and pair construction cover every $k\ge2$, so $\ell_N\le c_{k,N}\le1$. The physical endpoint construction further gives $c_{k,N}\le\beta_N$ and the vanishing size-optimized weak limit.

%% file: proofs/two_qubit_certificate.tex
\section{A complete two-qubit exclusion certificate}
\label{w:two-qubit-certificate}
For every static two-qubit Hamiltonian,
\begin{equation}
 K\le\tfrac{13}{20}\gamma\quad\Longrightarrow\quad\rho_*\in\STAB_2.
 \label{w:two-qubit-exclusion}
\end{equation}
Set $\gamma=1$, $\boldsymbol m=\kappa\boldsymbol v$. The control polytope
\begin{equation}
 \mathcal C_K=\{(\kappa,\boldsymbol m):0\le\kappa\le K,
                 \|\boldsymbol m\|_1\le\kappa\}
 \label{w:control-polytope}
\end{equation}
has vertices $(0,0)$ and $(K,\pm K\boldsymbol e_\alpha)$. Purity flow is affine in each local control and has zero Hamiltonian contribution. At stationarity, one of the 49 corner pairs has nonnegative flow. Local Clifford rotations and exchange reduce the nonzero pairs to
\begin{equation}
 (\kappa_1,\kappa_2)=(K,K),\quad(K,0),
 \label{w:two-corners}
\end{equation}
with $|0\rangle$ targets; the zero pair allows only $I/4$.
Order Pauli words lexicographically, $I<X<Y<Z$. In local Bloch vectors $\boldsymbol r_i$ and correlations $T$,
\begin{align}
 \tfrac12\dot q={}&-\sum_i(1+\kappa_i)|\boldsymbol r_i|^2
 -(2+\kappa_1+\kappa_2)\|T\|_F^2\nonumber\\
 &+\sum_i\kappa_i r_{iz}
 +\kappa_1\sum_\alpha T_{z\alpha}r_{2\alpha}
 +\kappa_2\sum_\alpha T_{\alpha z}r_{1\alpha}\nonumber\\
 ={}&-\boldsymbol a^TB\boldsymbol a+\boldsymbol b^T\boldsymbol a.
 \label{w:two-qubit-purity}
\end{align}
Here $B_{P,P}=|P|+\sum_{i:P_i\ne I}\kappa_i$, $B_{Z\alpha,I\alpha}=-\kappa_1/2$, $B_{\alpha Z,\alpha I}=-\kappa_2/2$, and $b_{ZI}=\kappa_1$, $b_{IZ}=\kappa_2$. At $K=13/20$, both $B$ are positive definite. Put $Q=B^{-1}$, $\boldsymbol c=Q\boldsymbol b$, $\chi=\boldsymbol b^TQ\boldsymbol b$. Completing the square gives
\begin{equation}
 \max_{\boldsymbol a^TB\boldsymbol a-\boldsymbol b^T\boldsymbol a\le0}
 \boldsymbol v^T\boldsymbol a
 =\frac{\boldsymbol v^T\boldsymbol c+
 \sqrt{\chi\boldsymbol v^TQ\boldsymbol v}}2.
 \label{w:ellipsoid-support}
\end{equation}
The complete polytope has 60 vertices and 22,320 facets in eight Clifford orbits~\cite{Reichardt2009}, Table~2, Eq.~(22), Sec.~5.3.3. Applying $H_1,S_1,H_2,S_2,\mathrm{CNOT}$ to the supplied integer representatives gives sizes
\begin{equation}
 192,\ 3840,\ 5760,\ 1920,\ 2304,\ 5760,\ 2304,\ 240.
 \label{w:facet-orbit-sizes}
\end{equation}
Each inequality $\boldsymbol v^T\boldsymbol a\le d$ is valid at all vertices. Writing $(Q,\boldsymbol c,\chi)=(Q_n,\boldsymbol c_n,\chi_n)/L$, $L>0$, reduces the exact support test to
\begin{align}
 g_{\boldsymbol v}&=2dL-\boldsymbol v^T\boldsymbol c_n\ge0,\nonumber\\
 g_{\boldsymbol v}^2&\ge\chi_n\boldsymbol v^TQ_n\boldsymbol v.
 \label{w:integer-facet-test}
\end{align}
\begin{table}[tbp]
\caption{Exact support certificates at $K/\gamma=13/20$. Minima run independently over all facets.}
\label{w:integer-certificate-table}
\centering\setlength{\tabcolsep}{2pt}
\begin{tabular}{@{}lrr@{}}
 & Both active & One active\\
 $L$ & 5901974265 & 47467860\\
 $\min g_{\boldsymbol v}$ & 7442661600 & 76236260\\
 Squared margin & 856109696614722000 & 3421483348800000
\end{tabular}
\end{table}
The supplied \texttt{data/certificates.json} contains the eight representatives, five signed Pauli permutations, 60 vertices, $40B$, $20\boldsymbol b$, and rational inverse/source certificates with positive LDL pivots. Integer matrix identities, orbit generation, and the displayed margins certify both ellipsoids. The corner argument extends this result to mixed targets and all heterogeneous or zero rates below the cap.

%% file: proofs/endpoint_threshold.tex
\section{The exact endpoint-family threshold}
\label{w:endpoint}
At fixed finite $N$, take common $|0\rangle$ resets and
\begin{align}
 H_{\rm e}&=\lambda(e^{-i\pi/4}|0^N\rangle\langle1^N|+\mathrm{h.c.}),
 \quad\lambda>0,\label{w:endpoint-H}\\
 u&=\kappa/\gamma,\quad a=\gamma+\kappa,\quad
 b=\frac1{2(1+u)},\nonumber\\
 p&=1-b,\qquad s=\lambda/(Na).\label{w:endpoint-variables}
\end{align}
The independent population chain jumps upwards at $ab$ and downwards at $ap$, with $\pi_0=p^N$, $\pi_1=b^N$. All diagonal populations and endpoint coherences form an invariant algebra. For its classical generator,
\begin{equation}
 \begin{aligned}
 \boldsymbol V&=\int_0^\infty e^{tQ_{\rm cl}}(\boldsymbol e_1-\boldsymbol e_0)\,dt,\\
 Q_{\rm cl}\boldsymbol V&=\boldsymbol e_0-\boldsymbol e_1.
 \end{aligned}\label{w:population-Green}
\end{equation}
Eliminating stationary coherence gives exchange rate $w=2\lambda^2/(Na)$ and
\begin{equation}
 \begin{aligned}
 \boldsymbol P&=\boldsymbol\pi+w\Delta\boldsymbol V,\qquad
 \Delta=\frac{\pi_0-\pi_1}{1+w(V_1-V_0)},\\
 P_1&=\pi_1+w\Delta V_1,\qquad c=e^{i\pi/4}s\Delta.
 \end{aligned}\label{w:full-populations}
\end{equation}
Independent-spin transitions give
\begin{align}
 A=NaV_1&=N\sum_{m=0}^{N-1}\frac{b^m}{N-m},\nonumber\\
 B=-NaV_0&=N\sum_{m=0}^{N-1}\frac{p^m}{N-m},\label{w:AB-sums}\\
 aV_1&=\int_0^1\frac{(b+pt)^N-b^N(1-t)^N}{t}\,dt.
 \label{w:Green-integral}
\end{align}
The integral $I_N$ obeys $I_N-bI_{N-1}=1/N$, $I_0=0$, proving $A$; exchanging $b,p$ gives $B$.
For this block structure, $P_0\ge P_1$, and freeness is equivalent to
\begin{equation}
 |\operatorname{Re}c|+|\operatorname{Im}c|\le P_1.
 \label{w:endpoint-STAB}
\end{equation}
Projecting the repetition-code sector with $Z_iZ_{i+1}$ and Clifford decoding prove necessity by the qubit octahedron. Encoding its Pauli decomposition and adding remaining diagonal populations prove sufficiency. The selected phase maximizes the left side at fixed $|c|$. With $r=(1+2u)^N$, $S=\sqrt2|c|-P_1$,
\begin{align}
 \frac S{\pi_1}&=\frac{(\sqrt2s-2As^2)(r-1)}{1+2(A+B)s^2}-1,
 \label{w:endpoint-gap}\\
 s_{\rm opt}&=\frac1{\sqrt2[A+\sqrt{A^2+A+B}]},\nonumber\\
 \max_s\frac S{\pi_1}&=\frac{r-1}{2[A+\sqrt{A^2+A+B}]}-1.
 \label{w:lambda-optimum}
\end{align}
The equivalent concave quadratic $-2(Ar+B)s^2+\sqrt2(r-1)s-1>0$ is feasible precisely when
\begin{equation}
 (r-1)^2>4(Ar+B).\label{w:endpoint-threshold}
\end{equation}
Put $z=1+2u>1$ and divide by $z^N$. Then $z^N+z^{-N}-2$ increases, $A(z)$ decreases, and every summand of
\begin{equation}
 \frac{B(z)}{z^N}=N\sum_{j=1}^N\frac{(1+z)^{j-N}}{jz^j}
 \label{w:monotone-B}
\end{equation}
decreases. Their difference rises from negative values to infinity, proving a unique positive threshold:
\begin{equation}
 \begin{aligned}
 {}[(1+2\beta_N)^N-1]^2
 =4\{&A_N(\beta_N)(1+2\beta_N)^N\\
 &+B_N(\beta_N)\}.
 \end{aligned}\label{w:beta-root}
\end{equation}
Clearing positive denominators gives an algebraic root. Equality is free; all larger $u$ admit magic. Here $\beta_2=1$, while $\beta_3$ obeys
\begin{align}
 16u^6+48u^5+60u^4+28u^3&-9u^2-15u-5=0,
 \label{w:beta3-polynomial}\\
 \frac{588363217683}{10^{12}}<\beta_3
 &<\frac{588363217684}{10^{12}}<\frac{13}{20}.
 \label{w:beta3-bracket}
\end{align}
At $u=C/N$, $r\to e^{2C}$ and $A,B\to2$: terms $m\le N/2$ are geometrically dominated, while the tail is bounded by $N\sum_{m>N/2}q_0^m\to0$ for fixed $q_0\in(1/2,1)$. The limiting equation $(e^{2C}-1)^2=8(e^{2C}+1)$ gives
\begin{equation}
 N\beta_N\to C_*=\tfrac12\ln(5+4\sqrt2)=1.1831016383204367\ldots.
 \label{w:beta-asymptotic}
\end{equation}
Signs at $C_*\pm\delta$ and monotonicity establish convergence of the roots.

%% file: proofs/endpoint_implementation.tex
\section{Physical two-body implementation with all populations}
\label{w:implementation}
Fix finite $N,\gamma,\kappa,\lambda$ and the open chain
\begin{align}
 H_0&=-\sum_{i=1}^{N-1}Z_iZ_{i+1}+\sum_{i=1}^Nh_iZ_i,
 \label{w:chain-parent}\\
 h_i&=2^{i-1-N}\ (i<N),\quad
 h_N=-\frac{2^{N-1}-1}{2^N}.\label{w:binary-fields}
\end{align}
The fields sum to zero, their positive sum is below $1/2$, and $\sum_i|h_i|<1$. Endpoints have energy $E_0=-(N-1)$; every proper nonempty flipped set $S$ has
\begin{equation}
 E_S-E_0=2d(S)-2\sum_{i\in S}h_i>1.
 \label{w:chain-gaps}
\end{equation}
Binary subset sums separate states within a domain-wall layer; field differences below two separate layers. Only endpoints are degenerate. Let $g>0$ be the smallest distinct-block gap. For
\begin{equation}
 G(\eta,\xi)=H_0+\eta\sum_iX_i+\xi Z_1,
 \label{w:perturbed-chain}
\end{equation}
analytic projections give a real orthogonal $U(\eta,\xi)\to I$ isolating the endpoint block, in gauge $U(0,\xi)=I$. Its diagonal difference is $2\xi$ at $\eta=0$; the implicit-function theorem supplies $\xi(\eta)=O(\eta^2)$ such that
\begin{align}
 B(\eta,\xi(\eta))&=E_c(\eta)I_2+t(\eta)X_L,
 \label{w:tuned-low-block}\\
 t(\eta)&=a_N\eta^N+O(\eta^{N+1}),\nonumber\\
 a_N&=\sum_{\pi\in S_N}\prod_{m=1}^{N-1}
 (E_0-E_{S_m(\pi)})^{-1}\ne0.
 \label{w:tunneling-coefficient}
\end{align}
All $N!$ leading paths require $N$ flips and share sign $(-1)^{N-1}$. None revisits an endpoint; lower-order normalization and diagonal tuning preserve their sum.
For small $\eta>0$, choose a fixed $Z_1$ rotation $V$ giving the endpoint phase. It preserves noise/reset covariance. The physical controller and charged budget are
\begin{align}
 \Lambda_\eta&=\lambda/|t(\eta)|,\quad
 H_\eta=V\Lambda_\eta G(\eta,\xi(\eta))V^\dagger,
 \label{w:amplified-chain}\\
 J_\eta&=\Lambda_\eta[\min(2,N-1)+\max_i|h_i|
                      +|\eta|+|\xi(\eta)|].
 \label{w:charged-budget}
\end{align}
This is finite for each chosen $\eta$ and diverges in the limit. In coordinates $U'_\eta=VU(\eta,\xi(\eta))V^\dagger\to I$, subtracting a scalar yields
\begin{equation}
 \begin{aligned}
 H'_\eta&=F_\eta+H_{\rm e},\\
 F_\eta&=\sum_{s\ne0^N,1^N}\Lambda_\eta
 (E_s(\eta)-E_c(\eta))|s\rangle\langle s|.
 \end{aligned}\label{w:fast-H}
\end{equation}
Let $\mathsf P$ retain all diagonals and endpoint coherences, $\mathsf Q=\mathrm{id}-\mathsf P$. The fast commutator vanishes on $\mathsf P$ and has inverse norm at most $2/(\Lambda_\eta g)$ on $\mathsf Q$. The transformed original dissipator converges to $\mathcal D$, so
\begin{equation}
 \mathcal S_\eta=-i[H_{\rm e},\cdot]+\mathcal D_\eta
 \longrightarrow\mathcal S_0=-i[H_{\rm e},\cdot]+\mathcal D.
 \label{w:slow-generator}
\end{equation}
The stationary equations and $\|\widetilde\rho_\eta\|_2\le1$ imply
\begin{align}
 \mathsf Q\widetilde\rho_\eta
 &=-\left(-i[F_\eta,\cdot]\right)_{\mathsf Q}^{-1}
       \mathsf Q\mathcal S_\eta\widetilde\rho_\eta,
 \label{w:Q-equation}\\
 \|\mathsf Q\widetilde\rho_\eta\|_2
 &\le\frac{2\|\mathcal S_\eta\|_{2\to2}}{\Lambda_\eta g}\to0,
 \label{w:fast-coherence-bound}
\end{align}
and $\mathsf P\mathcal S_\eta\widetilde\rho_\eta=0$. Every compactness limit therefore lies in the target algebra and satisfies $\mathsf P\mathcal S_0\rho_\infty=0$. This algebra is invariant under local replacements and the endpoint commutator; endpoint coherence decays at $N(\gamma+\kappa)$. Thus the full target stationary equation, including every intermediate population, holds. Finite-system primitivity gives a unique target state and a unique physical state. Every subsequence has the same limit, and $U'_\eta\to I$ gives
\begin{equation}
 \|\rho_*(H_\eta)-\rho_*(H_{\rm e})\|_1\to0.
 \label{w:complete-state-convergence}
\end{equation}

%% file: proofs/weak_size_advantage.tex
\section{Strict size advantage and the all-size weak limit}
\label{w:weak-conclusions}
For $u>\beta_N$, Eq.~\eqref{w:lambda-optimum} gives a strict witness that survives full-state convergence at sufficiently small finite $\eta$. The resulting physical controller has $K=u\gamma$ and finite $J_\eta$, and remains feasible at larger budgets. Taking $u\downarrow\beta_N$ gives
\begin{equation}
 c_N\le\beta_N,\qquad c_3\le\beta_3<\frac{13}{20}\le c_2.
 \label{w:strict-three-versus-two}
\end{equation}
For any $\delta>0$, choose finite $N$ with $\beta_N<\delta/2$, then $u\in(\beta_N,\delta)$ and a finite realizing budget $J_\delta$. For $J\ge J_\delta$,
\begin{equation}
 0\le K_{\rm on}/\gamma\le K_{\rm on}^{(N)}/\gamma<\delta.
 \label{w:all-size-quantifiers}
\end{equation}
Hence $F_\infty(\gamma/J)\to0$ and $\limsup_NNc_N\le C_*$. Size is chosen before the coherent budget and tuning precision; the three-spin construction beats the complete mixed-target, heterogeneous-rate two-spin exclusion.

%% file: proofs/finite_triangle_certificate.tex
\section{A finite three-spin certificate below the two-spin bound}
\label{sec:finite-triangle-certificate}
Take $\kappa=16\gamma/25$, $|0\rangle$ resets, and the exact decimal Hamiltonian
\begin{equation}
 \begin{aligned}
 H/\gamma={}&-20.0Z_a+10.0(Z_b+Z_c)+1.4X_a+0.3Y_a\\
 &+(1.4X_a+2.0Y_a-0.4Z_a)(Z_b+Z_c)\\
 &-1.5(X_bX_c-Y_bY_c).
 \end{aligned}\label{eq:finite-triangle-H}
\end{equation}
Grouping equal supports before charging their norm gives
\begin{equation}
 \begin{aligned}
 j_a/\gamma&=\sqrt{8041/20}+2\sqrt{153/25}<24.998912<25,\\
 j_b/\gamma=j_c/\gamma&=13+\sqrt{153/25}<15.473864<25.
 \end{aligned}\label{eq:finite-triangle-budget}
\end{equation}
The $bc$ term has norm $3\gamma$, so $J=25\gamma$ is admissible. Repetition-code decoding makes $S=\operatorname{Re}\rho_{000,111}+\operatorname{Im}\rho_{000,111}-\rho_{111,111}$ a global stabilizer witness with observable width $\sqrt3$. The rational stationary state obeys
\begin{equation}
 \begin{aligned}
 S&\simeq1.07905\times10^{-4}>\frac{1079}{10^7}>0,\\
 d_3(\rho_*)&\ge S/\sqrt3\simeq6.230\times10^{-5}.
 \end{aligned}\label{eq:finite-triangle-margin}
\end{equation}
Integer Bareiss elimination verifies all 64 stationary equations, trace one, and Hermiticity; positive rational LDL pivots of the $16\times16$ real embedding prove full rank. Rational radical bounds certify the budgets. Thus $16/25<13/20$ beats the complete two-spin exclusion at finite parameters. This triangle differs from the dynamical chain.

%% file: proofs/finite_ratio.tex
\section{Explicit finite-ratio bounds}
\label{sec:finite-proof}
Use the Pauli norm $\|2^{-N}\sum_Pa_PP\|_{P,1}=\sum_P|a_P|$ and its induced column norm. Targets may be mixed and rates heterogeneous or zero.
\subsection{A free affine transformation}
Put $u_i=\kappa_i/\gamma$, $A_i=1+u_i$, $u=K/\gamma$, $A=1+u$, and $r=u/(1+u)$. The invertible free channel
\begin{equation}
 \Phi=\bigotimes_i\Phi_i,\qquad
 \Phi_i=A_i^{-1}\id+(1-A_i^{-1})\mathcal R_{\tau_i}
 \label{eq:affine}
\end{equation}
has, for $\tau_i=(I+\bm v_i\cdot\bm\sigma)/2$, $\|\bm v_i\|_1\le1$, actions $\Phi_i(I)=I+(u_i/A_i)\bm v_i\cdot\bm\sigma$, $\Phi_i(P)=A_i^{-1}P$, $\Phi_i^{-1}(I)=I-u_i\bm v_i\cdot\bm\sigma$, and $\Phi_i^{-1}(P)=A_iP$ for $P\ne I$. Consequently $\widetilde{\mathcal D}=\Phi^{-1}\mathcal D\Phi=\gamma\sum_iA_i(\mathcal T_i-\id)$, with decay $d(P)=\gamma\sum_{i\in\operatorname{supp}P}A_i$.
For $s=|X|\le2$, at most $2^{2s-1}$ Hamiltonian Pauli coefficients anticommute with an input, so
\begin{equation}
 \|-i[h_X,\cdot]\|_{P,1\to1}
 \le2\sqrt{2^{2s-1}}\|h_X\|_\infty.
 \label{eq:comm-column}
\end{equation}
A commutator output has a nonidentity site $q\in X$: its inverse-channel factor $A_q$ cancels the decay denominator $\gamma A_q$, while each other site costs at most $A$. Hence $\|\widetilde{\mathcal D}^{-1}\Phi_X^{-1}Q\|_{P,1}\le A^{s-1}/\gamma$. External factors cancel, and external excitations increase decay. Input columns have norm at most $(1+r)^s$, with identity source at most $(1+r)^s-1$. The budget $\sum_{|X|=1}\|h_X\|+2\sum_{|X|=2}\|h_X\|\le NJ$ gives
\begin{equation}
 \begin{aligned}
 B_\beta&=\frac{2\sqrt2N}{y}A(1+r)^2,\\
 B_0&=\frac{2\sqrt2N}{y}A(2r+r^2).
 \end{aligned}\label{eq:beta-source}
\end{equation}
Writing $\Phi^{-1}\rho_*=2^{-N}(I+\sum_{P\ne I}a_PP)$ and $S=\sum_{P\ne I}|a_P|$, stationarity gives $S\le B_0+B_\beta S$. For $u>0$, $B_0>0$, so $B_0+B_\beta\le1$ implies $S\le1$ and
\begin{equation}
 \Phi^{-1}\rho_*=(1-S)\frac I{2^N}
 +\sum_{P\ne I}|a_P|\frac{I+\operatorname{sgn}(a_P)P}{2^N}.
 \label{eq:affine-free}
\end{equation}
Applying the free channel $\Phi$ proves freeness. For $u=0$, $\rho_*=I/2^N$ directly. Since $B_\beta+B_0=(2\sqrt2N/y)(1+6u+7u^2)/(1+u)$ and the rational factor has derivative $7-2/(1+u)^2\ge5$, equality with $T=y/(2\sqrt2N)$ gives
\[
 G_N(y)=\left[\frac{T-6+\sqrt{T^2+16T+8}}{14}\right]_+.
\]
Equality is free. The bound is positive for $y>2\sqrt2N$ and has expansion $G_N=y/(14\sqrt2N)+1/7+O(N/y)$.
\subsection{Pair-family upper bound}
For $H=h(XX+YX)/\sqrt2$, $0<h\le J$, with pure common $Z$ resets, the witness sign is that of $2\sqrt2h\kappa-\gamma^2-2h^2$. Its onset infimum $(\gamma^2+2h^2)/(2\sqrt2h)$ is minimized at $h=\min(J,\gamma/\sqrt2)$, yielding $U_2(y)$. Positive gaps above this infimum give strict magic; noisy idle spins extend the construction to every $N\ge2$.

%% file: proofs/entropy_budget.tex
% Source: research_next/cost/entropy_budget.tex; condensed proof.
\section{Entropy supply and stationary information}
\label{sec:entropy-supply-new}
Write $a_i=\gamma+\kappa_i$, $u_i=\kappa_i/\gamma$, $u=K/\gamma$,
$R=\sum_i\kappa_i$ and $a_{\min}=\min_i a_i$. The physical dissipator is
$\mathcal D_i=a_i(E_i-\mathrm{id})$, with
$E_i\rho=\theta_i\otimes\rho_{\bar i}$ and
$\theta_i=(\gamma I/2+\kappa_i\tau_i)/a_i$.
For the faithful stationary state define
$\mathcal I=N\ln2-S(\rho_*)=D(\rho_*\Vert I/2^N)$.
All logarithms are natural. Put
\[
 \begin{aligned}
 f(r)&=\tfrac12[(1+r)\ln(1+r)+(1-r)\ln(1-r)],\\
 g(u)&=\frac{1+u}{u}f\!\left(\frac{u}{1+u}\right),\qquad 0\le r<1,
 \end{aligned}
\]
with $g(0)=0$. Convexity of $f$, with $f(0)=0$, makes $f(r)/r$
increasing; hence $g$ increases to $\ln2$ and $g(u)=u/2+O(u^2)$.

\begin{theorem}[Entropy supply]
For any static Hamiltonian and physical one-qubit reset targets,
\begin{equation}
 \begin{aligned}
 \mathcal I&\le\frac{\sum_i a_i[\ln2-S(\theta_i)]}{a_{\min}}\\
 &\le\frac{\sum_i\kappa_i g(u_i)}{a_{\min}}\le\frac R\gamma g(u).
 \end{aligned}
 \label{eq:newentropy-budget}
\end{equation}
At common reset rate $\kappa$, with $u=\kappa/\gamma$,
\begin{equation}
 \mathcal I\le\sum_i[\ln2-S(\theta_i)]
 \le Nf\!\left(\frac{u}{1+u}\right).
 \label{eq:common-information-sharp}
\end{equation}
The last information bound is attained by $H=0$ and pure targets;
for stabilizer targets the attaining product state is free.
\end{theorem}
\begin{proof}
Let $c_i=\ln2-S(i|\bar i)_{\rho_*}\ge0$ and
$e_i=\ln2-S(\theta_i)$, and $\Delta_i=D(E_i\rho_*\Vert\rho_*)$.
The Hamiltonian preserves entropy, while
\[
 \begin{aligned}
 \left.\dot S\right|_i
 &=a_i\{S(E_i\rho_*)-S(\rho_*)+\Delta_i\}\\
 &=a_i(c_i-e_i+\Delta_i).
 \end{aligned}
\]
Stationarity therefore gives
\begin{equation}
 \sum_i a_i\{c_i+D(E_i\rho_*\Vert\rho_*)\}=\sum_i a_i e_i.
 \label{eq:exact-entropy-supply}
\end{equation}
Strong subadditivity gives $S(i|\bar i)\le S(i|1,\ldots,i-1)$;
the chain rule then yields $\mathcal I\le\sum_i c_i$.
Thus $a_{\min}\mathcal I\le\sum_i a_i e_i$.
The Bloch length of $\theta_i$ is at most $u_i/(1+u_i)$, so
$a_i e_i\le\kappa_i g(u_i)$. Monotonicity of $g$ and
$a_{\min}\ge\gamma$ prove Eq.~\eqref{eq:newentropy-budget}.
For common $a_i$, cancel this factor before estimating. Product states
saturate the entropy chain bound; pure targets saturate the Bloch bound.
\end{proof}

The nonnegative correlation slack is
$\mathcal B=\sum_iS(\rho_{\bar i})-(N-1)S(\rho_*)=\sum_i c_i-\mathcal I$.
It vanishes on products. The exact accounting identity is
\begin{equation}
 \begin{aligned}
 a_{\min}(\mathcal I+\mathcal B)&+\sum_i(a_i-a_{\min})c_i\\
 &+\sum_i a_iD(E_i\rho_*\Vert\rho_*)=\sum_i a_i e_i.
 \end{aligned}
 \label{eq:entropy-budget-decomposition}
\end{equation}
At common rates this reduces to
$\mathcal I+\mathcal B+\sum_iD(E_i\rho_*\Vert\rho_*)=\sum_i e_i$.
The supply is shared by global information, correlation slack and replacement mismatch.

Since $I/2^N$ is free, $M_{\rm rel}\le\mathcal I$; Pinsker gives
$d_N\le\sqrt{Rg(u)/(2\gamma)}$.
Thus $d_N\ge d_0>0$ requires
\begin{equation}
 R/\gamma\ge2d_0^2/g(u),\qquad N\ge2d_0^2/[ug(u)],
 \label{eq:entropy-distance-cost}
\end{equation}
whereas $M_{\rm rel}\ge m_0>0$ requires
\begin{equation}
 R/\gamma\ge m_0/g(u),\qquad N\ge m_0/[ug(u)].
 \label{eq:entropy-relative-cost}
\end{equation}
A density $M_{\rm rel}/N\ge m_0$ therefore requires $ug(u)\ge m_0$,
or $u\gtrsim\sqrt{2m_0}$ at small $m_0$, independently of the Hamiltonian.
Along common rates $u=C/N$,
\[
 \begin{aligned}
 M_{\rm rel}&\le Nf\!\left(\frac C{N+C}\right)
 =\frac{C^2}{2N}+O(N^{-2}),\\
 d_N&\le\frac C{2\sqrt N}[1+O(N^{-1})].
 \end{aligned}
\]
The endpoint family's exponential magic loss is stronger than this universal bound.
For heterogeneous rates, polarization transfer by an exchange Hamiltonian
can exceed the uncoupled product information; the weighted bound
in Eq.~\eqref{eq:newentropy-budget} accommodates that transfer.

%% file: proofs/endpoint_resources.tex
% Source: research_followup/weak/weak_resource_appendix.tex; condensed proof.
\section{Endpoint resource amount and filtering cost}
\label{sec:weak-new-resource}
\subsection{Exact resource geometry}
Consider a physical state
\begin{equation}
 \rho=\sum_xP_x|x\rangle\langle x|
 +c|0^N\rangle\langle1^N|+c^*|1^N\rangle\langle0^N|.
 \label{wn:state}
\end{equation}
Write $P_0=P_{0^N}\ge P_1=P_{1^N}$ and
\begin{equation}
 \begin{aligned}
 p&=P_0+P_1,\quad \boldsymbol a=(2\Re c,-2\Im c,P_0-P_1),\\
 S&=|\Re c|+|\Im c|-P_1.
 \end{aligned}
 \label{wn:geometry-variables}
\end{equation}
Signed stabilizer robustness, whose value on free states is one, is
\begin{equation}
 \mathcal R_N(\rho)=\inf_{\rho=\sum_jw_j\sigma_j}
 \left\{\sum_j|w_j|:w_j\in\mathbb R,\ \sigma_j\in\STAB_N\right\}.
 \label{wn:rom-convention}
\end{equation}
Its mixing operators are stabilizer states; this is distinct from generalized robustness.
For the full global stabilizer set,
\begin{align}
 \mathcal R_N&=1+2[S]_+,\label{wn:rom}\\
 d_N&=\tfrac12\operatorname{dist}_2
 (\boldsymbol a,\{\boldsymbol b:\|\boldsymbol b\|_1\le p\}).
 \label{wn:distance}
\end{align}
Clifford decoding with $\prod_{j>1}\mathrm{CNOT}_{1\to j}$ and a
computational-basis syndrome flag maps the coherent sector to
$A=(pI+\boldsymbol a\cdot\boldsymbol\sigma)/2$ and leaves other blocks diagonal.
The image of the free set consists exactly of flagged qubit stabilizer cones:
conditional stabilizer closure gives one inclusion, and encoding Pauli eigenstates
with fixed flags gives the converse. The signed norm of
$(qI+\boldsymbol b\cdot\boldsymbol\sigma)/2$, $q\ge0$, is
$\max\{q,\|\boldsymbol b\|_1\}$, attained by signed Pauli mixtures.
Blockwise decompositions and the triangle inequality after flag measurement
make this norm additive. Other blocks contribute $1-p$, proving Eq.~\eqref{wn:rom}.

For distance, allow the free comparison block weight to be $q$ and put
$D(q)=\operatorname{dist}_2(\boldsymbol a,\{\|\boldsymbol b\|_1\le q\})$.
The qubit trace norm is $\max\{|q-p|,\|\boldsymbol a-\boldsymbol b\|_2\}$;
other blocks contribute at least $|q-p|$. Hence
\begin{equation}
 d_N=\tfrac12\min_{0\le q\le1}
 [\max\{|q-p|,D(q)\}+|q-p|].
 \label{wn:variable-mass}
\end{equation}
The radius-$q$ and radius-$p$ octahedra have Euclidean Hausdorff distance
at most $|q-p|$, giving $D(q)\ge D(p)-|q-p|$. Thus $q=p$ attains the minimum.
In the magic region, soft thresholding determines the unique $\tau>0$
with $\sum_j\max\{|a_j|-\tau,0\}=p$, and
\begin{equation}
 d_N=\tfrac12\sqrt{\sum_j\min\{|a_j|,\tau\}^2}.
 \label{wn:soft-threshold}
\end{equation}
For the endpoint family at phase $\pi/4$ and witness-optimal coupling,
$c=e^{i\pi/4}s_{\rm opt}\Delta$, $P_0-P_1=\Delta$, where
\begin{equation}
 s_{\rm opt}=\{\sqrt2[A+\sqrt{A^2+A+B}]\}^{-1}.
 \label{wn:optimal-s}
\end{equation}
The finite sums below satisfy $A,B\ge1$. Consequently
$\sqrt2s_{\rm opt}\le(1+\sqrt3)^{-1}$, and the absolute Bloch coordinates
are $(a,a,z)$ with $a=\sqrt2s_{\rm opt}\Delta<z=\Delta$.
When $S=a-P_1>0$, $\tau=2S/3<a<z$, so all three coordinates remain active:
\begin{equation}
 d_N=[S]_+/\sqrt3,\qquad \mathcal R_N-1=2[S]_+.
 \label{wn:exact-optimal-distance}
\end{equation}
For other endpoint coherences use Eq.~\eqref{wn:soft-threshold}.

\subsection{Rare logical sectors at fixed accumulated bias}
Fix $C>C_*:=\tfrac12\log(5+4\sqrt2)$ and $u=\kappa/\gamma=C/N$.
Set $b=[2(1+u)]^{-1}$, $p_+=1-b$ and $s=s_{\rm opt}$. The exact steady state obeys
\begin{align}
 A&=N\sum_{m=0}^{N-1}\frac{b^m}{N-m},\quad
 B=N\sum_{m=0}^{N-1}\frac{p_+^m}{N-m},\nonumber\\
 \Delta&=\frac{p_+^N-b^N}{1+2(A+B)s^2},\nonumber\\
 P_0&=p_+^N-2Bs^2\Delta,\quad P_1=b^N+2As^2\Delta,\nonumber\\
 |c|&=s\Delta.
 \label{wn:finite-endpoint}
\end{align}
At fixed $C$,
\begin{equation}
 \begin{aligned}
 A,B&\to2,\quad s\to(2-\sqrt2)/4,\\
 2^Np_+^N&\to e^C,\quad2^Nb^N\to e^{-C}.
 \end{aligned}
 \label{wn:basic-asymptotic}
\end{equation}
Put $g(C)=\tfrac{1+\sqrt2}{2}\sinh C-\cosh C>0$. Then
\begin{align}
 p&=2^{-N}[2\cosh C+o(1)],\label{wn:success-probability}\\
 |c|&=2^{-N}[\tfrac14\sinh C+o(1)],\label{wn:coherence-amount}\\
 S&=2^{-N}[g(C)+o(1)],\nonumber\\
 d_N&=2^{-N}[g(C)/\sqrt3+o(1)],\label{wn:distance-asymptotic}\\
 \mathcal R_N-1&=2^{-N}[2g(C)+o(1)].\label{wn:rom-asymptotic}
\end{align}
The total reset rate is $C\gamma$, and $(p_+/b)^N\to e^{2C}$.
Conditioning on zero syndrome yields $\chi_N=A/p$, with limiting Bloch vector
\begin{equation}
 \boldsymbol r_{\chi,\infty}
 =\tanh C\left(\frac1{4\sqrt2},-\frac1{4\sqrt2},\frac{2+\sqrt2}{4}\right).
 \label{wn:logical-limit}
\end{equation}
For $W=X-Y+Z$, whose stabilizer bound is one,
\begin{equation}
 h_N=\operatorname{Tr}(W\chi_N)-1=2S/p
 \longrightarrow \tfrac{1+\sqrt2}{2}\tanh C-1=:h(C)>0.
 \label{wn:conditional-gap}
\end{equation}
Thus conditional logical magic stays finite while its heralding probability is
$\Theta(2^{-N})$. A continuous output rate additionally depends on preparation and reloading time.

\subsection{Matching syndrome-filtered copy cost}
For fixed $C>C_*$ and $0<d_0<1$, the least copy count in this family satisfies
\begin{equation}
 m_N(d_0)=\Theta(2^N).
 \label{wn:replication-theorem}
\end{equation}
Necessity follows by tensor telescoping against product free approximants:
$d_{mN}(\rho_*^{\otimes m})\le m d_N(\rho_*)$, and therefore
\begin{equation}
 m_N(d_0)\ge\frac{\sqrt3d_0}{g(C)+o(1)}2^N.
 \label{wn:replication-lower}
\end{equation}
For sufficiency, choose $N$ with $h_N>0$ and set
\begin{equation}
 \begin{aligned}
 \epsilon&=(1-d_0)/3,\quad r=\left\lceil72h_N^{-2}\log(1/\epsilon)\right\rceil,\\
 m&=\lceil2r/p\rceil.
 \end{aligned}
 \label{wn:test-parameters}
\end{equation}
Process blocks sequentially by the syndrome instrument. On each zero-syndrome
logical qubit, measure a uniformly selected $X,Y,Z$ and multiply its outcome
by $3,-3,3$, respectively. Accept if at least $r$ blocks succeed and their
first $r$ scores have mean at least $1+h_N/2$.
For the product source, successes are independent with probability $p$.
The missing-success probability is at most $e^{-mp/8}\le e^{-r/4}$;
Hoeffding bounds the score error by $e^{-rh_N^2/72}$.
Both are at most $\epsilon$, since $h_N\le\sqrt3-1$.
For an arbitrary entangled global stabilizer alternative, conditional closure
bounds every successful logical score mean by one. The supermartingale
Hoeffding bound therefore gives acceptance at most $\epsilon$.
If fewer than $r$ successes occur, append dummy scores equal to one;
actual acceptance is a subset of the resulting tail event, covering the stopping times.
The acceptance gap is at least $1-3\epsilon=d_0$. Since $h_N\to h(C)>0$,
$r=O(1)$ and $m=O(2^N)$.
The replicated source consequently uses $\Theta(N2^N)$ qubits and
$\Theta(\gamma2^N)$ total reset rate. Filtering improves the elementary
unfiltered-score estimate $O(S^{-2})=O(4^N)$; these costs concern this source family.

\subsection{Transfer to physical two-body devices}
The fixed-$N$ tunnelling construction of Sec.~\ref{w:implementation} permits
a diagonal sequence with
$e_N=\tfrac12\|\rho_N^{\rm phys}-\rho_N^{\rm end}\|_1=o(2^{-N})$.
Trace-distance Lipschitz continuity preserves Eq.~\eqref{wn:distance-asymptotic};
$e_N/p\to0$ preserves syndrome probabilities and logical limits.
For $m=O(2^N)$, tensor telescoping gives total approximation error
$me_N=o(1)$. This existence argument leaves the required coherent budget,
preparation time and precision unquantified uniformly in $N$.
The robustness asymptotic in Eq.~\eqref{wn:rom-asymptotic} is exact for the endpoint state;
its transfer requires a robustness error estimate beyond trace-distance accuracy.

%% file: proofs/dynamics_k2.tex
% Source: physics_rewrite/dynamics_k2/chain_example.tex; condensed.
\subsection{Nearest-neighbor two-body chain}
The connected three-site example has
\[
 H=\frac{0.9J}{\sqrt2}(X_1X_2+Y_1X_2)+0.1JZ_2Z_3,
 \quad\gamma=J,\quad\kappa_i=2\gamma,
\]
with resets to $|0\rangle$. Exact-support edge norms $.9J,.1J$ give
site budgets $(.9J,J,.1J)$. Full GKLS propagation and stabilizer LPs over
all $1080$ vertices give stationary robustness excess
$0.1162132624676704$ from both $I/8$ and $|000\rangle$.
The pure input peaks at $.2278184$ near $\gamma t=.26$.
Without resets its peak is $.3359219$ near $.30$, followed by decay to
$I/8$. At $H=0$, all sampled LP values equal one.
The fixed dual witness
\[
 \begin{aligned}
 W=\tfrac12[&I+Z_1+Z_2-Z_1Z_2+X_1X_2\\
 &-X_1Y_2-Y_1X_2-Y_1Y_2]\otimes I_3
 \end{aligned}
\]
has $|\operatorname{Tr}(W\sigma)|\le1$ on every stabilizer vertex.
Its excess $\max\{0,\operatorname{Tr}(W\rho_t)-1\}$ matches all
903 trajectory LP values within $7\times10^{-15}$.
It detects the pair condition
$\Re(\rho_{12})_{00,11}+\Im(\rho_{12})_{00,11}>(\rho_{12})_{11,11}$;
the certificate concerns pair magic in a connected chain.

%% file: proofs/dynamics_k34.tex
% Source: physics_rewrite/dynamics_k34/model.tex; condensed.
\subsection{Three- and four-body modules}
For $N=k=3,4$, take
\begin{equation}
 H_k=\frac J{\sqrt2}(X_1\cdots X_k+Y_1X_2\cdots X_k).
 \label{eq:k34-H}
\end{equation}
The strings anticommute on the same contiguous support, so the support
norm and every site's coherent load equal $J$. The propagated equation is
\begin{equation}
 \begin{aligned}
 \dot\rho={}&-i[H_k,\rho]\\
 &+\sum_{j=1}^k[\gamma(\mathcal T_j-\mathrm{id})+\kappa(\mathcal R_j-\mathrm{id})]\rho,
 \end{aligned}
 \label{eq:k34-master}
\end{equation}
where $\mathcal R_j$ replaces site $j$ by $|0\rangle$.
Use $J=\gamma=1$, $\kappa=2$ and free inputs $I/2^k,|0\rangle^{\otimes k}$.
The signed robustness is
\begin{equation}
 \mathcal R(\rho)=\min_{\rho=\sum_sq_s|s\rangle\langle s|}\sum_s|q_s|,
 \quad |s\rangle\ \text{a global stabilizer state}.
 \label{eq:k34-RoM}
\end{equation}
The evolution preserves X states. With $b=\bar a$, $a_1=0$,
$p_a=\rho_{aa},q_a=\rho_{bb},c_a=\rho_{ab}$, their exact full-polytope robustness is
\begin{equation}
 \mathcal R-1=2\sum_{a:a_1=0}
 [|\Re c_a|+|\Im c_a|-\min(p_a,q_a)]_+.
 \label{eq:k34-exact-X-RoM}
\end{equation}
Indeed the Clifford decoder $\prod_{j>1}\mathrm{CNOT}_{1\to j}$ flags
single-qubit blocks with norms
$\max\{p_a+q_a,2|\Re c_a|+2|\Im c_a|+|p_a-q_a|\}$.
Per-flag decompositions give an upper bound; flag measurement and conditional
stabilizer closure give the matching lower bound. Add the block norms and subtract trace one.

The formula is evaluated at 1001 times over $0\le\gamma t\le20$.
Selected transient and stationary points are checked by complete LPs over
1080 and 36720 vertices. Resetting gives stationary excesses
$0.1128221$ and $0.0925417$ for $k=3,4$.
Without resets the pure input has transient magic and approaches $I/2^N$;
the maximally mixed input stays fixed. Nonzero local Pauli jumps on every
site generate the full matrix algebra, ensuring a faithful unique steady
state attracting all inputs. The examples use different sizes and Hamiltonians;
they illustrate generation and maintenance within each specified device.

%% file: proofs/allsector_source.tex
% Source: research_next/architecture/polynomial_allsector_source.tex; condensed.
\section{An all-sector source with polynomial resource cost}
\label{sec:allsector-source}
\subsection{Exact stationary source}
Reset every spin to $|0\rangle$ and put $a=\gamma+\kappa$, $u=\kappa/\gamma$.
The analytical source is
\begin{equation}
 \begin{aligned}
 H_{\rm all}&=\frac\lambda{\sqrt2}(X_1\cdots X_N+Y_1X_2\cdots X_N),\\
 s&=\lambda/(Na),\qquad r=2\lambda^2/(Na^2).
 \end{aligned}
 \label{eq:allH}
\end{equation}
Its norm is $\lambda$ and every complement matrix element has modulus
$\lambda$ and odd-$\pi/4$ phase. Its $N$-body coupling is implemented by
the physical two-body bridge below. In the invariant diagonal--complement algebra,
each coherence decays at rate $Na$ and obeys
\begin{equation}
 c_x=i(H_{\rm all})_{x\bar x}(P_x-P_{\bar x})/(Na).
 \label{eq:allcoherence}
\end{equation}
Elimination gives the exact population generator
\begin{equation}
 Q=Q_0+w(F-I),\qquad w=2\lambda^2/(Na)=ra,
 \label{eq:allQ}
\end{equation}
where $Q_0$ has independent rates $\gamma/2$ for $0\to1$ and
$\gamma/2+\kappa$ for $1\to0$, and $F$ complements all bits.
Permutation symmetry reduces the computation to $N+1$ Hamming weights;
interleaving complementary weights makes the chain block tridiagonal.

\subsection{An algebraic resource margin}
Let $\pi$ be the $Q_0$ product law, $v=u/(1+u)$, $L=1+\ln N$, and choose
\begin{equation}
 r=\frac1{256NL^2},\quad s=\frac1{16\sqrt2NL},\quad
 \lambda=\frac a{16\sqrt2L}.
 \label{eq:allparameters}
\end{equation}
For $N\ge2$, $0<u\le1$ and
\begin{equation}
 Nv^2\ge32\ln N,
 \label{eq:allsizecondition}
\end{equation}
the complete stabilizer-set bounds are
\begin{equation}
 d_N(\rho_*)\ge\frac1{32\sqrt3NL},\qquad
 \mathcal R_N(\rho_*)-1\ge\frac1{16NL}.
 \label{eq:allpolynomialsource}
\end{equation}
To prove them, couple classical copies with shared replacement clocks.
Their zero-mass total-variation contraction is at most
$\min\{1,Ne^{-at}\}$. Stationarity then yields
\begin{equation}
 P-\pi=w\int_0^\infty e^{tQ_0}(F-I)P\,dt,\quad
 \|P-\pi\|_{\rm TV}\le rL.
 \label{eq:allTV}
\end{equation}
Choose $A=\{x:M(x)\ge Nv/2\}$, $M(x)=\sum_i(-1)^{x_i}$.
Hoeffding and Eq.~\eqref{eq:allsizecondition} give
\begin{equation}
 \pi(A)\ge1-N^{-4},\qquad \pi(\bar A)\le N^{-36}.
 \label{eq:alltails}
\end{equation}
The phase-aligned facet witness on pairs intersecting $A$, zero on other
pairs, has spectral width $\sqrt3$ and is nonpositive on $\STAB_N$.
Its margin is
\begin{equation}
 S_A=\sqrt2sP(A)-(1+\sqrt2s)P(\bar A).
 \label{eq:allmargin}
\end{equation}
Writing $t=\sqrt2s=1/(16NL)$, $\delta=rL=t/16$,
\begin{equation}
 \begin{aligned}
 S_A&\ge t(1-N^{-4})-(1+t)N^{-36}-(1+2t)\delta\\
 &\ge t/2=1/(32NL).
 \end{aligned}
 \label{eq:allmarginbound}
\end{equation}
Here $t<1/2$, $1-N^{-4}\ge15/16$ and $(1+t)N^{-36}\le t/16$.
Trace duality gives $d_N\ge S_A/\sqrt3$; the flagwise proof of
Eq.~\eqref{eq:k34-exact-X-RoM} gives $\mathcal R_N-1\ge2S_A$.
One may take $N=O(u^{-2}\ln(1/u))$, with
$R=O(\gamma u^{-1}\ln(1/u))$ and $\lambda=O(\gamma/\ln N)$.

\subsection{Amplification against global stabilizer alternatives}
Decode the complement-pair syndrome and select the full pair when it
intersects $A$. Orient the logical basis by its member in $A$; measure a
uniformly chosen logical Pauli with its facet sign. Assign score
$(3b-1)/2\in\{-2,1\}$ for signed outcome $b$, or zero outside
$A\cup\bar A$. Its source mean is $S_A$; every conditionally free state
has mean at most zero. For $m$ copies, accept if the average score exceeds
$S_{\min}/2$, where $S_{\min}=1/(32NL)$.
The product source obeys Hoeffding independently. Conditional stabilizer
closure makes the same scores a supermartingale for arbitrary globally
entangled free alternatives. Their range is three, so
\begin{equation}
 m\ge\frac{18}{S_{\min}^2}\ln\frac2{1-d_0}
 \quad\Longrightarrow\quad d_{mN}(\rho_*^{\otimes m})\ge d_0.
 \label{eq:allamplification}
\end{equation}
Thus this source uses $m=O(N^2\ln^2N)$ copies,
$O(N^3\ln^2N)$ source qubits and total reset
$O(\gamma uN^3\ln^2N)$. Physical implementation and measurement ancillas
are charged separately.

\subsection{Trace-norm resolvent}
For any diagonal $F$, let
$\Omega=\bigotimes_i\operatorname{diag}((1+v)/2,(1-v)/2)$ and
$\mathcal L_F=\mathcal D-i[F,\cdot]$.
Since $[F,\Omega]=0$, replacement entropy and tensorization give
\begin{equation}
 \frac{d}{dt}D(\rho\Vert\Omega)
 \le-a\sum_iD(\rho\Vert\theta_i\otimes\rho_{\bar i})
 \le-aD(\rho\Vert\Omega).
 \label{eq:allentropy}
\end{equation}
Tensorization follows from $\sum_iS(\rho_{\bar i})\ge(N-1)S(\rho)$.
Set $D_{\max}=-\ln\lambda_{\min}(\Omega)=N\ln[2(1+u)]$.
Pinsker, triangle inequality and Jordan decomposition give, for traceless Hermitian $X$,
\begin{equation}
 \|e^{t\mathcal L_F}X\|_1
 \le\min\{1,\sqrt{2D_{\max}}e^{-at/2}\}\|X\|_1.
 \label{eq:alltracemixing}
\end{equation}
Integrating on this space yields
\begin{equation}
 \|\mathcal L_F^{-1}\|_{1\to1}\le
 T_N=[\ln(2D_{\max})+2]/a.
 \label{eq:allresolventF}
\end{equation}
For Eq.~\eqref{eq:allparameters}, $T_N\le3L/a$ and
$2\lambda T_N\le3/(8\sqrt2)<1$. The commutator norm is $2\lambda$,
so the resolvent Neumann series gives
\begin{equation}
 \begin{aligned}
 \|(\mathcal D-i[F+H_{\rm all},\cdot])^{-1}\|_{1\to1,\operatorname{Tr}=0}
 &\le A_N,\\
 A_N:=T_N/(1-2\lambda T_N)&\le5L/a.
 \end{aligned}
 \label{eq:allresolventfull}
\end{equation}
Pair-dependent Pauli signs preserve this bound. A diagonal $F_x=F_{\bar x}$
commutes with the exact source state and leaves it stationary.

%% file: proofs/boundary_source.tex
% Source: research_stage3/cost/boundary_sector_source.tex; condensed.
\section{A boundary-sector source}
\label{sec:boundary-sector-source}
\subsection{Parameters and margin}
For $0<u\le1$, put $a=\gamma+\kappa$, $v=u/(1+u)$, $L=1+\ln N$ and
\begin{equation}
 \begin{aligned}
 \ell&=\ln(16/v),&N&=\lceil256v^{-2}\ell\rceil,\\
 r&=v^2/2^{16},&\lambda&=\frac{av\sqrt N}{256\sqrt2},\\
 t&=\sqrt{r/N}=\frac v{256\sqrt N}.&&
 \end{aligned}
 \label{eq:boundary-parameters}
\end{equation}
Use the exact all-complement source, with coherence parameter $s=t/\sqrt2$,
and select
\begin{equation}
 A=\{x:M(x)\ge Nv-2\sqrt N\},\qquad M(x)=\sum_i(-1)^{x_i}.
 \label{eq:boundary-selection}
\end{equation}
Since $v\sqrt N>4$, $A$ contains no complement pair. The selected facet
witness is nonpositive on $\STAB_N$ and has spectral width $\sqrt3$.
\begin{proposition}
The ideal steady state satisfies
\begin{equation}
 \begin{aligned}
 S_A&=tP(A)-(1+t)P(\bar A)\ge t/8,\\
 d_N(\rho_*)&\ge\frac v{2048\sqrt{3N}}.
 \end{aligned}
 \label{eq:boundary-margin}
\end{equation}
\end{proposition}

\subsection{Conditional population law and proof}
In time units $a^{-1}$, condition on the common-flip history. The spins
are iid with conditional mean $vY$, where $\dot Y=1-Y$ between
rate-$r$ flips $Y\mapsto-Y$:
\begin{equation}
 \dot Y=1-Y,\qquad Y\mapsto-Y\quad\text{at rate }r.
 \label{eq:boundary-PDMP}
\end{equation}
The stationary law of $(1+Y)/2$ is $\operatorname{Beta}(r+1,r)$, with
\begin{equation}
 h_r(y)=\frac{[(1+y)/2]^r[(1-y)/2]^{r-1}}{2B(r+1,r)},\quad -1<y<1.
 \label{eq:boundary-Y-density}
\end{equation}
Substitution verifies
$-\partial_y[(1-y)h_r(y)]+r[h_r(-y)-h_r(y)]=0$.
Local replacements erase past spin dependence, giving the unique population law.
Its useful bounds are
\begin{equation}
 \mathbb E(1-Y)=\frac{2r}{1+2r}\le2r,\qquad
 h_r(y)\le4r\quad(-1<y\le0).
 \label{eq:boundary-density-bounds}
\end{equation}
The first follows also from $0=1-(1+2r)\mathbb EY$.
For $0<r\le1$, $B(r+1,r)\ge4^{-r}/r\ge1/(4r)$;
on $x\le1/2$, $x^r\le1$ and $(1-x)^{r-1}\le2$, proving the density bound.

Write $q=v\sqrt N$. Conditional Hoeffding gives
\begin{equation}
 P(A)\ge(1-e^{-1/2})P(Y\ge1-1/q)
 \ge(1-e^{-1/2})(1-2rq).
 \label{eq:boundary-PA}
\end{equation}
It also gives $P(\bar A|Y=y)\le
\exp[-\{q(1+y)-2\}_+^2/2]$. Integrate over $y\le0$ using
Eq.~\eqref{eq:boundary-density-bounds}, and over $y\ge0$ using $q\ge4$:
\begin{equation}
 \begin{aligned}
 P(\bar A)&\le\frac{4r}{q}\int_0^q e^{-(z-2)_+^2/2}\,dz
             +e^{-(q-2)^2/2}\\
 &\le14r/q+e^{-q^2/8}.
 \end{aligned}
 \label{eq:boundary-PB}
\end{equation}
Here $4[2+\sqrt{\pi/2}]<14$. The cutoff restricts negative-mean
contamination to a layer of width $O(q^{-1})$.
The explicit size implies
\begin{equation}
 q\ge4,\quad q\le32/v,\quad rq\le1/8,\quad
 t\le1/2,\quad e^{-q^2/8}\le t/64.
 \label{eq:boundary-scalar-conditions}
\end{equation}
Indeed $256\ell\le q^2\le256\ell+v^2$ and
$v^2\ln(16/v)<1$ for $v\le1/2$, so $rq\le v/2048$.
Moreover $e^{-q^2/8}\le(v/16)^{32}\le2^{-19}v^3\le t/64$.
Thus $P(A)\ge1/4$, $P(\bar A)\le9t/128$, since $r/q=t/256$.
Consequently $S_A\ge t/4-(3/2)(9t/128)=37t/256\ge t/8$.

\subsection{Physical transfer and amplification}
The Poisson lemma in Sec.~\ref{sec:witness-poisson-new} applies because
$\lambda\le Na/4$, $F_x=F_{\bar x}$ and the phases are unchanged.
Its generic physical error is at most $18NL\epsilon_U$. Therefore
\begin{equation}
 \epsilon_U\le t/(288NL)=v/(73728N^{3/2}L)
 \label{eq:boundary-physical-accuracy}
\end{equation}
retains $S_{\rm phys}\ge t/16$.
The Gaussian bridge below attains this accuracy with
$\ln(J/\gamma)=O(N\ln N)$; the transfer uses the witness Poisson equation.
For the same bounded estimator, conditional stabilizer closure covers
entangled global alternatives. Applying Eq.~\eqref{eq:allamplification}
with $S_{\min}=t/16$ gives
$m\ge18\cdot16^2t^{-2}\ln[2/(1-d_0)]$ and hence
\begin{equation}
 \begin{aligned}
 m&=O(N/v^2),\qquad N_{\rm tot}=O(v^{-6}\ln^2(1/v)),\\
 R_{\rm tot}/\gamma&=O(u^{-5}\ln^2(1/u)).
 \end{aligned}
 \label{eq:boundary-total-costs}
\end{equation}
These count physical source qubits in decoupled static two-body devices;
external measurement ancillas are separate. The ideal-source copy bound
below matches $m=\Theta(N/v^2)$. The construction improves the previous
three logarithms while leaving a gap between total-size powers $u^{-6}$
and the universal lower bound $u^{-2}$.

%% file: proofs/copy_lower_bound.tex
% Source: research_stage3/cost/collective_copy_obstruction.tex; condensed.
\subsection{Copy lower bound for the ideal complement source}
\label{sec:collective-copy-obstruction}
\begin{proposition}
Suppose $\rho$ is block diagonal on disjoint pairs $(x,\bar x)$, with
$\rho_x=\left(\begin{smallmatrix}P_x&c_x\\c_x^*&P_{\bar x}\end{smallmatrix}\right)$
and $|c_x|=s|P_x-P_{\bar x}|$ for a common $s\ge0$ and arbitrary phases.
Put $q_s=[1+(1+4s^2)^{-1/2}]/2$. For every integer $m\ge1$,
\begin{equation}
 d_{mN}(\rho^{\otimes m})\le\sqrt{1-q_s^m}\le s\sqrt m.
 \label{eq:all-copy-upper}
\end{equation}
For $s>0$, a target $d_{mN}\ge d_0>0$ requires
\begin{equation}
 m\ge\frac{\ln(1-d_0^2)}{\ln q_s}=\Omega(s^{-2}),\qquad
 m\ge d_0^2/s^2.
 \label{eq:all-copy-necessary}
\end{equation}
\end{proposition}
\begin{proof}
For a nonzero block, put $p_x=P_x+P_{\bar x}$ and
$z_x=(P_x-P_{\bar x})/p_x$. Its Bloch vector has form
$z_x(2s\cos\phi_x,2s\sin\phi_x,1)$.
Each block is diagonalized by a rotation through $\theta=\arctan(2s)$,
whose diagonal entries may both be chosen $\cos(\theta/2)$.
Their direct sum gives $\rho=U\sigma U^\dagger$ with computationally
diagonal free $\sigma$ and $\operatorname{Tr}(\sigma U)=\cos(\theta/2)$.
Purifications give squared fidelity
$F(\rho,\sigma)\ge|\operatorname{Tr}(\sigma U)|^2=q_s$.
Fidelity multiplication and Fuchs--van de Graaf imply
$d_{mN}\le\tfrac12\|\rho^{\otimes m}-\sigma^{\otimes m}\|_1
\le\sqrt{1-q_s^m}$.
Finally $(1+x)^{-1/2}\ge1-x/2$ gives $q_s\ge1-s^2$ and
$1-q_s^m\le m(1-q_s)\le ms^2$.
\end{proof}
For the small-drive source $s=1/[16\sqrt2N(1+\ln N)]$,
this matches $m=O(N^2\ln^2N)$; for constant $\lambda/a$ it requires
$m=\Omega(N^2)$. At boundary-source parameters $s^2=r/(2N)$,
$r=v^2/2^{16}$, the matched order is $m=\Theta(N/v^2)$.
The explicit free comparison proves necessity even for unrestricted
collective tests and globally entangled alternatives, within this source family.

A physical approximation transfers this order if
$\tfrac12\|\widetilde\rho-\rho\|_1\le Cs^2$:
purified-distance triangle inequality gives
$\sqrt{1-F(\widetilde\rho,\sigma)}\le(1+\sqrt{2C})s$, so
$d_{mN}(\widetilde\rho^{\otimes m})\le(1+\sqrt{2C})s\sqrt m$.
An accuracy of order $s$ alone leaves that transfer unproved.

%% file: proofs/witness_poisson.tex
% Source: research_stage3/source/witness_poisson_bridge.tex; condensed.
\section{A witness Poisson bridge}
\label{sec:witness-poisson-new}
Put $a=\gamma+\kappa$, $L=1+\ln N$, $h_x=(H_s)_{x\bar x}$,
$|h_x|=\lambda$, $h_{\bar x}=h_x^*$, and $F_x=F_{\bar x}$.
The physical dissipator is $\mathcal D=a\sum_i(E_i-\mathrm{id})$.
For a set $A$ containing at most one member per complement pair, define
\begin{equation}
 \begin{aligned}
 (W_A)_{xx}&=w_x=-\mathbf1_{\bar A}(x),\\
 (W_A)_{x\bar x}&=ih_x\beta_x,\qquad
 \beta_x=\frac{\mathbf1_A(x)-\mathbf1_{\bar A}(x)}{\sqrt2\lambda}.
 \end{aligned}
 \label{eq:poisson-witness-form}
\end{equation}
For odd-$\pi/4$ phases this is the stabilizer-facet witness of width $\sqrt3$.
Let $w_{\rm flip}=2\lambda^2/(Na)$, $t=\sqrt2\lambda/(Na)$ and
$b_x=w_x+w_{\rm flip}\beta_x$. Thus $b$ equals $t,-1-t,0$ on
$A,\bar A,(A\cup\bar A)^c$. With stationary population $P$ for
$Q=Q_0+w_{\rm flip}(\mathsf F-I)$, its margin is $S_A=\sum_xP_xb_x$.
\begin{proposition}
For $N\ge2$ and $0<\lambda\le Na/4$, a Hermitian $O_A$ satisfies
\begin{equation}
 \mathcal L^\dagger O_A=W_A-S_AI,\quad
 \|O_A\|\le2L/a,
 \label{eq:poisson-observable-bound}
\end{equation}
where $\mathcal L=\mathcal D-i[F+H_s,\cdot]$.
\end{proposition}
\begin{proof}
Couple classical copies using shared global flips, replacement clocks and
sampled bits. Sitewise agreement persists, so
\begin{equation}
 \begin{aligned}
 \sup_{x,y}\|e^{\tau Q}\delta_x-e^{\tau Q}\delta_y\|_{\rm TV}
 &\le\min\{1,Ne^{-a\tau}\},\\
 \int_0^\infty\min\{1,Ne^{-a\tau}\}\,d\tau&=L/a.
 \end{aligned}
 \label{eq:poisson-classical-coupling}
\end{equation}
The semigroup integral solves $Q^\dagger f=b-S_A$.
Centering $f$ gives $\operatorname{osc}f\le(1+2t)L/a$ and
$\|f\|_\infty\le(1+2t)L/(2a)$.
Set $(O_A)_{xx}=f_x$, $(O_A)_{x\bar x}=ih_x\alpha_x$, where
$\alpha_x=-(\beta_x+f_x-f_{\bar x})/(Na)$.
Antisymmetry of $\alpha$ gives Hermiticity. Direct block multiplication yields
\[
 \begin{aligned}
 (\mathcal L^\dagger O_A)_{xx}&=(Q_0^\dagger f)_x+2\lambda^2\alpha_x,\\
 (\mathcal L^\dagger O_A)_{x\bar x}
 &=ih_x[-Na\alpha_x-(f_x-f_{\bar x})].
 \end{aligned}
\]
Local replacements damp full-complement entries by $Na$ and $F$ commutes
with this algebra; substitution proves the identity. The off-diagonal
norm is the largest weighted-complement entry, giving
\begin{equation}
 \|O_A\|\le\frac{(1+2t)L}{2a}+\frac1{\sqrt2Na}
 +\frac{\lambda(1+2t)L}{Na^2}.
 \label{eq:poisson-norm-detailed}
\end{equation}
With $q=\lambda/(Na)\le1/4$, the coefficient of $L/a$ is
$(1+2\sqrt2q)(1/2+q)+1/(\sqrt2NL)<2$.
\end{proof}

For an exact Gaussian normal form $U^\dagger H_{\rm phys}U=F+H_s$, let
$\epsilon_U=\|U-I\|$ and
$\delta_{\mathcal D}=\|\mathcal D_U-\mathcal D\|_{1\to1}$ on Hermitian inputs.
Stationarity and Eq.~\eqref{eq:poisson-observable-bound} bound the
transformed witness error by $(2L/a)\delta_{\mathcal D}$.
Returning to physical coordinates gives
\begin{equation}
 |\langle W_A\rangle_{\rm phys,*}-S_A|
 \le(2L/a)\delta_{\mathcal D}+\sqrt3\epsilon_U.
 \label{eq:new-poisson-physical-bridge}
\end{equation}
The generic defect $\delta_{\mathcal D}\le8Na\epsilon_U$ gives loss
at most $18NL\epsilon_U$. Thus $S_A\ge S_{\min}>0$ retains half its margin if
\begin{equation}
 \epsilon_U\le S_{\min}/(36NL).
 \label{eq:new-poisson-accuracy}
\end{equation}
Local defect estimates can strengthen this interface. It transfers the actual
bounded syndrome--Pauli test, including amplification against global
stabilizer alternatives, for drives up to $Na/4$ without a condition on
$\lambda L/a$. Finite preparation uses the separate dynamical proof below.

%% file: proofs/gaussian_foundation.tex
\subsection{Gaussian normal form and spin lift}
\label{sec:gaussian-bridge}
\begin{proposition}
For $N\ge2$, $J_i=1+i/N$ and
\begin{equation}
 H_\eta=-\sum_{i<N}J_iZ_iZ_{i+1}+\eta\sum_iX_i,
 \quad0<\eta\le\frac1{128N^2},
 \label{eq:bridge-chain}
\end{equation}
there are a unitary $U$, positive $\widetilde J_i$ and $s_0>0$ such that
\begin{align}
 U^\dagger H_\eta U&=-\sum_{i<N}\widetilde J_iZ_iZ_{i+1}+s_0G_N,
 \label{eq:bridge-exact-normal}\\
 G_N&=(-1)^NY_1X_2\cdots X_{N-1}Y_N,\nonumber\\
 |\widetilde J_i-J_i|&\le\eta,\quad s_0=\eta^N/\prod_i\widetilde J_i,
 \label{eq:bridge-exact-edge}\\
 \eta^N/3^{N-1}&\le s_0\le\eta^N,\quad \|U-I\|\le24N^3\eta.
 \label{eq:bridge-cost-bounds}
\end{align}
\end{proposition}
\begin{proof}
Take Majoranas $A_i=(\prod_{j<i}X_j)Z_i$,
$B_i=(\prod_{j<i}X_j)Y_i$ and $c=(A_1,B_1,\ldots,A_N,B_N)^T$.
In the convention $H(\mathsf A)=(i/2)c^T\mathsf A c$,
$(\mathsf A_\eta)_{A_i,B_i}=\eta$,
$(\mathsf A_\eta)_{A_{i+1},B_i}=J_i$, with antisymmetric partners and
$\|\mathsf A_\eta-\mathsf A_0\|=\eta$.
The matrix $K_0=i\mathsf A_0$ has eigenvalues $\pm J_i$ and a
rank-two zero cluster, separated by $1/N$.
Its real planes are $E_0=\operatorname{span}(A_1,B_N)$ and
$E_i=\operatorname{span}(A_{i+1},B_i)$.
Riesz circles of radius $\rho=1/(3N)$ retain their ranks and satisfy
\begin{equation}
 \|P(\eta)-P(0)\|\le\eta/(\rho-\eta)\le6N\eta.
 \label{eq:bridge-projector-bound}
\end{equation}
Combining conjugate $\pm J_i$ projectors gives real orthogonal $P_i$,
since $\overline{K_\eta}=-K_\eta$, with
$\|P_i-P_i^{(0)}\|\le12N\eta$.
Put $S=\sum_iP_iP_i^{(0)}$, $t=12N^2\eta\le3/32$ and
$O=S(S^TS)^{-1/2}$. Then $\|S-I\|\le t$, $\det S>0$ and
$P_iO=OP_i^{(0)}$. Singular values in $[1-t,1+t]$ imply
\begin{equation}
 O\in SO(2N),\quad \|O-S\|\le t,\quad \|O-I\|\le24N^2\eta.
 \label{eq:bridge-orthogonal-bound}
\end{equation}
Thus $O^T\mathsf A_\eta O$ is scalar antisymmetric on each fixed oriented
plane. Continuity fixes positive bulk coefficients; Weyl gives
$|\widetilde J_i-J_i|\le\eta$, hence $1<\widetilde J_i<3$.
For $Q=\log O$, $\|Q\|\le2\|O-I\|$; the Gaussian lift
$U=e^T$, $T=\tfrac14\sum_{rs}Q_{rs}c_rc_s$, obeys
$[T,c]=-Qc$, $U^\dagger cU=Oc$ and
$U^\dagger H(\mathsf A_\eta)U=H(O^T\mathsf A_\eta O)$.
Its rotation angles give
$\|T\|\le\tfrac12\sum_k|\theta_k|\le N\|Q\|/2$,
so $\|U-I\|\le24N^3\eta$.

The alternating open Majorana chain has a unique perfect matching,
forcing every transverse edge; hence $\operatorname{Pf}\mathsf A_\eta=\eta^N$.
Reordering to $(A_1,B_N,A_2,B_1,\ldots,A_N,B_{N-1})$ has sign
$(-1)^{N-1}$. Pfaffian preservation by $O$ therefore fixes the edge
coefficient $e=(-1)^{N-1}\eta^N/\prod_i\widetilde J_i$.
Finally $iA_{i+1}B_i=-Z_iZ_{i+1}$ and
$iA_1B_N=-Y_1X_2\cdots X_{N-1}Y_N$, proving the stated sign and normal form.
\end{proof}

For the staggered refinement, $\|O-I\|<1/2$ also gives
\begin{equation}
 \|T\|\le\tfrac14\|Q\|_*\le\tfrac{\sqrt{2N}}4\|Q\|_F,
 \quad\|Q\|_F\le2\|O-I\|_F.
 \label{eq:foundation-frobenius-spin}
\end{equation}
Disjoint input planes imply
\begin{equation}
 \|S-I\|_F^2=\sum_j\|(P_j-P_j^{(0)})P_j^{(0)}\|_F^2.
 \label{eq:foundation-kato-frobenius}
\end{equation}
The closest-orthogonal polar factor satisfies $\|O-I\|_F\le2\|S-I\|_F$,
which yields the $O(N^2\eta^2)$ spin-lift remainder below.

%% file: proofs/staggered_bridge.tex
% Source: research_stage3/architecture/staggered_gaussian_bridge.tex; condensed.
\subsection{Staggered bonds and local first-order dressing}
\label{sec:stage3-staggered-bridge}
\begin{proposition}
For $N\ge2$, set
\begin{equation}
 \begin{aligned}
 J_i&=1+(i\bmod2)+i/(4N),\quad 0<\eta\le(2048N)^{-1},\\
 H_\eta&=-\sum_{i<N}J_iZ_iZ_{i+1}+\eta\sum_iX_i.
 \end{aligned}
 \label{eq:stage3-staggered-H}
\end{equation}
An exact Gaussian $U$ gives the normal form of Eq.~\eqref{eq:bridge-exact-normal},
with the same $G_N$, $|\widetilde J_i-J_i|\le\eta$,
$s_0=\eta^N/\prod_i\widetilde J_i$ and
\begin{equation}
 \begin{aligned}
 \eta^N/3^{N-1}&\le s_0\le2^{N-1}\eta^N,\\
 \|U-I\|&\le4N\eta+1536N^2\eta^2\le5N\eta.
 \end{aligned}
 \label{eq:stage3-staggered-bound}
\end{equation}
The sharper upper edge estimate $(1-\eta)^{1-N}\eta^N$ also holds.
\end{proposition}
\begin{proof}
Write $\mathsf A_\eta=\mathsf A_0+V$, $K_0=i\mathsf A_0$, $W=iV$;
$\|V\|=\|W\|=\eta$. Distinct clusters have spacing $1/(2N)$,
but onsite transverse edges connect only adjacent dimer planes or an
edge plane to its neighboring bulk plane. Every coupled eigenvalue
difference has modulus at least $3/4$: adjacent positive frequencies
differ by $1-1/(4N)$ or more, opposite signs sum to at least two,
and zero-to-bulk gaps exceed one.
In the $K_0$ eigenbasis define $Q_{ab}=-W_{ab}/(e_a-e_b)$ for
$e_a\ne e_b$, and zero otherwise. Then $[K_0,Q]=-W$.
Each row and column of $W$ has absolute sum at most $4\eta$;
therefore $q=\|Q\|\le8\eta$.
Hermiticity and conjugation pairing make $Q$ real antisymmetric.
With $O_1=e^Q$, the commutator series gives
\begin{equation}
 \begin{aligned}
 O_1^T\mathsf A_\eta O_1&=\mathsf A_0+R,\\
 \|R\|&\le\eta(e^{2q}-1)\le4q\eta\le32\eta^2.
 \end{aligned}
 \label{eq:stage3-first-SW}
\end{equation}
Here $R=\sum_{k\ge1}k\,\operatorname{ad}_Q^k(V)/(k+1)!$,
with $\operatorname{ad}_Q(X)=[X,Q]$. Its spin lift obeys
$\|U_1-I\|\le Nq/2\le4N\eta$.
Put $r=\|R\|$. Riesz circles of radius $1/(6N)$ give projector
errors $12Nr$ and real-plane errors $24Nr$.
The Kato sum satisfies $\|S-I\|_F\le24\sqrt2N^{3/2}r$;
polar minimization gives $\|O_2-I\|_F\le48\sqrt2N^{3/2}r$.
Equation~\eqref{eq:foundation-frobenius-spin} then gives
$\|U_2-I\|\le48N^2r\le1536N^2\eta^2$;
also $\|O_2-I\|\le48N^2r$.
The continuous lift $U=U_1U_2$ preserves all original real planes.
The quadratic contribution is at most $3N\eta/4$ in the stated range.
Weyl's bound and the same oriented Pfaffian identity give the bulk and
edge coefficients. Since $1-\eta\le\widetilde J_i<3$, the edge bounds follow.
\end{proof}

For a required dressing accuracy $\varepsilon_{\rm req}\le1/(8192N)$,
choose $\eta=\varepsilon_{\rm req}/(5N)$ and $\Lambda=\lambda/s_0$.
Conjugation by $V_Z=e^{-i\pi Z_1/8}$ supplies the source phase.
The controller has nearest-neighbor and onsite terms with original physical dissipators, and
\begin{equation}
 \begin{aligned}
 J_{\rm budget}/\gamma
 &\le5(\lambda/\gamma)3^{N-1}(5N/\varepsilon_{\rm req})^N,\\
 \ln(J_{\rm budget}/\gamma)
 &\le N\ln(15N/\varepsilon_{\rm req})+\ln(5\lambda/3\gamma).
 \end{aligned}
 \label{eq:stage3-staggered-cost}
\end{equation}
For $\varepsilon_{\rm req}=1/(4096N^2L^2)$, $L=1+\ln N$,
this gives $\eta=1/(20480N^3L^2)$ and
$\ln(J_{\rm budget}/\gamma)\le3N\ln N+2N\ln L+O(N)+\ln(\lambda/\gamma)$,
improving the original bridge's leading coefficient five while preserving order $N\ln N$.

%% file: proofs/local_noise_bridge.tex
% Source: research_stage3/architecture/local_noise_bridge.tex; condensed.
\subsection{Local control of the physical noise defect}
\label{sec:stage3-local-noise}
For $U=U_1U_2$ in the staggered bridge, the first lift is $U_1=e^T$, $T=\sum_iT_i$, with
\begin{equation}
 \begin{aligned}
 T_i&=\frac{i\eta}{2}Y_iD_i^{-1}\quad(1<i<N),\\
 D_i&=J_{i-1}Z_{i-1}+J_iZ_{i+1},\\
 T_1&=\frac{i\eta}{2J_1}Y_1Z_2,\quad
 T_N=\frac{i\eta}{2J_{N-1}}Z_{N-1}Y_N.
 \end{aligned}
 \label{eq:stage3-local-SW}
\end{equation}
Here $D_i^{-1}=(J_{i-1}Z_{i-1}-J_iZ_{i+1})/(J_{i-1}^2-J_i^2)$.
Each $T_i$ is a sum of two adjacent Pauli terms, supported on at most three
sites, with $[H_0,T_i]=-\eta X_i$ and $\|T_i\|\le2\eta/3$.
The zero-within-eigenspace commutator equation identifies this with the preceding
Gaussian generator; hence $\|U_1-I\|\le2N\eta/3$.

For $\mathcal D=\sum_i\mathcal D_i$, $\|\mathcal D_i\|_{1\to1}\le2a$,
the commutator $[\mathcal D_i,\operatorname{ad}_{T_j}]$ vanishes off
$\operatorname{supp}T_j$ and otherwise has norm at most $8a\|T_j\|$.
Integrating unitary conjugations gives
\begin{equation}
 \|\mathcal D_{U_1}-\mathcal D\|_{1\to1}
 \le24a\sum_j\|T_j\|\le16Na\eta.
 \label{eq:stage3-local-D1}
\end{equation}
With $\epsilon_2=\|U_2-I\|\le1536N^2\eta^2$,
\begin{equation}
 \delta_{\mathcal D}:=\|\mathcal D_U-\mathcal D\|_{1\to1}
 \le16Na\eta+12288N^3a\eta^2.
 \label{eq:stage3-local-Dtotal}
\end{equation}
Both terms retain the original physical local channels.

\paragraph{Small-drive full-state transfer.}
For $\lambda=a/(16\sqrt2L)$, $L=1+\ln N$, use $A_N\le5L/a$.
Stationarity in the transformed frame and return to physical coordinates give
\begin{equation}
 \tfrac12\|\rho_{\rm phys,*}-\rho_{\rm source,*}\|_1
 \le41NL\eta+32256N^3L\eta^2.
 \label{eq:stage3-local-traceerror}
\end{equation}
Choose
\begin{equation}
 \eta=(2^{15}N^2L^2)^{-1}.
 \label{eq:stage3-local-eta}
\end{equation}
The error is less than $1/(512NL)$; if $Nv^2\ge32\ln N$ this retains
$\langle W_A\rangle_{\rm phys,*}\ge1/(64NL)$ and
$d_N\ge1/(64\sqrt3NL)$.
The charged budget satisfies
\begin{equation}
 \begin{aligned}
 J_{\rm budget}/\gamma
 &\le5(\lambda/\gamma)3^{N-1}(2^{15}N^2L^2)^N,\\
 \ln(J_{\rm budget}/\gamma)
 &\le2N\ln N+2N\ln L+O(N)+\ln(\lambda/\gamma).
 \end{aligned}
 \label{eq:stage3-local-J}
\end{equation}
Thus the leading logarithmic coefficient is two, with the same size and reset costs.

\paragraph{Witness and preparation interfaces.}
Let a source witness have margin $S$, width $b$, and adjoint Poisson solution
$\mathcal L_0^*Q=W-SI$ with $\|Q\|\le B_Q$.
Scalar shifts of $Q$ leave the trace-preserving defect unchanged. Exact stationarity gives
\begin{equation}
 \begin{aligned}
 \langle W\rangle_{\rm phys,*}\ge S
 &-b(2N\eta/3+1536N^2\eta^2)\\
 &-B_Q(16Na\eta+12288N^3a\eta^2).
 \end{aligned}
 \label{eq:stage3-local-Poisson-interface}
\end{equation}
For the boundary witness, $B_Q\le2L/a$ and $b=\sqrt3$, so
\begin{equation}
 \begin{aligned}
 E_W\le{}&32NL\eta+24576N^3L\eta^2\\
 &+\sqrt3(2N\eta/3+1536N^2\eta^2).
 \end{aligned}
 \label{eq:stage3-local-boundary-loss}
\end{equation}
For $0<S_{\min}\le\min\{1,L/N\}$, take
\begin{equation}
 \eta=S_{\min}/(1024NL).
 \label{eq:stage3-local-boundary-eta}
\end{equation}
It lies in the staggered range and gives $E_W<S_{\min}/8$:
the two noise fractions are at most $1/32,3/128$, and the coordinate fraction is below $1/256$.

For preparation from $I/2^N$, the dynamical theorem below gives
$T_{\rm prep}=4LB/a$, $B=\ln(40\sqrt3/S_{\min})$, under
$8rL\le1/4$ and $\lambda/(Na)\le1/16$.
The choice
\begin{equation}
 \eta=S_{\min}/(1024NLB)
 \label{eq:stage3-local-prep-eta}
\end{equation}
satisfies $\delta_{\mathcal D}\le aS_{\min}/(16\sqrt3LB)$ and
$\|U-I\|\le S_{\min}/(8\sqrt3)$; the first fraction is bounded by
$1/64+3/256<1/(16\sqrt3)$. It retains physical margin $S_{\min}/2$ at $T_{\rm prep}$.
For either choice, using $B=1$ for stationarity,
\begin{equation}
 J_{\rm budget}/\gamma\le5(\lambda/\gamma)3^{N-1}
 (1024NLB/S_{\min})^N.
 \label{eq:stage3-local-optimized-cost}
\end{equation}
At boundary scaling $N=\Theta(u^{-2}\ln(1/u))$,
$S_{\min}=\Theta(u/\sqrt N)$, $\lambda/a=\Theta(u\sqrt N)$,
this gives $\ln(J_{\rm budget}/\gamma)\le2N\ln N+O(N\ln\ln N)$.

\paragraph{Longitudinal fields and joint defects.}
For $V=\sum_i h_iZ_i$, $H_1=\sum_i|h_i|$, only $Y_i$ in $T_i$
fails to commute with $V$. Hence $\|[T,V]\|\le4\eta H_1/3$ and
\begin{equation}
 \delta_V:=\|U^\dagger VU-V\|
 \le(4\eta/3+3072N^2\eta^2)H_1.
 \label{eq:stage3-local-longitudinal-error}
 % alias: eq:robust-local-field-coordinates -> eq:stage3-local-longitudinal-error
\end{equation}
The Poisson bound for a detuned source includes the residual
\[
 \begin{aligned}
 E_{\rm joint}\le&\sqrt3(2N\eta/3+1536N^2\eta^2)\\
 &+(2L/a)[16Na\eta+12288N^3a\eta^2\\
 &\hspace{18mm}+2(4\eta/3+3072N^2\eta^2)H_1+2\|K\|],
 \end{aligned}
\]
where $K$ is an additional uncontrolled Hamiltonian error. The ideal detuned
source must separately satisfy its detuning assumptions.
A simultaneous boundary-source choice is
$t=v/(256\sqrt N)$, $S_{\min}=\sigma=t/8$,
$B=\ln(320\sqrt3/t)$ and
$\eta=t/(2^{15}NLB)=\sigma/(4096NLB)$.
Writing $q=v\sqrt N=\Theta(\sqrt L)$ gives
$\eta^{-1}=2^{23}N^2LB/q$ and
\[
 \ln(J_{\rm budget}/\gamma)
 \le2N\ln N+\tfrac32N\ln\ln N+O(N).
\]
This choice meets stationary and free-input preparation conditions with room for the detuning defects below.

%% file: proofs/static_robustness.tex
% Source: research_stage3/robustness/static_error_robustness.tex; condensed.
\section{Static detuning and control precision}
\label{sec:stage3-static-robustness}
Put $a=\gamma+\kappa$, $L=1+\ln N$, $|h_x|=\lambda$,
$t=\sqrt2\lambda/(Na)$, $w=2\lambda^2/(Na)$ and $r=w/a$.
Let $F$ be diagonal with $F_x=F_{\bar x}$ and add a diagonal $V$.
Write $\Delta_x=V_x-V_{\bar x}$, $d_x=\Delta_x/(Na)$,
$D=\max_x|d_x|$ and $q_x=(1+d_x^2)^{-1}$.
Scalar identity shifts of $V$ are irrelevant.

\paragraph{Exact detuned source and comparison.}
Stationary elimination in the diagonal--complement algebra gives
\begin{equation}
 c_x=\frac{ih_x(P_x-P_{\bar x})}{Na+i\Delta_x},\quad
 Q_D=Q_0+w_x(\mathsf F-I),\quad w_x=wq_x=w_{\bar x}.
 \label{eq:robust-exact-detuned}
\end{equation}
Here $(Q_D^\dagger f)_x=(Q_0^\dagger f)_x+w_x(f_{\bar x}-f_x)$.
The original phase-aligned Pauli facet has margin
\begin{equation}
 S_D=t\sum_{x\in A}q_xP_D(x)-\sum_{x\in\bar A}(1+tq_x)P_D(x).
 \label{eq:robust-detuned-facet}
\end{equation}
Its coherent projection attenuates by $q_x$, giving quadratic detuning loss
on linewidth $Na$. For undetuned stationary $P$ with margin $S\ge\sigma>0$,
the common-flip inverse bound $L/a$ and
$\|(Q_D-Q)P_D\|_{\rm TV}\le wD^2$ give
\begin{equation}
 \|P_D-P\|_{\rm TV}\le rLD^2.
 \label{eq:robust-population-comparison}
\end{equation}
The undetuned population witness has oscillation $1+2t$; its pointwise
change is at most $tD^2$. Therefore
\begin{equation}
 |S_D-S|\le BD^2,\qquad B=t+(1+2t)rL,
 \label{eq:robust-general-comparison}
\end{equation}
and the sufficient choice
\begin{equation}
 D^2\le\sigma/(4B)\quad\Longrightarrow\quad S_D\ge3\sigma/4
 \label{eq:robust-detuning-choice}
\end{equation}
preserves the fixed original Pauli witness.

\paragraph{Detuned Poisson observable.}
Assume $\lambda\le Na/4$, $0<\sigma\le t$ and
Eq.~\eqref{eq:robust-detuning-choice}. On functions modulo constants,
the classical perturbation has oscillation norm at most $2wD^2$.
The Poisson inverse is bounded by
\begin{equation}
 \frac{L}{a(1-2rLD^2)},\qquad
 2rLD^2\le\frac{\sigma}{2(1+2t)}<1/8.
 \label{eq:robust-detuned-classical-inverse}
\end{equation}
Let $\beta_x=(\mathbf1_A(x)-\mathbf1_{\bar A}(x))/(\sqrt2\lambda)$.
Set $b_D(x)=tq_x\mathbf1_A(x)-(1+tq_x)\mathbf1_{\bar A}(x)$.
Solve $Q_D^\dagger f=b_D-S_D$ and center $f$.
Define the Hermitian observable
\begin{equation}
 (O_D)_{xx}=f_x,\qquad
 (O_D)_{x\bar x}=-\frac{ih_x(\beta_x+f_x-f_{\bar x})}{Na-i\Delta_x}.
 \label{eq:robust-detuned-poisson-observable}
\end{equation}
Block substitution gives
\begin{equation}
 (\mathcal D-i[F+V+H_s,\cdot])^\dagger O_D=W_A-S_DI.
 \label{eq:robust-detuned-poisson-identity}
\end{equation}
The diagonal contribution is $-w_x(\beta_x+f_x-f_{\bar x})$;
the off-diagonal identity follows on multiplying by $Na-i\Delta_x$.
Consequently
\begin{equation}
 \begin{aligned}
 \|O_D\|&\le\frac{(1+2t)L}{a(1-2rLD^2)}
 (\tfrac12+\lambda/(Na))+\frac1{\sqrt2Na}\\
 &\le2L/a.
 \end{aligned}
 \label{eq:robust-detuned-poisson-norm}
\end{equation}
Use $t\le\sqrt2/4$, $(1-2rLD^2)^{-1}\le8/7$ and
$NL\ge2(1+\ln2)$ for the last bound.

\paragraph{Physical local noise and additional errors.}
Suppose $U^\dagger H_{\rm phys}U=F+H_s$, $\epsilon=\|U-I\|$,
and add physical $V+K$, with arbitrary Hermitian $K$.
Let $\delta_{\mathcal D}\ge\|\mathcal D_U-\mathcal D\|_{1\to1}$ and
$\delta_V\ge\|U^\dagger VU-V\|$.
The stationary Poisson identity gives
\begin{equation}
 \langle W_A\rangle_{\rm phys}\ge3\sigma/4-\sqrt3\epsilon
 -(2L/a)(\delta_{\mathcal D}+2\delta_V+2\|K\|).
 \label{eq:robust-local-interface}
\end{equation}
In particular, $\delta_{\mathcal D}\le8Na\epsilon$ and
$\delta_V\le2\epsilon\|V\|$ yield
\begin{equation}
 \begin{aligned}
 \langle W_A\rangle_{\rm phys}\ge3\sigma/4
 &-(16NL+\sqrt3)\epsilon\\
 &-8L\epsilon\|V\|/a-4L\|K\|/a.
 \end{aligned}
 \label{eq:robust-physical-general}
\end{equation}
A positive right side divided by $\sqrt3$ lower-bounds $d_N$ and transfers
the same bounded test to actual physical copies.

\paragraph{Boundary-source tolerances.}
Use
\begin{equation}
 \begin{aligned}
 N&=\lceil256v^{-2}\ln(16/v)\rceil,\quad r=v^2/2^{16},\\
 t&=v/(256\sqrt N),\quad\lambda=av\sqrt N/(256\sqrt2),\quad\sigma=t/8.
 \end{aligned}
 \label{eq:robust-boundary-source-parameters}
\end{equation}
The detuning condition holds for $D\le D_*$, where
\begin{equation}
 D_*=[32+(1+2t)v\sqrt N\,L/8]^{-1/2}.
 \label{eq:robust-boundary-D}
\end{equation}
For $V=\sum_i h_iZ_i$, sufficient worst-case field bounds are
\begin{equation}
 \sum_i|h_i|\le NaD_*/2,\quad\text{or sitewise }|h_i|\le aD_*/2.
 \label{eq:robust-boundary-fields}
\end{equation}
The sum form also covers sparse fields. Since $D_*\le1/\sqrt{32}$,
the simultaneous choices
\begin{equation}
 \epsilon\le t/(1024NL),\qquad \|K\|\le at/(512L)
 \label{eq:robust-boundary-joint-tolerances}
\end{equation}
retain
\begin{equation}
 \langle W_A\rangle_{\rm phys}\ge t/16,\qquad
 d_N(\rho_*^{\rm phys})\ge t/(16\sqrt3).
 \label{eq:robust-boundary-result}
\end{equation}
Indeed the dressing term is at most $19NL\epsilon$, and
$3t/32-19t/1024-t/128>t/16$.
Here $D_*=\Theta(L^{-3/4})$ as $v\downarrow0$.
These specify sufficient absolute tolerances: generic static errors satisfy
$\|K\|\le av/(2^{17}\sqrt N L)$, or each of $M$ Pauli coefficients may
be bounded by $at/(512LM)$.

The local field-coordinate estimate in
Eq.~\eqref{eq:robust-local-field-coordinates} follows by integrating
$\|[T,V]\|\le(4\eta/3)\sum_i|h_i|$ and adding the remainder lift.
Together with Eqs.~\eqref{eq:stage3-local-Dtotal} and
\eqref{eq:robust-boundary-fields}, the full loss in
Eq.~\eqref{eq:robust-local-interface} is at most
\begin{equation}
 35NL\eta+32768N^3L\eta^2+4L\|K\|/a.
 \label{eq:robust-local-total-loss}
\end{equation}
Thus
\begin{equation}
 \eta=t/(4096NL),\quad \|K\|\le at/(512L),\quad
 \sum_i|h_i|\le NaD_*/2
 \label{eq:robust-local-joint-choice}
\end{equation}
retains Eq.~\eqref{eq:robust-boundary-result} with the improved coherent cost.
The explicit size gives $Nt\le L$, so the quadratic loss is at most $t/512$
and the complete loss at most $75t/4096<t/32$;
$\eta\le1/(2048N)$ also holds. A smaller $\eta$ can impose the separate
finite-preparation condition.

\paragraph{Relative precision.}
The physical scale is $\Lambda=\lambda/s_0$, with
$s_0\ge\eta^N/3^{N-1}$. A sufficient relative longitudinal tolerance is
\begin{equation}
 |h_i|/\Lambda\le\frac{128\sqrt2D_*}{v\sqrt N}\frac{\eta^N}{3^{N-1}}.
 \label{eq:robust-relative-longitudinal}
\end{equation}
Thus a sufficient number of fractional binary digits is
\begin{equation}
 b_Z=\left\lceil N\log_2(1/\eta)+(N-1)\log_2 3
 +\log_2\frac{v\sqrt N}{128\sqrt2D_*}\right\rceil_+,
 \label{eq:robust-longitudinal-bits}
\end{equation}
where $\lceil x\rceil_+=\max\{0,\lceil x\rceil\}$; rounding may add a safety bit.
For each of $M$ unrestricted coefficients it suffices to use
\begin{equation}
 b_K=\left\lceil N\log_2(1/\eta)+(N-1)\log_2 3
 +\log_2(256\sqrt2NLM)\right\rceil.
 \label{eq:robust-generic-bits}
\end{equation}
Both sufficient resolutions are $O(N\ln N)=O(v^{-2}\ln^2(1/v))$
for the stated polynomial $\eta^{-1}$.

\paragraph{Earlier small-drive source.}
For $\lambda=a/(16\sqrt2L)$, $t=1/(16NL)$ and $r=1/(256NL^2)$,
comparison directly with $\pi$ gives $\|P_D-\pi\|_{\rm TV}\le rL=t/16$.
If $D\le1/2$ and $Nv^2\ge32\ln N$, then $q_x\ge4/5$ and
\begin{equation}
 \begin{aligned}
 S_D&\ge\tfrac45t(1-N^{-4})-(1+t)N^{-36}
 -(\tfrac45+1+t)t/16\\
 &\ge t/2=1/(32NL).
 \end{aligned}
 \label{eq:robust-old-source-Oa-fields}
\end{equation}
Thus $\sum_i|h_i|\le Na/4$, or $|h_i|\le a/4$ sitewise, suffices
for this small-drive source. The entropy baseline with diagonal $F+V$
retains $A_N\le5L/a$, and the stationary identity yields
\begin{equation}
 \tfrac12\|\rho_{\rm phys}-\rho_D\|_1
 \le(1+4NaA_N+2A_N\|V\|)\epsilon+A_N\|K\|.
 \label{eq:robust-old-fullstate}
\end{equation}
With $\epsilon\le1/(4096N^2L^2)$, $\|V\|\le Na/4$ and
$\|K\|\le a/(2048NL^2)$, this is at most
$(24/4096+5/2048)/(NL)<1/(64\sqrt3NL)$, retaining physical margin $1/(64NL)$.
All inverses here act on traceless Hermitian inputs.
At $t_0=[\ln(2D_{\max})+\ln16]/a\le4L/a$, entropy contraction and
Duhamel give physical trace contraction at most
$1/4+1/(2\sqrt2)+13/(1024NL)<3/4$.
Drive calibration up to $4/3$ replaces $1/(2\sqrt2)$ by $2/(3\sqrt2)$
and preserves the bound. This small-drive full-state preparation takes
$O(L\ln(1/\varepsilon)/a)$, independently of the large diagonal Ising norm.

%% file: proofs/preparation_time.tex
% Source: research_stage3/source/mask_volterra_mixing.tex; condensed.
\section{Finite-time preparation and full-state relaxation}
\label{sec:mask-volterra-prep}
Let $\mathcal L=\mathcal D-i[F+H_s,\cdot]$, where $F$ is any diagonal
Hamiltonian, $L=1+\ln N$, $r=2\lambda^2/(Na^2)$, and
\begin{equation}
 8rL\le1/4,\qquad \lambda/(Na)\le1/16.
 \label{eq:volterra-conditions}
\end{equation}
The boundary-source parameters satisfy these conditions.

\paragraph{Mixing in the diagonal--complement algebra.}
For a traceless Hermitian difference, let $p(t),c(t)$ denote the trace norms
of its diagonal and full-complement parts. Under
$\mathcal L_0=\mathcal D-i[F,\cdot]$ these contract by
$k(t)=\min\{1,Ne^{-at}\}$ and $e^{-Nat}$, respectively.
The source commutator exchanges them with norm at most $2\lambda$.
Duhamel therefore gives
\begin{equation}
 \begin{aligned}
 p(t)&\le k(t)p(0)+2\lambda\int_0^tk(t-s)c(s)\,ds,\\
 c(t)&\le e^{-Nat}c(0)+2\lambda\int_0^te^{-Na(t-s)}p(s)\,ds.
 \end{aligned}
 \label{eq:volterra-system}
\end{equation}
For $\beta=a/(4L)$, the weighted kernel bounds are
$\|e^{\beta t}k\|_\infty\le2$,
$\|e^{\beta t}k\|_1\le2L/a$ and
$\|e^{\beta t}e^{-Nat}\|_1\le2/(Na)$.
For the middle bound split at $\ln N/a$ and use
$e^{\beta\ln N/a}\le e^{1/4}<4/3$,
$[1-1/(4L)]^{-1}\le6/5$.
Substitution in the feedback convolution gives norm at most $8rL$.
Weighted suprema, first on finite intervals, satisfy
\[
 p_\beta\le\frac{2p(0)+8\lambda c(0)/(Na)}{1-8rL},\qquad
 c_\beta\le c(0)+\frac{4\lambda}{Na}p_\beta.
\]
Their sum is at most $5[p(0)+c(0)]$.
Both mask projections have induced norm at most one, proving
\begin{equation}
 \|e^{t\mathcal L}X\|_1\le10e^{-at/(4L)}\|X\|_1
 \label{eq:volterra-X-mixing}
\end{equation}
inside this algebra, with arbitrary diagonal detuning phases.

\paragraph{Preparation from a free input.}
Both $I/2^N$ and the ideal stationary state belong to the algebra, so a
width-$\sqrt3$ witness satisfies
$|\operatorname{Tr}[W_A(\rho_t-\rho_*)]|\le10\sqrt3e^{-at/(4L)}$.
For stationary margin $S_{\min}>0$, choose
\begin{equation}
 B=\ln(40\sqrt3/S_{\min}),\qquad T=4LB/a.
 \label{eq:free-prep-time}
\end{equation}
The ideal time error is at most $S_{\min}/4$.
For the physical two-body device, let $\epsilon_U=\|U-I\|$ and
$\delta_{\mathcal D}=\|\mathcal D_U-\mathcal D\|_{1\to1}$.
The common input $I/2^N$ has no coordinate error. CPTP contraction and
Duhamel give $\|\rho'_t-\rho_t\|_1\le\delta_{\mathcal D}t$;
returning to physical coordinates adds witness error at most $\sqrt3\epsilon_U$.
Thus the sufficient choices
\begin{equation}
 \delta_{\mathcal D}\le\frac{aS_{\min}}{16\sqrt3LB},\qquad
 \epsilon_U\le\frac{S_{\min}}{8\sqrt3}
 \label{eq:finite-prep-dressing}
\end{equation}
bound the two physical errors by $S_{\min}/8$ each and retain
$\operatorname{Tr}(W_A\rho_T^{\rm phys})\ge S_{\min}/2$.
For $S_{\min}=v/(2048\sqrt N)$ and
$N=\lceil256v^{-2}\ln(16/v)\rceil$, this gives
$\gamma T=O(\ln^2N)=O(\ln^2(1/u))$.
The local bridge allows $\eta$ of order $S_{\min}/(NLB)$;
the same test amplifies actual independently prepared finite-time outputs.
For a detuned ideal source with a separately proved margin, replace
$\delta_{\mathcal D}$ by $\delta_{\mathcal D}+2\delta_V+2\|K\|$.
The diagonal field norm itself does not enter the Volterra estimate.

\paragraph{Simultaneous robust preparation.}
For $t=v/(256\sqrt N)$ and detuned margin at least $3t/32$, take
$B=\ln(320\sqrt3/t)$, $T=4LB/a$ and
\[
 \begin{gathered}
 \eta=\frac{t}{2^{15}NLB},\qquad
 \|K\|\le\frac{at}{2^{16}LB},\\
 \sum_i|h_i|\le NaD_*/2,\quad D_*\le1.
 \end{gathered}
\]
Use $\epsilon_U\le4N\eta+1536N^2\eta^2$,
$\delta_{\mathcal D}\le16Na\eta+12288N^3a\eta^2$ and
$\delta_V\le(4\eta/3+3072N^2\eta^2)\sum_i|h_i|$.
Since $Nt\le L$, $t\le1/2$ and $LB\ge1$, these imply
\[
 \begin{aligned}
 \epsilon_U&\le t/(4096LB),\quad \delta_{\mathcal D}/a\le t/(1024LB),\\
 \delta_V/a&\le t/(32768LB),\\
 \frac{\delta_{\mathcal D}+2\delta_V+2\|K\|}{a}&\le\frac{35t}{32768LB}.
 \end{aligned}
\]
The ideal time loss is at most $t/32$; the physical loss is below
$[70\sqrt3/32768+\sqrt3/(4096LB)]t<t/128$.
Thus the actual margin exceeds $t/32$, jointly with the stated field errors,
original local noise and free-input preparation time.

\paragraph{Arbitrary initial states.}
The mask projector
$\Pi_SX=2^{-N}\sum_z(-1)^{z\cdot S}Z^zXZ^z$
selects $x\oplus y=S$, preserves Hermiticity and has induced norm at most one.
On mask $S$, $\mathcal L_0$ contracts at rate $a|S|$: sites in $S$
annihilate the mask on replacement; other replacements and $F$ give a CPTP contraction.
The source couples only $S$ and $\bar S$. For $1\le k=|S|<N$,
their norms are bounded by the positive comparison matrix
$M_k=\left(\begin{smallmatrix}-ak&2\lambda\\2\lambda&-a(N-k)\end{smallmatrix}\right)$.
If also $\lambda\le a\sqrt N/8$, then $M_k\le-(a/2)I$:
its shifted determinant is nonnegative since
$4\lambda^2\le a^2N/16\le a^2(N/2-3/4)$.
Summing mask pairs gives
\begin{equation}
 \begin{aligned}
 \|e^{t\mathcal L}X\|_1\le{}&[10e^{-at/(4L)}\\
 &+\sqrt2(2^N-2)e^{-at/2}]\|X\|_1.
 \end{aligned}
 \label{eq:full-mask-mixing}
\end{equation}
At $T_0=20N/a$ the bracket is below $1/16$ for $N\ge2$
(the first term is maximal at $N=2$, below $.028$; the second is below $1.3\times10^{-8}$).
If $\delta_{\mathcal D}+2\delta_V+2\|K\|\le a/(160N)$,
Duhamel adds at most $1/8$ per block, giving contraction at most $1/4$.
Hence, for every input,
$d(\rho_t^{\rm phys},\rho_*^{\rm phys})\le
(1/4)^{\lfloor at/(20N)\rfloor}$.
This full-state bound has a different size dependence from the free-input resource preparation time.

%% file: proofs/dressing_barrier.tex
% Source: research_stage3/architecture/global_closeness_barrier.tex; condensed.
\subsection{Cost of the globally close dressing strategy}
\label{sec:stage3-dressing-barrier}
\begin{proposition}
Consider an open Ising chain with $1\le J_i\le B$, where $B$ is size
independent, and onsite $0<\eta_i\le1/4$. Suppose a Gaussian $U$ exactly
maps its Majorana matrix to the original dimer and edge planes, with the
unique small singular coefficient on the edge plane and
$\varepsilon=\|U-I\|\le1/16$. Then
\begin{equation}
 \max_i\eta_i\le8B\varepsilon.
 \label{eq:stage3-method-eta-barrier}
\end{equation}
Amplifying the edge coefficient to $\lambda$ requires per-site budget
\begin{equation}
 J_{\rm budget}/\lambda\ge[3/(32B\varepsilon)]^N.
 \label{eq:stage3-method-J-barrier}
\end{equation}
Thus $\varepsilon\le N^{-c}$ implies
$\ln(J_{\rm budget}/\lambda)\ge cN\ln N-O(N)$ within this strategy.
\end{proposition}
\begin{proof}
For the orthogonal Majorana action $O$ and real unit $v$,
$\|(O-I)v\|_2=\|U^\dagger c(v)U-c(v)\|\le2\varepsilon$.
The onsite entry $(A_i,B_i)$ links distinct original planes and vanishes
in $\mathsf A_{\rm eff}=O^T\mathsf A O$. Hence
$\eta_i\le2\|\mathsf A\|\|O-I\|\le4(B+1/4)\varepsilon\le8B\varepsilon$.
Every bulk singular coefficient is at least $3/4$ by Weyl, so the unique-matching
Pfaffian identity gives
$s_0=\prod_i\eta_i/\prod_j\widetilde J_j
\le(4/3)^{N-1}(8B\varepsilon)^N\le(32B\varepsilon/3)^N$.
The physical scale $\Lambda=\lambda/s_0$ contributes at least $\Lambda$
to a site's budget because an unscaled bond is at least one.
\end{proof}

For uniform $J_i=1$, the bulk frequencies are degenerate at $\eta=0$.
On the $A_2,\ldots,A_N$ bulk subspace, the first-order term of
$\mathsf A_\eta\mathsf A_\eta^T$ is $\eta$ times the path adjacency matrix.
Its sine-wave modes have squared coordinate overlap at most $2/N$;
for fixed $N\ge3$, exact modewise dimerization therefore requires a rotation
that stays away from the identity as $\eta\downarrow0$.
Keeping this degenerate mixing instead connects, already at $N=3$,
$|001\rangle$ and $|011\rangle$, whose Ising energies agree but reset-product
weights differ. It fails to commute with $\Omega$ and source populations,
so the preceding entropy-resolvent argument does not apply.
At criticality $\eta=J$, the smallest frequency is
$2J\sin[\pi/(4N+2)]=\Theta(J/N)$, but the extended normal-form rotation
also transforms local physical noise into nonlocal channels.
A polynomial-budget construction using that spectrum requires an open-system
source analysis for those transformed channels. The proposition constrains
the globally close dressing strategy, leaving other static two-body architectures open.